\documentclass[11pt]{article}
\RequirePackage{amsthm,amsmath}
\RequirePackage{natbib}
\usepackage{amsgen,amsmath,amstext,amsbsy,amsopn,amssymb,latexsym,float}
\usepackage{array}
\usepackage{multirow}\usepackage{graphicx}\usepackage{caption}
\usepackage{subcaption}
\usepackage{amssymb}
\usepackage{hyperref}
\hypersetup{hidelinks}
\usepackage{placeins}
\usepackage{mathtools}

\DeclareMathOperator*{\argmin}{arg\,min}
\DeclareMathOperator{\supp}{supp}
\usepackage{bm}
\usepackage{bbm}
\usepackage{booktabs}
\usepackage{adjustbox}
\usepackage{algorithm}
 \usepackage{algpseudocode}
 \usepackage{algorithmicx}
 \algdef{SE}[DOWHILE]{Do}{doWhile}{\algorithmicdo}[1]{\algorithmicwhile\ #1}
\algrenewcommand\algorithmicrequire{\textbf{Input:}}
\algrenewcommand\algorithmicensure{\textbf{Output:}}

 \usepackage[dvipsnames,table,dvipsnames*, svgnames*]{xcolor}
  
\newcommand{\E}{\mathbb{E}}
\newcommand{\prob}{\mathbb{P}}
\newcommand{\I}{\mathbb{I}}
\newcommand{\Var}{{\rm Var}}

\newcommand{\tr}{^{\mkern-1.5mu\mathsf{T}}}

\newcommand{\given}{\;\middle|\;}
\newcommand{\dto}{\overset{d}{\to}}
\newcommand{\pto}{\overset{p}{\to}}

\newcommand{\bbmd}{\mathbbm{d}}
\newcommand{\bbmb}{\mathbbm{b}}
\newcommand{\calH}{\mathcal{H}}
\newcommand{\Hhat}{\widehat{\calH}}
\newcommand{\thetad}{\theta^{\bbmd}}
\newcommand{\taud}{\tau^{\bbmd}}
\newcommand{\thetal}{\theta_l}
\newcommand{\thetau}{\theta_u}
\newcommand{\thetain}{\theta^{\rm in}}
\newcommand{\thetainH}{\thetain(\calH)}

\newcommand{\thetaout}{\theta^{\rm out}}
\newcommand{\thetaoutH}{\thetaout(\calH)}

\newcommand{\thetahat}{\widehat{\theta}}

\newcommand{\sigmahat}{\widehat{\sigma}}
\newcommand{\bfa}{\mathbf{a}}
\newcommand{\bfx}{\mathbf{x}}
\newcommand{\bfA}{\mathbf{A}}
\newcommand{\bfX}{\mathbf{X}}
\newcommand{\bfc}{\mathbf{c}}
\newcommand{\bfd}{\mathbf{d}}
\newcommand{\bfh}{\mathbf{h}}
\newcommand{\bfu}{\mathbf{u}}
\newcommand{\bfC}{\mathbf{C}}

\newcommand{\bfAd}{\bfA^{\bbmd}}
\newcommand{\bfW}{\mathbf{W}}
\newcommand{\bfWd}{\mathbf{W}^{\bbmd}}
\newcommand{\bfw}{\mathbf{w}}
\newcommand{\bfwd}{\mathbf{w}^{\bbmd}}
\newcommand{\bfCd}{\mathbf{C}^{\bbmd}}
\newcommand{\bfcd}{\mathbf{c}^{\bbmd}}
\newtheorem{Theorem}{Theorem}
\newtheorem{Corollary}{Corollary}
\newtheorem{Proposition}{Proposition}
\newtheorem{Lemma}{Lemma}

\newtheorem{Definition}{Definition}

\newtheorem{Assumption}{Assumption}

\usepackage{authblk}
\title{Causal Effects of Modified Treatment Policies under Positivity Violations: Partial Identification and Inference}
\author[1]{Taehyeon Koo\thanks{Correspondence to Taehyeon Koo (\texttt{tk3077@cumc.columbia.edu}). Rudolph and Miles are co-senior authors. This work was funded by   R01DA056407 to investigator Rudolph from the National Institute on Drug Abuse, R01DA059824 to Miles and Rudolph from the National Institute on Drug Abuse, and R01AG069090 to investigator Torres from the National Institute on Aging.
The authors are grateful to Jacqueline Torres and Lucia Calderon for their guidance in using and interpreting the CHAMACOS data, and gratefully acknowledge Hyunseung Kang, Dana Goin, Oliver Hines, and participants in the Causal Inference Learning Group Seminar at the Columbia University Mailman School of Public Health for helpful feedback. No potential conflict of
interest was reported by the authors.}}
\author[2]{Elizabeth A. Stuart}
\author[1]{Kara E. Rudolph}
\author[1]{Caleb H. Miles}
\affil[1]{Columbia University Mailman School of Public Health}
\affil[2]{Johns Hopkins Bloomberg School of Public Health}
\date{}
\begin{document}
\maketitle
\begin{abstract}
Multivariate treatments are common, particularly in environmental epidemiology. Common causal effects of interest measure how expected outcome distributions would change under hypothetical interventions shifting the exposure mixture up or down. Such interventions are examples of modified treatment policies (MTPs), {which depend on the natural value of treatment and possibly covariates}. {A key condition for nonparametrically identifying MTP effects is positivity: observed data support MTP-generated treatment values given covariates.} However, multivariate treatments often lead to positivity challenges, which we propose to address with partial identification. We decompose the estimand into terms based on portions inside and outside a fixed region of adequate positivity. The inside portion is point identified. We bound the unsupported portion assuming the conditional mean potential outcome is Lipschitz continuous, treating the Lipschitz constant as a sensitivity parameter.
{We use smooth approximations to address non-regular inference challenges without changing the estimand, yielding influence functions, efficiency conditions, and policy-mean confidence intervals.}
In simulations, our confidence intervals attain nominal coverage where those from a naive approach assuming positivity undercover. We analyze pesticide exposure among a cohort of women in California's Salinas Valley. The proposed and naive analyses yield different conclusions about whether a poorly supported pesticide-exposure shift affects hypertension risk.

\end{abstract}

\vfill

\newpage
\section{Introduction}
{Multivariate treatments are common, particularly in environmental epidemiology, where individuals are exposed to mixtures of pollutants \citep[e.g.,][]{carlin2013unraveling,bobb2015bayesian}. A common causal question is how outcomes would change under hypothetical interventions that shift one or more components of the mixture up or down.
Such shifts modify each individual's natural value of treatment \citep{robins2004effects,sarvet2026natural}.
{For a single time point, as considered here, the natural value of treatment is the treatment's observed value.} A modified treatment policy (MTP) maps this natural value,  possibly together with covariates, to a policy-determined value \citep{diaz2012population,haneuse2013estimation,young2014identification}.}
{MTPs have been developed and applied across a range of settings, including longitudinal settings \citep{diaz2023nonparametric}, mediation analysis \citep{diaz2020causal}, and network interference \citep{balkus2026causal}.}

Positivity is a standard sufficient condition for nonparametrically identifying the mean outcome under an MTP while avoiding extrapolation. Under positivity, treatment values generated by the policy are represented by the observed treatment distribution at the same covariate values \citep{diaz2012population,haneuse2013estimation}. Because positivity concerns the joint treatment--covariate distribution, {it becomes less likely to hold as the number of exposures or covariates increases.\footnote{Throughout the paper, we focus on moderate-dimensional settings. Our aim is to address positivity violations arising from multivariate exposures and covariates, not to overcome the curse of dimensionality.}
Even with few variables, policies that change only some highly correlated components can generate sparsely represented or unsupported combinations \citep{antonelli2024causal}.}
This challenge arises in our application, where five of the seven pesticide classes that make up our exposure mixture have pairwise correlations between 0.70 and 0.88 (Supplementary Figure~\ref{fig: CHAMACOS corr t3}).

Positivity problems arise in two forms: structural and practical. A structural positivity violation occurs when a policy-generated value lies outside the support of the population conditional distribution of exposures given covariates. The conditional mean of the outcome given the exposures and covariates is not defined at such points. Even when a policy-generated value lies within the conditional support, a practical positivity problem can arise when few observations lie nearby. The conditional mean remains formally identified but can be difficult to estimate reliably with so few observations. Structural positivity violations mean that the causal estimand is not identified by the observed data distribution, and practical positivity violations mean that the causal estimand is only weakly identified, compromising estimation and inference \citep{robins2007comment}.

Some approaches address positivity problems by restricting inference to a population with better support, sometimes described as ``moving the goalposts'' \citep{crump2009dealing}, or by redefining the intervention to generate only adequately supported values \citep{kennedy2019nonparametric,bao2025addressing}. Such changes in the estimand inherently change the scientific question, and one then needs to consider the extent to which the revised target estimand remains scientifically meaningful.  Herein, we instead retain both the original population and original hypothetical MTP and ask what can be concluded based on the observed data.
\subsection{Results and Contributions}
In this article, we develop a partial identification approach for the counterfactual mean outcome under an MTP that may generate values outside a prespecified region of adequate positivity, denoted $\calH$. We decompose this target estimand into portions based on the supported (inside $\calH$) and unsupported regions. The inside portion is point identified.
{If all policy-generated values lie in $\calH$, the target is point identified under positivity.}
For the unsupported portion, i.e., the portion outside $\calH$, we invoke a Lipschitz continuity assumption that restricts the extent to which {the conditional mean potential
outcome} can vary across the extrapolation region.
Using points inside $\calH$ as anchors, {where this mean is point identified, we bound the conditional mean potential outcome at each policy-generated value outside $\calH$.}
{Taking expectations of the resulting lower and upper bounds yields bounds for the unsupported portion.}
{We combine these bounds with the identified inside portion to obtain a partial-identification interval for the target estimand.}

Under a structural positivity violation, the resulting interval contains every value of the estimand that is compatible with the observed data distribution
and the Lipschitz restriction. Under a practical positivity violation, the estimand remains point identified, but we deliberately report a partial-identification interval to not rely on conditional-mean outcome estimates from areas of sparse support. Instead, the interval's endpoints use observed-data information only where positivity is adequate.
For each policy-generated value outside $\calH$, every anchor inside $\calH$ yields an interval containing the conditional mean potential outcome at that outside value. Marginalizing the intersected endpoints and adding the inside portion yields the sharp interval based on the information within $\calH$ and the Lipschitz restriction.
 {The sharp endpoints can fail to be pathwise differentiable.} We therefore consider the simpler class of intervals that use one anchor for each outside value.  The nearest anchor, obtained by projecting the outside value onto $\calH$, gives the narrowest interval in this class.
We call this nearest-point map the hard projection. The resulting interval is not always sharp, but its width has a direct geometric interpretation. Supplementary Material~\ref{append: sharpness} quantifies its gap from the sharp interval.

The hard projection sends outside, unsupported values to a lower-dimensional boundary, concentrating their anchors on a set
that has probability zero under the observed treatment--covariate distribution. Supplementary Proposition~\ref{prop: hard nondiff} shows that the hard-projection endpoints {can fail to be} pathwise differentiable, {precluding regular root-$n$ inference}. We address this obstacle by instead using an interior-displaced projection that moves each boundary projection into $\calH$ by an amount that depends on the exterior value's distance from the boundary. Exterior values sharing the same boundary projection are therefore assigned different anchors in a full-dimensional interior layer (Figure~\ref{fig: projection}).

Under suitable conditions, the interior-displaced projection restores pathwise differentiability of the bound endpoints. For a fixed region $\calH$, we derive the influence functions of the bound endpoints and characterize when these influence functions are efficient. We use them to construct cross-fitted one-step estimators and establish asymptotic normality.
The interior displacement level trades interval width against sampling precision; smaller displacements tighten the bounds but can make estimation less stable because the anchors occupy a thinner interior layer. Because the value that best balances these goals is unknown, we construct confidence intervals (CIs) with simultaneous coverage over a prespecified finite grid of displacement levels using the multiplier bootstrap \citep[e.g.,][]{kennedy2019nonparametric,susmann2025non}; see Section~\ref{subsec: estimators}.

Our theory conditions on a fixed region of adequate positivity, $\mathcal{H}$, which would be unlikely to be known in advance, and so would be constructed from data. Section~\ref{sec: practical} provides a construction and the conditions under which the fixed-region results continue to apply. In Section~\ref{sec: sim}, we study finite-sample performance under structural and practical positivity violations. The CIs estimated from our proposed method attain at least nominal coverage in settings where a standard estimator that naively assumes positivity undercovers. Section~\ref{sec: real data} then re-estimates the effect of hypothetical reductions in pesticide exposures on maternal hypertension risk in the Center for the Health Assessment of Mothers and Children of Salinas (CHAMACOS) cohort in California \citep{rudolph2026everything}.

To summarize, our contributions are as follows:
\begin{itemize}
    \item[(1)] We derive partial-identification bounds for the mean outcome under an MTP for which positivity would be violated or nearly violated {under a Lipschitz continuity assumption on the conditional mean potential outcome}.     \item[(2)] {To address non-regular inference challenges posed by hard projection, we use a smooth approximation to construct confidence intervals for the original policy mean.}\end{itemize}

\subsection{Related Work}
\noindent\textbf{Positivity violations and alternative targets.}
Existing approaches to estimation in the presence of positivity violations often change the target population or intervention. Trimming restricts the target to a subpopulation in which positivity is adequate \citep{crump2009dealing,branson2023causal}. {Incremental interventions define alternative targets by exponentially
tilting the observed conditional treatment distribution, allowing
identification without the usual positivity assumption \citep{kennedy2019nonparametric,schindl2026incremental,huang2026multivariate}.} Feasible interventions use projection to replace each poorly supported policy value with a nearby value inside a region of adequate positivity \citep{antonelli2024causal,bao2025addressing}. In contrast, we retain the original population and policy, and we use projection only to select, for each policy-generated value outside the supported region, the nearest anchor from which we generate partial identification bounds.

\noindent\textbf{Smoothness restrictions and extrapolation.}
Related methods address extrapolation without imposing a parametric outcome model by {restricting conditional mean potential outcomes}. For binary treatments, \citet{armstrong2021finite} impose smoothness or shape restrictions over the covariate space and derive finite-sample optimal linear estimators and CIs calibrated to the maximum bias allowed by the prespecified function class. {Building on their approach, \citet{ma2025sensitivity} use standard inference for the portion retained after propensity-score trimming and bias-aware inference for the excluded portion, calibrated to the largest bias allowed by a Lipschitz continuity assumption.}
{Lipschitz restrictions have also been used in policy learning and evaluation with finitely many treatment levels.
\citet{ben2025safe} learn policies by optimizing lower bounds on policy performance under Lipschitz restrictions over covariates. They derive finite-sample bounds on the learned policy’s population performance relative to the policy under which outcomes were observed and an optimal policy.}
Most similar to our approach, \citet{khan2024off} derive sharp {bounds for the mean potential outcome under a given policy}. They develop point estimators attaining the minimax rate, {but do not address uncertainty quantification for these bounds.}
\citet{pfister2026extrapolation} {derive related extrapolation bounds under restrictions on directional derivatives, including higher orders. They address uncertainty quantification but do not provide inference for nonsmooth bounds.}
Our work differs from this set of prior work in that we study mean outcomes under MTPs for continuous, possibly multivariate treatments. We impose a Lipschitz restriction over the joint treatment--covariate space, characterize the corresponding sharp interval, and develop influence-function-based inference for the interior-displaced projection bounds.

\noindent\textbf{Restoring pathwise differentiability.}
\citet{yang2018asymptotic} and \citet{branson2023causal} smooth propensity-score trimming indicators for binary and continuous treatments, respectively.  \citet{susmann2025non} smooth propensity-score threshold indicators that separate the point-identified and bounded contributions to ATE bounds. \citet{levis2025covariate} use margin conditions or smooth approximations for the pointwise extrema in instrumental-variable bounds. \citet{bibaut2017data} give a general approach for approximating a target that is not pathwise differentiable by using a family of pathwise differentiable targets indexed by a smoothing scale.
 {As in \citet{susmann2025non}, we construct pathwise differentiable bounds that still contain the original causal estimand. However, we modify the projection itself by moving anchors from the boundary into the interior of $\mathcal H$, rather than smoothing an indicator or a pointwise extremum.}

\section{Modified Treatment Policies: Identification and Its Challenges}\label{sec: MTP}\label{subsec: id with pos}\label{subsec: challenge pos}
Suppose that $n$ independent observations of $O=(Y,\bfA,\bfX)\sim P$ are available, where $Y\in\mathbb{R}$ is an outcome, $\bfA\in\mathbb R^q$ is a vector of continuous treatments (i.e., the naturally occurring treatment vector value), and $\bfX\in\mathbb R^{d_X}$ contains pre-treatment covariates. A modified treatment policy is a prespecified map $\bbmd$ that assigns the treatment $\bfAd=\bbmd(\bfA,\bfX)$ \citep{diaz2012population,haneuse2013estimation}. Examples include a proportional reduction, $\bfAd=\kappa\bfA$ for $0<\kappa<1$, or a policy that changes one component of a mixture while leaving the others at their naturally occurring values.

{Let} $Y(\bfa)$ denote the potential outcome under treatment $\bfa$. Our primary target is the mean outcome under the policy, $\thetad = \E[Y(\bfAd)]$, and we also consider its contrast with the observed outcome mean, $\taud = \E[Y(\bfAd)-Y] = \thetad-\E[Y]$.
Throughout, we write $\bfW=(\bfA\tr,\bfX\tr)\tr$ and $\bfWd=((\bfAd)\tr,\bfX\tr)\tr$. We split the covariates into continuous and discrete parts, $\bfX=(\bfX_C\tr,\bfX_D\tr)\tr$, and collect all continuous variables except for $Y$ in $\bfC=(\bfA\tr,\bfX_C\tr)\tr\in\mathbb R^p$, where $\bfX_C\in\mathbb R^{d_C}$ and $p=q+d_C$. Its policy-generated counterpart is $\bfCd=((\bfAd)\tr,\bfX_C\tr)\tr$, so that $\bfWd=((\bfCd)\tr,\bfX_D\tr)\tr$.

Identification of $\thetad$ proceeds in two steps. First, the following two causal assumptions connect potential outcomes to the observed data distribution.
\begin{Assumption}\label{assump: basic}
(i) Consistency: if $\bfA=\bfa$, then $Y=Y(\bfa)$. (ii) {Conditional mean exchangeability: $\mu(\bfA,\bfX)=\E[Y\mid\bfA,\bfX]$ and $\E[Y(\bfAd)\mid\bfA,\bfX]=\mu(\bfAd,\bfX)$, both almost surely, where $\mu(\bfa,\bfx)=\E[Y(\bfa)\mid\bfX=\bfx]$.}
\end{Assumption}
Consistency connects observed and potential outcomes, and would fail under interference, where one unit's outcome responds to another's treatment. {Conditional mean exchangeability} holds by design in a {randomized} trial but requires {appropriate} covariates in an observational study.

Second, positivity provides a standard route to identifying the MTP mean
without relying on extrapolation of the outcome regression \citep[e.g.,][]{diaz2012population,haneuse2013estimation,young2014identification}. \begin{Assumption}[MTP positivity]\label{assump: MTP pos}
{If $(\bfa,\bfx)\in{\rm supp}(\bfA,\bfX)$, then $(\bbmd(\bfa,\bfx),\bfx)\in{\rm supp}(\bfA,\bfX)$.}
\end{Assumption}
Since {conditional mean exchangeability} and positivity {involve the same covariates}, adding covariates may strengthen confounding control while making positivity less likely to hold.
{Assumptions~\ref{assump: basic} and \ref{assump: MTP pos} provide key conditions for identification by the g-formula \citep{haneuse2013estimation}. When the observed distribution uniquely determines the outcome regression at policy-generated values almost surely, this formula gives}
\begin{equation}
    \thetad = \E\!\left[Q(\bfWd)\right]= \E\!\left[Q(\bfAd,\bfX)\right],\quad\text{where}\quad  Q(\bfw) = Q(\bfa,\bfx)=\E\!\left[Y\given\bfA=\bfa,\bfX=\bfx\right],\label{eq: id under pos}
\end{equation}
and $\taud=\E[Q(\bfWd)-Y]$.
We consider the setting in which Assumption \ref{assump: basic} holds but the positivity condition of Assumption~\ref{assump: MTP pos} does not.
Structural and practical positivity problems become more pronounced in the joint space of treatments and covariates, for two related reasons. The first is due to dimension, since each treatment may be well-supported marginally, while the joint distribution concentrates on a narrow part of the space. Intervening on one element can then create an exposure combination the data do not contain---the difficulty \citet{antonelli2024causal} identify for exposure mixtures. Conditioning on several continuous covariates reproduces the problem in a single-treatment setting. The second reason is high correlation among the exposures, which can confine the bulk of the joint exposure distribution to a narrow subset of its domain. Small shifts can then more readily move values outside this subset \citep{d2021overlap}; See Supplementary Figure~\ref{fig: hdr illus} for an illustration.

\section{Partial Identification of MTP Causal Effects}\label{sec: partial id}

In the previous section, we discussed how an MTP can generate unsupported or sparsely represented combinations of covariates and intervened-on treatment values. To retain the original research question and its policy, we introduce a region of adequate positivity, $\calH$, and decompose the expected mean outcome under the policy into
{portions with policy-generated values inside and outside $\calH$.} Because information within $\calH$ alone does not point identify the latter portion, we impose a Lipschitz condition on conditional mean potential outcomes, under which we can recover partial identification. We then derive the sharp partial identification interval as well as an alternative interval based on projections into the nearest point in $\calH$, where the latter is more tractable for regular statistical inference.
\subsection{Decomposition and the Lipschitz continuity assumption}\label{subsec: decomposition}\label{subsec: outside bound}

For each stratum of discrete covariates $\bfd\in\supp(\bfX_D)$, we let $\calH(\bfd)\subseteq\mathbb R^p$ denote a fixed region of adequate positivity in the continuous treatment--covariate space of $\bfC$. The following definition states the requirements for each $\calH(\bfd)$.
\begin{Definition}[Region of adequate positivity]\label{def: positivity region}
    For each $\bfd\in\supp(\bfX_D)$, $\calH(\bfd)$ is a nonempty, closed, {convex subset of $\supp(\bfC\mid\bfX_D=\bfd)$.} \end{Definition}

{We define the region of adequate positivity in the joint treatment--covariate space as} $\calH=\bigcup_{\bfd\in\supp(\bfX_D)}\calH(\bfd)\times\{\bfd\}$. In this section and the next, we treat $\calH$ as known and fixed, so uncertainty from estimating the region does not enter the identification or inference results. Section~\ref{sec: practical} discusses data-driven construction of $\calH$.

The target estimand decomposes into portions of policy-generated values inside and outside $\calH$, $\thetad=\thetainH+\thetaoutH$,
where
\begin{equation}\label{def: theta in and out}
\begin{aligned}
\thetainH=\E\left[Y(\bfAd)\I(\bfWd\in\calH)\right],\quad \thetaoutH=\E\left[Y(\bfAd)\I(\bfWd\notin\calH)\right].
\end{aligned}
\end{equation}

To identify the inside portion, {write} $\bfc=(\bfa\tr,\bfx_C\tr)\tr$ and {$\bfx=(\bfx_C\tr,\bfd\tr)\tr$. In these coordinates, the conditional mean potential outcome is}
\begin{equation}\label{eq: mu}
     \mu(\bfc,\bfd){=\mu(\bfa,\bfx)}=\E\left[Y(\bfa)\mid\bfX_C=\bfx_C,\bfX_D=\bfd\right].
\end{equation}
{Similarly, we write $Q(\bfc,\bfd)=Q(\bfa,\bfx)$.} By the law of iterated expectation and {conditional mean exchangeability}, the inside portion $\thetainH$ equals $\E[\mu(\bfCd,\bfX_D)\allowbreak \I(\bfWd\in\calH)]$. {Assumption~\ref{assump: basic} implies that $\mu(\bfc,\bfd)=Q(\bfc,\bfd)$ almost everywhere under the observed
distribution.
Then for the region $\mathcal H$ in Definition~\ref{def: positivity region}, the identification formula for the inside portion is
}
\begin{equation}\label{eq: id of theta in}
\thetainH
=
\E\left[Q(\bfWd)\I(\bfWd\in\calH)\right].
\end{equation}
{We discuss below a sufficient condition under which the observed distribution uniquely determines both expressions for the inside portion.}
If $\bfWd\in\calH$ almost surely, then $\thetaoutH=0$ and \eqref{eq: id of theta in} recovers the g-formula in \eqref{eq: id under pos}. Otherwise, $\thetainH$ accounts for only part of the mean outcome under the policy.

The outside portion depends on $\mu$ at policy-generated values outside $\calH$. Information about $Q$ within $\calH$ alone does not determine these outside means, so an additional restriction is needed. A global parametric outcome model could supply the outside means by extrapolation, but conclusions would be sensitive to its functional form. We instead impose the following Lipschitz restriction on $\mu$ in \eqref{eq: mu}, relating its values outside $\calH$ to those within the region.
\begin{Assumption}[$\mathbf{V}$-norm Lipschitz continuity]\label{assump: weight Lip}
       For each stratum of discrete covariates $\bfd$, let $L(\bfd)\geq0$ and let $\mathbf{V}$ be diagonal with positive entries. Then, for every \emph{anchor} $\bfc'\in\calH(\bfd)$ and every $\bfc\in\mathbb R^p$,
        $\left|\mu(\bfc,\bfd)-\mu(\bfc',\bfd)\right|\leq L(\bfd)\|\bfc'-\bfc\|_{\mathbf{V}}$, where $\|\bfc'-\bfc\|_{\mathbf{V}}\coloneqq\sqrt{(\bfc'-\bfc)\tr \mathbf{V}(\bfc'-\bfc)}$.
\end{Assumption}

In Assumption~\ref{assump: weight Lip}, {the diagonal matrix $\mathbf{V}$ specifies the coordinate weights in the weighted Euclidean distance  between $\bfc$ and the anchor $\bfc'$. Its diagonal entries can reflect the relative scale or importance of each continuous coordinate.} The constant $L(\bfd)$ limits how much $\mu$ may change per unit distance and serves as a sensitivity parameter. For fixed $\mathbf V$, larger values of $L(\bfd)$ allow more variation in $\mu$. Section~\ref{subsec: practical algorithm} discusses practical choices of $\mathbf V$ and $L(\bfd)$. Although we use a weighted-norm formulation, other Lipschitz restrictions can be imposed; {see Supplementary Assumption~\ref{assump: coord Lip} for a coordinate-wise version.}

{
The continuity imposed by Assumption~\ref{assump: weight Lip} ensures that the observed distribution uniquely determines the inside portion in \eqref{eq: id of theta in}. Without further restrictions, changing $Q$ on a subset of $\calH$ that has probability zero under the observed distribution leaves that distribution unchanged. But if $\bfWd$ falls in that set with positive probability, the change can affect \eqref{eq: id of theta in} \citep[see][Section~2.1]{haneuse2013estimation}.
Under Assumption~\ref{assump: basic}, $\mu$ agrees with $Q$ almost surely under the observed distribution. Because each anchor lies in the observed support, continuity uniquely determines $\mu$ at each anchor from this agreement. These anchor values enter both the expectation below \eqref{eq: mu} and the identification formula in \eqref{eq: id of theta in}.
}

{For the outside portion, we apply Assumption~\ref{assump: weight Lip} with $\mathbf \bfcd\notin\mathcal H(\mathbf d)$. This bounds the difference between the unknown mean $\mu(\bfcd,\bfd)$ and the  observed data distribution conditional mean $Q(\bfc',\bfd)$ at any anchor $\bfc'\in\calH(\bfd)$.
Section~\ref{subsec: sharp projection} uses these comparisons to derive partial-identification bounds for the outside portion.}

{In practice, the Lipschitz restriction used to bound the outside portion may be difficult to justify over the full treatment range. Although our theoretical guarantees rely on Assumption~\ref{assump: weight Lip}, our CIs attained at least nominal coverage in a simulation design where a treatment interaction produced unbounded slopes. One possible explanation is that the restriction held on a bounded region containing nearly all policy-generated values. Supplementary Material~\ref{append: impl details} gives details.}

\subsection{Sharp and projection intervals}\label{subsec: sharp projection}

We now use the pointwise comparisons supplied by Assumption~\ref{assump: weight Lip} to derive a partial-identification interval for $\thetad$. Consider a policy-generated value $\bfwd=(\bfcd,\bfd)$ with $\bfcd\notin\calH(\bfd)$. Any $\bfc'\in\calH(\bfd)$ can serve as a reference point, which we call an anchor. At this anchor, $\mu(\bfc',\bfd)$ is identified by $Q(\bfc',\bfd)$. Assumption~\ref{assump: weight Lip} applied to $\bfc'$ and $\bfcd$ gives
\begin{align}
    \mu(\bfcd,\bfd) \in \left[Q(\bfc',\bfd)-L(\bfd)\|\bfc'-\bfcd
    \|_{\mathbf{V}},\ Q(\bfc',\bfd)+L(\bfd)\|\bfc'-\bfcd
    \|_{\mathbf{V}}\right].\label{eq: bound mu c'}
\end{align}
For a fixed $\bfwd\notin\calH$, Assumption~\ref{assump: weight Lip} places the same conditional mean $\mu(\bfcd,\bfd)$ in the interval \eqref{eq: bound mu c'} for every anchor $\bfc'\in\calH(\bfd)$ such that the intersection of all the intervals also contains $\mu(\bfcd,\bfd)$. Denoting its lower and upper endpoints by $\ell(\bfwd)$ and $u(\bfwd)$, this intersection is
\begin{equation}\label{eq: envelopes main}
\left[
\sup_{\bfc'\in\calH(\bfd)}
\left\{Q(\bfc',\bfd)-L(\bfd)\|\bfc'-\bfcd\|_{\mathbf V}\right\},
\inf_{\bfc'\in\calH(\bfd)}
\left\{Q(\bfc',\bfd)+L(\bfd)\|\bfc'-\bfcd\|_{\mathbf V}\right\}
\right].
\end{equation}
Integrating $\ell$ and $u$ over the outside values gives the sharp interval
\begin{equation}\label{eq: sharp PI main}
    {\rm PI}^{\star}=\left[\thetainH+\E\left\{\ell(\bfWd)\I(\bfWd\notin\calH)\right\},\ \thetainH+\E\left\{u(\bfWd)\I(\bfWd\notin\calH)\right\}\right].
\end{equation}
Under Assumption~\ref{assump: weight Lip}, one possible $\mu$ equals $Q$ on $\calH$ and $\ell$ at every point outside, so the corresponding policy mean attains the lower endpoint in \eqref{eq: sharp PI main}. Replacing $\ell$ with $u$ attains the upper endpoint, so \eqref{eq: sharp PI main} is sharp given $Q$ on $\calH$.
Computing the endpoints in \eqref{eq: sharp PI main} requires optimization over all anchors at each outside value. Inference for these intersection bounds is difficult \citep[e.g.,][]{chernozhukov2013intersection,semenova2025debiased} and we do not pursue here.

To avoid this inferential challenge, we instead construct bounds by selecting one anchor-specific interval from \eqref{eq: bound mu c'} at each outside value. Let $g$ be a deterministic, measurable anchor map that assigns each outside value $\bfwd=(\bfcd,\bfd)$ an anchor $g(\bfwd)=(g_{\bfd}(\bfcd),\bfd)$ in the same stratum, where $g_{\bfd}(\bfcd)\in\calH(\bfd)$. Because $g$ preserves the stratum of the discrete variables, we write $\|\bfwd-g(\bfwd)\|_{\mathbf V}$ for $\|\bfcd-g_{\bfd}(\bfcd)\|_{\mathbf V}$. Applying \eqref{eq: bound mu c'} with $\bfc'=g_{\bfd}(\bfcd)$ at every outside value, integrating the endpoints, and adding $\thetainH$ gives a valid interval for $\thetad$. For every $\bfwd\notin\calH$, the selected interval contains the all-anchor intersection $[\ell(\bfwd),u(\bfwd)]$ in \eqref{eq: envelopes main}. Thus, the resulting interval for $\thetad$ contains ${\rm PI}^{\star}$.

For $\bfcd\notin\calH(\bfd)$ {and $\bfc'\in\calH(\bfd)$}, \eqref{eq: bound mu c'} has width $2L(\bfd)\|\bfc'-\bfcd\|_{\mathbf V}$. With fixed $L(\bfd)$ and $\mathbf V$, {the anchor nearest} to $\bfcd$ in $\bf V$-norm minimizes this width. {This pointwise choice gives the narrowest interval for $\thetad$ among maps assigning one anchor per outside value.} We denote this nearest point by $\mathbf p_{\bfd}(\bfcd)$ and call the resulting map the \emph{hard projection}:
\begin{equation}\label{eq: proj}
\mathbf p_{\bfd}(\bfcd)
=\argmin_{\bfc'\in\calH(\bfd)}\|\bfc'-\bfcd\|_{\mathbf V},
\quad
\Pi_{\calH}(\bfwd)
=\left(\mathbf p_{\bfd}(\bfcd)\tr,\bfd\tr\right)\tr.
\end{equation}
By Definition~\ref{def: positivity region}, each $\calH(\bfd)$ is nonempty, and closed, so the projection in $\mathbf V$-norm exists; convexity makes it unique. Thus, both $\mathbf p_{\bfd}$ and $\Pi_{\calH}$ are well-defined. We write $\Pi$ for $\Pi_{\calH}$ when the region is clear.

Proposition~\ref{prop: partial gen} gives the resulting bounds and establishes the optimality of the hard projection among anchor maps.

\begin{Proposition}[Bounds and the optimal anchor]\label{prop: partial gen}
    Let $\calH$ be a region as in Definition~\ref{def: positivity region} and let $g$ be an anchor map. Suppose Assumptions~\ref{assump: basic} and \ref{assump: weight Lip} hold, $\E[|Y(\bfAd)|]<\infty$, and $\E[L(\bfX_D)\|\bfWd-g(\bfWd)\|_{\mathbf V}\I(\bfWd\notin\calH)]<\infty$. Then $\thetad\in{\rm PI}^{g}=[{\rm PI}^{g}_l,{\rm PI}^{g}_u]$, where
    \begin{equation}\label{eq: partial theta out}
    {\rm PI}^{g}_{l,u} = \thetainH + \E\left[\left\{Q(g(\bfWd))\mp L(\bfX_D)\left\|\bfWd-g(\bfWd)\right\|_{\mathbf{V}}\right\}\I\left(\bfWd\notin\calH\right)\right],
    \end{equation}
    and $\thetainH$ is identified as in \eqref{eq: id of theta in}. Among all anchor maps satisfying the stated integrability condition, the hard projection $\Pi_{\calH}$ of \eqref{eq: proj} minimizes the length of ${\rm PI}^{g}$.
\end{Proposition}
We write ${\rm PI}^{\Pi}$ for the interval in \eqref{eq: partial theta out} obtained with $g=\Pi_{\calH}$. At each outside value, the hard projection selects the shortest of the anchor-specific intervals in \eqref{eq: bound mu c'}. However, pointwise optimality does not imply sharpness. The sharp interval ${\rm PI}^{\star}$ in \eqref{eq: sharp PI main} can be narrower because it intersects all anchor-specific intervals in \eqref{eq: bound mu c'}. The intervals coincide when there is only one continuous variable. In higher dimensions, a more distant anchor may have a lower value of $Q$ for the upper endpoint or a higher value for the lower endpoint, offsetting its greater distance. Supplementary Material~\ref{append: sharpness} formalizes and quantifies this gap.

The choice of $\calH$ also affects the width of {both ${\rm PI}^{\Pi}$ and ${\rm PI}^{\star}$}. When $\calH$ is smaller than the conditional support, the interval leaves unused information about $Q$ at supported values outside $\calH$ and instead bounds the contribution of these values according to their distance from $\calH$. A smaller fixed region reduces reliance on outcome-regression information in sparse areas but applies the Lipschitz bounds to more policy values.

\section{Pathwise Differentiability and Inference}\label{sec: est inf}
The hard projection gives the shortest single-anchor-based interval, but it does not generally support regular influence-function-based inference. We explain how its boundary-valued anchors create this problem, introduce {a smooth approximation to the hard projection}, and develop influence functions, estimators, and confidence intervals for the resulting bounds.
{Throughout this section, we assume that, for every $\bfd\in\supp(\bfX_D)$, the conditional distribution of $\bfC$ given $\bfX_D=\bfd$ has a positive density almost everywhere on $\calH(\bfd)$.}

\subsection{Interior-displaced projection}\label{subsec: reg proj}
The hard projection $\Pi_{\calH}$ in \eqref{eq: proj} maps many exterior values to the same boundary point. When $\prob(\bfWd\notin\calH)>0$, its anchors place positive probability on this lower-dimensional boundary. {Under the density condition above and boundary conditions stated below, the observed distribution assigns the boundary probability zero, so the anchor distribution cannot be obtained by reweighting the observed distribution.} Supplementary Proposition~\ref{prop: hard nondiff} shows that the hard-projection endpoints {can fail to be} pathwise differentiable. Regular root-$n$ inference is then unavailable \citep{hirano2012impossibility}. {If $\prob(\bfWd\notin\calH)=0$, policy-generated values lie in $\calH$ almost surely, where Definition~\ref{def: positivity region} ensures the positivity condition in Assumption~\ref{assump: MTP pos}, so no outside portion or projection is needed.}

\begin{figure}[htp!]
    \centering
    \includegraphics[width=.8\linewidth]{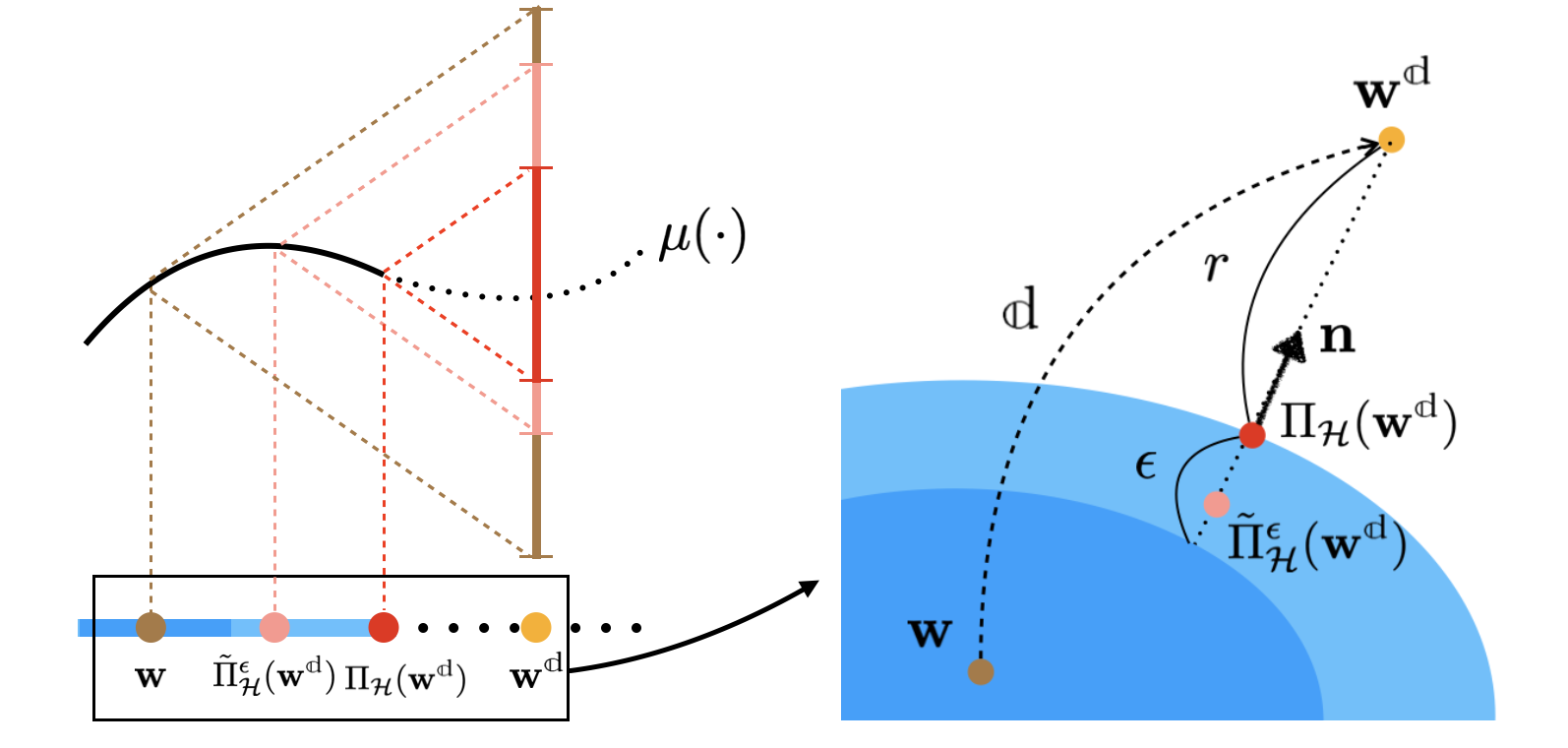}
    \caption{Hard and interior-displaced projections. Left: the observed value $\bfw$, policy value $\bfwd$ outside $\calH$, boundary anchor $\Pi_{\calH}(\bfwd)$, and interior-displaced anchor $\tilde\Pi_{\calH}^{\epsilon}(\bfwd)$. Dashed lines are the associated Lipschitz cones. Right: the two-dimensional geometry, with $\mathbf n$ and $r$ defined in \eqref{eq: proj notation}. The lighter band is the interior layer where exterior points land.}
    \label{fig: projection}
\end{figure}

We address this problem by projecting each exterior value onto the boundary and then moving the resulting anchor a short distance inside $\calH$, which we refer to as an \emph{interior-displaced projection}. We tie the displacement to the exterior distance such that values sharing the same hard projection are sent to different interior locations rather than collapsed onto one boundary point. The displaced anchors thus occupy a full-dimensional interior layer, allowing the bound endpoints to be pathwise differentiable under the conditions stated below. Figure~\ref{fig: projection} illustrates the two projection maps.

In this section, we present the construction when $\mathbf V$ is the identity matrix. The construction for a general diagonal $\mathbf V$ follows by rescaling the continuous coordinates by $\mathbf V^{1/2}$.
The following condition makes inward displacement well-defined in a thin interior layer.
\begin{Assumption}[Boundary regularity]\label{assump: boundary reg}
     For every $\bfd\in\supp(\bfX_D)$, $\calH(\bfd)$ has nonempty interior and a continuously differentiable boundary. There exists $\epsilon_{\max}>0$ such that every interior point $\bfc$ satisfying $\operatorname{dist}\{\bfc,\partial\calH(\bfd)\}<\epsilon_{\max}$ has a unique nearest point on $\partial\calH(\bfd)$.
\end{Assumption}
Assumption~\ref{assump: boundary reg} is needed only for uniqueness of the interior displacement, not for the partial-identification results in Section~\ref{sec: partial id}. Nonempty interior allows a projection point to be moved inside the region. The remaining conditions allow each point in the interior layer to be uniquely represented by a boundary point and a corresponding inward normal vector.
{For bounded $\calH(\bfd)$ in dimension $p\geq2$, the uniqueness condition in Assumption~\ref{assump: boundary reg} holds when the boundary is twice continuously differentiable and the largest eigenvalue of its curvature matrix is bounded above by $1/\epsilon_{\max}$.}

For $\bfwd=(\bfcd,\bfd)\notin\calH$, let $\mathbf p=\mathbf p_{\bfd}(\bfcd)$ be the projection point defined in \eqref{eq: proj}. Since $\bfcd$ lies outside the closed region, $\mathbf p\in\partial\calH(\bfd)$. In the rescaled coordinates where $\mathbf V=\mathbf I$, write
\begin{equation}\label{eq: proj notation}
\bfcd=\mathbf p+r\,\mathbf n,
\quad
r = \|\bfcd-\mathbf p\|,
\quad
\mathbf n = \frac{\bfcd-\mathbf p}{r}.
\end{equation}
Here $r$ is the distance from $\bfcd$ to $\calH(\bfd)$ and $\mathbf n$ is the outward unit normal at $\mathbf p$.

To separate exterior values sharing a projection point, let the inward displacement of $\mathbf p$ vary with $r$. For $\epsilon\in(0,\epsilon_{\max})$, define $\phi^\epsilon(r)=\epsilon\{1-\exp(-r/\epsilon)\}$. This function is smooth and strictly increasing in $r$, has range $[0,\epsilon)$, and increases with $\epsilon$ for every $r>0$. Larger displacement levels thus place anchors farther inside $\calH$. See Supplementary Material~\ref{pf: reg proj cov} for other possible choices of $\phi^\epsilon$.
In each stratum, define the interior-displaced projection as
\begin{equation}\label{eq: reg proj}
    \tilde\Pi_{\calH}^{\epsilon}(\bfwd)
    =\left(\{\mathbf p-\phi^\epsilon(r)\mathbf n\}\tr,\bfd\tr\right)\tr,
    \quad\text{for}\ \bfwd\notin\mathcal H.
\end{equation}
The subtraction moves the projection point along the inward normal vector. If two exterior values share a projection point, strict monotonicity of $\phi^\epsilon$ sends them to different interior locations. Conversely, Assumption~\ref{assump: boundary reg} allows the boundary point and inward distance to be recovered from a displaced anchor. The map is consequently one-to-one and spreads exterior values through a full-dimensional layer. Supplementary Lemma~\ref{lem: reg proj cov} states this result precisely. When the region is clear, we write $\tilde\Pi^{\epsilon}$ for $\tilde\Pi_{\calH}^{\epsilon}$.

The displacement level $\epsilon$ trades partial-identification width against precision in estimating the endpoints. Smaller levels keep anchors near the boundary and yield intervals closer to hard projection, whereas larger levels spread the exterior policy mass over a thicker interior layer and can improve precision in estimation. Supplementary Proposition~\ref{prop: eps tradeoff} bounds each endpoint's discrepancy from its hard-projection counterpart by order $\epsilon$ and, under additional conditions, gives an order $\epsilon^{-1}$ lower bound for its influence-function variance. Because the value that best balances width and precision depends on unknown distributional features, we consider a prespecified finite grid rather than choose one arbitrary level. Section~\ref{subsec: estimators} constructs a confidence interval with simultaneous coverage over this grid.

\subsection{Estimation and inference}\label{subsec: eif}\label{subsec: estimators}

We discuss estimating the lower and upper endpoints ${\rm PI}_l^g$ and ${\rm PI}_u^g$ in \eqref{eq: partial theta out} and constructing CIs for $\thetad$. Our proposal takes the anchor map $g$ to be $\tilde\Pi_{\calH}^\epsilon$, but we state the influence-function result for any anchor map satisfying the conditions below. Following the decomposition in \eqref{def: theta in and out}, each endpoint influence function has one component for the identified inside portion $\thetainH$ and another for the lower or upper bound on the outside portion $\thetaoutH$.

{The influence function for the inside portion uses the policy-induced treatment density.
We write the observed conditional treatment density as $f_{\bfA\mid\bfX}(\bfa\mid\bfx)$, abbreviated $f(\bfa\mid\bfx)$.} The following assumption is a multivariate analogue of the piecewise smooth invertibility condition for continuous-treatment MTPs \citep{haneuse2013estimation,diaz2023nonparametric}.

\begin{Assumption}[Piecewise smooth invertibility]\label{assump: smooth inv}
  For each $\bfx$, $\supp(\bfA\mid\bfX=\bfx)$ can be partitioned, up to a set of conditional probability zero, into finitely many {open} cells $\{\mathcal{A}_j(\bfx)\}_{j=1}^{J(\bfx)}$. For each $j$, there exists a one-to-one function $\bbmd_j(\cdot,\bfx)$ with inverse $\bbmb_j(\cdot,\bfx)$ such that $\bbmd(\bfa,\bfx)=\bbmd_j(\bfa,\bfx)$ on $\mathcal{A}_j(\bfx)$,
  {and both maps are continuously differentiable with nonzero Jacobian determinants on the interiors of $\mathcal A_j(\mathbf x)$ and $\bbmd_j(\mathcal A_j(\mathbf x),\mathbf x)$, respectively.}
\end{Assumption}

In particular, Assumption~\ref{assump: smooth inv} excludes MTPs that map a positive-probability range of continuous treatment values to a single value. Under this condition, {the policy-induced density $f_{\bfAd\mid\bfX}(\bfa\mid\bfx)$, abbreviated $f^{\bbmd}(\bfa\mid\bfx)$, is obtained by change of variables on each cell, as shown in Supplementary Material~\ref{pf: EIF}. For the inside portion, we define the policy density ratio $r^{\rm in}(\bfw)=r^{\rm in}(\bfa,\bfx)=f^{\bbmd}(\bfa\mid\bfx)/f(\bfa\mid\bfx)$ for $\bfw\in\calH$.}

The outside portion evaluates $Q$ at anchors assigned to policy-generated values outside $\calH$. Deriving its influence function requires these anchors to admit a density supported where the observed density is positive. \begin{Assumption}[Regularity of the anchor map]\label{assump: anchor reg}
Let $g$ be an anchor map. For every $\bfd$, the following conditions hold. (i) Writing $\bfx=(\bfx_C\tr,\bfd\tr)\tr$, the set of $(\bfa,\bfx_C)$ satisfying $\{\bbmd(\bfa,\bfx)\tr,\bfx_C\tr\}\tr\notin\calH(\bfd)$ admits a finite partition refining the cells of Assumption~\ref{assump: smooth inv}. On each cell, the map
$(\bfa,\bfx_C)\mapsto
g_{\bfd}\!\left(\{\bbmd(\bfa,\bfx)\tr,\bfx_C\tr\}\tr\right)$
is one-to-one and locally bi-Lipschitz, with nonzero Jacobian determinant almost everywhere. (ii) Conditional on $\bfX_D=\bfd$, anchors assigned to policy-generated values outside $\calH$ admit an unnormalized density $f^g(\cdot\mid\bfd)$. \end{Assumption}

{Assumption~\ref{assump: anchor reg} provides the anchor density needed to weight observations within $\mathcal H$ when estimating the outside bounds. We can define the anchor ratio}
$r^g(\bfw)=f^g(\bfa,\bfx_C\mid\bfx_D)/f(\bfa,\bfx_C\mid\bfx_D)$ {since the density $f^g(\bfa,\bfx_C\mid\bfd)$ is zero almost everywhere that $f(\bfa,\bfx_C\mid\bfd)$ is zero.}
The integral of $f^g$ over $(\bfa,\bfx_C)$ equals
$\prob(\bfWd\notin\calH\mid\bfX_D)$, which explains why it is unnormalized. We define $r^g$ to be zero wherever its denominator is zero.

When exterior values occur with positive probability, the hard projection violates (i) and (ii). It is many-to-one along each normal ray, and its anchors concentrate on the boundary rather than admitting a $p$-dimensional density. {Supplementary Lemma~\ref{lem: reg proj cov} shows that if exterior policy values have a density within a stratum, so do their displaced anchors.} The following theorem gives the influence functions for the bound endpoints.
\begin{Theorem}[Influence functions for the bound endpoints]\label{thm: EIF}
{Suppose the conditions of Proposition~\ref{prop: partial gen} and Assumptions~\ref{assump: smooth inv} and \ref{assump: anchor reg} hold.} Let $\calH$ and $g$ be known and fixed. Assume that differentiation along regular submodels can be passed through the conditional means and expectations defining the endpoints. Define
    {\small
    \begin{align*}
    \psi^{\rm in}(O)&=(Y-Q(\bfW))r^{{\rm in}}(\bfW)\I(\bfW\in\calH)+Q(\bfWd)\I(\bfWd\in\calH),\\
    \psi_{l,u}^{g}(O)&=(Y-Q(\bfW))r^{g}(\bfW)\I(\bfW\in\calH)+\{Q(g(\bfWd))\mp L(\bfX_D)\|\bfWd-g(\bfWd)\|_{\mathbf{V}}\}\I(\bfWd\notin\calH).
    \end{align*}}
For $*\in\{l,u\}$, let $\Psi_*^g=\psi^{\rm in}+\psi_*^g$. If both $\Psi_l^g$ and $\Psi_u^g$ belong to $L_2(P)$, then $\Psi_*^g-{\rm PI}_*^g$ is an influence function of ${\rm PI}_*^g$ for each $*\in\{l,u\}$.
\end{Theorem}

The inside portion retains the standard MTP influence-function structure through the policy ratio $r^{\rm in}$. The outside portion introduces the anchor ratio $r^g$, which accounts for estimation of the outcome regression at the assigned anchors. If $\prob(\bfWd\in\calH)=1$, the outside portion vanishes and Theorem~\ref{thm: EIF} recovers the standard MTP influence function \citep[e.g.,][]{diaz2012population,hejazi2022efficient}.

We use Theorem~\ref{thm: EIF} with $g=\tilde\Pi^\epsilon$ to construct cross-fitted one-step estimators \citep{chernozhukov2018double,kennedy2024semiparametric}. Cross-fitting allows flexible nuisance estimators without Donsker conditions.
Below, a superscript $(\epsilon)$ denotes evaluation at $g=\tilde\Pi^\epsilon$.
Independently of the data, form $K$ nearly equal-sized validation folds $\mathcal I_k$. On each training sample $\mathcal T_k=\{1,\ldots,n\}\setminus\mathcal I_k$, estimate $Q$ by $\widehat Q_k$ and estimate $r^{\rm in}$ and $r^{(\epsilon)}$ by probabilistic classification \citep[Section~5.4]{diaz2023nonparametric}. The classifiers compare observed values with policy-generated values inside $\calH$ for $r^{\rm in}$ and with interior-displaced anchors assigned to policy-generated values outside $\calH$ for $r^{(\epsilon)}$. Assumption~\ref{assump: reg nuisance} states the required convergence conditions, and Supplementary Material~\ref{append: impl nuisance} gives the exact construction.

We estimate each endpoint and its asymptotic variance by aggregating across folds. Let $n_k=|\mathcal I_k|$, and let $\widehat\Psi_{*,k}^{(\epsilon)}$ be the uncentered representation in Theorem~\ref{thm: EIF} with nuisance functions fitted on $\mathcal T_k$. Using the empirical mean $\widehat{\E}_{k}$ and variance $ \widehat{\Var}_{k}$ over $\mathcal I_k$, we obtain \begin{equation}\label{eq: aggregation of theta and sigma}
\widehat{{\rm PI}}_{*}^{(\epsilon)}=\sum_{k=1}^K\frac{n_k}{n}\widehat{\E}_{k}\left[\widehat\Psi_{*,k}^{(\epsilon)}\right],
\quad
\big(\sigmahat_{*}^{(\epsilon)}\big)^2=\sum_{k=1}^K\frac{n_k}{n}\widehat{\Var}_{k}\left(\widehat\Psi_{*,k}^{(\epsilon)}\right),
\quad *\in\{l,u\}.
\end{equation}
Supplementary equation~\eqref{eq: theta k} gives the observation-level form of $\widehat\Psi_{*,k}^{(\epsilon)}$.
For the causal contrast $\taud=\thetad-\E[Y]$, replace $\widehat\Psi_{*,k}^{(\epsilon)}$ by $\widehat\Psi_{*,k}^{\tau,(\epsilon)}=\widehat\Psi_{*,k}^{(\epsilon)}-Y${, with the conditions below applied to the contrast representations.}

For fixed $\epsilon$, Theorem~\ref{thm: asy-normal} in the following subsection establishes joint asymptotic normality of the endpoint estimators, permitting a CI for $\thetad$ using standard methods for partially identified parameters \citep[e.g.,][]{imbens2004confidence}; see Supplementary equation~\eqref{eq: CI theta}. Since no displacement level is known to best balance bound width and sampling variability, we consider a prespecified finite grid and report the intersection of the CIs across the grid, which is no wider than any constituent CI.
However, fixed-$\epsilon$ coverage alone does not guarantee coverage after taking this intersection. Following \citet{susmann2025non}, we use a multiplier bootstrap to calibrate the level-specific CIs jointly over both endpoints and all grid points, allowing their intersection to retain coverage. For each grid point $\epsilon_j$, a superscript $(j)$ denotes evaluation at that level. We write $\widehat{\rm PI}_*^{(j)}$ and $\widehat\sigma_*^{(j)}$ in \eqref{eq: aggregation of theta and sigma} for {estimates of the endpoints and asymptotic standard deviations}, and $\widehat\Psi_{*,i}^{(j)}$ as the corresponding observation-level representation. For each bootstrap draw $m=1,\ldots,M$, generate one vector of i.i.d.~mean-zero, variance-one multipliers $\{\xi_i^{[m]}\}_{i=1}^n${, independent of the data,} and define $\widehat T_{*,j}^{[m]}
=
n^{-1/2}\sum_{i=1}^n
\xi_i^{[m]}
({\widehat\Psi_{*,i}^{(j)}-\widehat{\rm PI}_{*}^{(j)}})/
{\widehat\sigma_{*}^{(j)}}$,
for $*\in\{l,u\}$, and $j=1,\ldots,J$.
Set $\widehat D^{[m]}=\max_{1\le j\leq J}\{\widehat T_{l,j}^{[m]}\vee(-\widehat T_{u,j}^{[m]})\}$, and let $\widehat q_{1-\alpha}$ be the empirical $(1-\alpha)$ quantile of $\{\widehat D^{[m]}\}_{m=1}^M$ conditional on the data. Using this common critical value, we intersect the endpoint-adjusted intervals across the grid. The resulting CI is
\begin{equation}\label{eq: unif CI}
    \widehat{\rm CI}_{\alpha}=\left[\max_j\left\{\widehat{\rm PI}_l^{(j)}-\widehat{q}_{1-\alpha}\widehat{\sigma}_l^{(j)}/\sqrt{n}\right\},\ \min_j\left\{\widehat{\rm PI}_u^{(j)}+\widehat{q}_{1-\alpha}\widehat{\sigma}_u^{(j)}/\sqrt{n}\right\}\right].
\end{equation}

Corollary~\ref{cor: uniform set coverage} in Section~\ref{subsec: theo} establishes that \eqref{eq: unif CI} covers the intersection of the population bound intervals over $j$, and hence $\thetad$, at the appropriate rate. For bounded outcomes, intersecting this CI with the known target range preserves coverage. Supplementary Material~\ref{append: impl bounded} gives details and Supplementary Algorithm~\ref{alg: est tau(H)} summarizes the procedure.

\subsection{Asymptotic theory}\label{subsec: theo}

We now establish the large-sample properties of the proposed estimators and confidence intervals.

The following assumption collects the nuisance requirements standard for cross-fitted one-step estimation \citep{chernozhukov2018double,kennedy2024semiparametric}.

\begin{Assumption}[Nuisance estimation]\label{assump: reg nuisance}
Fix $K$ and an anchor map $g$. For $k=1,\ldots,K$, let $\widehat\Psi_{*,k}^{g}$ be the estimated uncentered representation based on nuisances trained on $\mathcal T_k$. Set $d_k=\|\widehat Q_k-Q\|_2$ and $e_k=\|\widehat r_k^{\rm in}-r^{\rm in}\|_2+\|\widehat r_k^{g}-r^{g}\|_2$. Assume {finite plug-in expectations and} the following conditions. (i) $\max_k d_k=o_p(1)$ and $\max_k e_k=o_p(1)$. (ii) $\max_k d_ke_k=o_p(n^{-1/2})$. (iii) For some $\delta>0$ and each $*\in\{l,u\}$, $\max_k\|\widehat\Psi_{*,k}^{g}-\Psi_*^{g}\|_2=o_p(1)$, $\|\Psi_*^{g}\|_{2+\delta}<\infty$, and $\max_k\|\widehat\Psi_{*,k}^{g}\|_{2+\delta}=O_p(1)$.
\end{Assumption}

Condition (i) requires nuisance consistency. Condition (ii) controls the mixed-bias remainder through a product rate. Condition (iii) gives $L_2(P)$ convergence and uniform moment control for the estimated representations. The next theorem gives joint asymptotic normality for a fixed anchor map.

\begin{Theorem}[Asymptotic normality]\label{thm: asy-normal}
Suppose the conditions of Theorem~\ref{thm: EIF} and Assumption~\ref{assump: reg nuisance} hold. Let $\widehat{\rm PI}_*^{g}$ denote the corresponding cross-fitted estimator defined in \eqref{eq: aggregation of theta and sigma}, and define
$\boldsymbol\theta=({\rm PI}_l^{g},{\rm PI}_u^{g})\tr$,
$\widehat{\boldsymbol\theta}=(\widehat{\rm PI}_l^{g},\widehat{\rm PI}_u^{g})\tr$, and
$\boldsymbol\Psi=(\Psi_l^{g},\Psi_u^{g})\tr$.
Then $\sqrt n(\widehat{\boldsymbol\theta}-\boldsymbol\theta)\dto\mathcal N(\mathbf 0_2,\Sigma)$, where $\Sigma=\Var(\boldsymbol\Psi)$. The variance estimator in \eqref{eq: aggregation of theta and sigma} is consistent for the diagonal of $\Sigma$.
\end{Theorem}

For the proposed map $g=\tilde\Pi^\epsilon$, Theorem~\ref{thm: asy-normal} gives joint asymptotic normality at each fixed $\epsilon\in(0,\epsilon_{\max})$ under its stated conditions. {As $\epsilon$ decreases to zero, the bound endpoints approach their hard-projection counterparts, but their influence-function variances can grow (Supplementary Proposition~\ref{prop: eps tradeoff}).} We leave {the extension of Theorem~\ref{thm: asy-normal} to sequences $\epsilon_n\to0$} to future work.

Theorem~\ref{thm: EIF} establishes valid influence functions but does not by itself establish their efficiency in the maintained Lipschitz model. Assumption~\ref{assump: weight Lip} constrains $Q$ wherever $\mu=Q$, making this model smaller than the unrestricted nonparametric model. The following proposition gives sufficient conditions under which the influence functions are efficient.

\begin{Proposition}[Efficiency under a nonbinding Lipschitz restriction]\label{prop: efficiency}
Suppose the conditions of Theorem~\ref{thm: EIF} hold, and there is a constant $\eta>0$ such that $\mu$ satisfies Assumption~\ref{assump: weight Lip} with $L(\bfd)$ replaced by $L(\bfd)-\eta$ in every stratum with $\prob(\bfX_D=\bfd)>0$. Suppose also that $\Var(Y\mid\bfW)\geq\sigma_{\min}^2>0$ almost surely and that the conditional squared residuals are uniformly integrable. Then, for each endpoint, the influence function in Theorem~\ref{thm: EIF} is efficient in the maintained Lipschitz model.
\end{Proposition}
The smallest values of $L(\bfd)$ for which Assumption~\ref{assump: weight Lip} holds are unknown. Values below them need not yield valid bounds.
{Proposition~\ref{prop: efficiency} establishes efficiency when every $L(\bfd)$ can be reduced by a common $\eta>0$ without violating Assumption~\ref{assump: weight Lip}.
Without this margin, efficiency is not established; see Supplementary Material~\ref{pf: efficiency} for further discussion. However, the influence functions in Theorem~\ref{thm: EIF} remain valid and the endpoint estimators remain asymptotically normal under the conditions of Theorem~\ref{thm: asy-normal}.}

{We now turn from a} fixed displacement level {to simultaneous inference over the grid.} The following corollary establishes {asymptotic coverage for the CI} in \eqref{eq: unif CI}.
\begin{Corollary}[Simultaneous set coverage]\label{cor: uniform set coverage}
Let $\{\epsilon_j\}_{j=1}^J\subset(0,\epsilon_{\max})$ be a prespecified finite grid that remains fixed as $n\to\infty$. Suppose the conditions of Theorem~\ref{thm: asy-normal} hold for $g=\tilde\Pi^{\epsilon_j}$ at every grid point, the stacked centered endpoint influence functions have a nonsingular covariance matrix, and $M=M_n\to\infty$. Then
$\liminf_{n\to\infty}
\prob\!\left\{
[\max_j{\rm PI}_l^{(j)},\min_j{\rm PI}_u^{(j)}]
\subseteq\widehat{\rm CI}_{\alpha}
\right\}
\geq1-\alpha$,
and consequently
$\liminf_{n\to\infty}\prob(\thetad\in\widehat{\rm CI}_{\alpha})\geq1-\alpha$.
\end{Corollary}
The guarantee is conservative for two reasons. The interval covers both endpoints of the intersection of the bound intervals, and the bootstrap quantile is calibrated simultaneously over the fixed grid.
The nonsingularity condition excludes $L=0$, where the endpoints coincide. Section~\ref{sec: real data} uses that value only as a benchmark.

\section{Practical Construction of the Region and Implementation}\label{sec: practical}\label{subsec: pos region}\label{subsec: boundary smoothing}\label{subsec: data adaptive H}\label{subsec: practical algorithm}

The preceding results are for a fixed region $\calH$ and treat $L(\cdot)$ and $\mathbf V$ as prespecified. This section presents two data-dependent constructions of $\calH(\bfd)$ within each stratum of $\bfX_D$ with smoothing to satisfy Assumption~\ref{assump: boundary reg}. We then describe how these regions enter the cross-fitted estimators and give practical choices for $\mathbf V$, $L(\cdot)$, and the displacement grid.

For the first construction, we form $\calH(\bfd)$ as the convex hull of all observed $\bfC$ values in stratum $\bfd$. This applies the exposure-space convex-hull diagnostic used by \citet{antonelli2024causal} to the joint continuous treatment--covariate space. Under convex-support and positive-density conditions, Supplementary Lemma~\ref{lem: hull admissible} shows that these hulls satisfy Definition~\ref{def: positivity region} across finitely many strata with probability tending to one. If the conditional support is nonconvex, however, the sample convex hull can extend beyond that support and thus may not satisfy Definition~\ref{def: positivity region}. Supplementary Material~\ref{append: impl region} discusses this limitation.

{The convex hull includes every observed $\mathbf{C}$ value and may therefore contain policy-generated values in sparse parts of the support. Their contribution is identified but can be unstable to estimate. This motivates our second construction: a density-trimmed hull.}
We first estimate a highest-density region of $\bfC$ within each stratum \citep{hyndman1996computing,bao2025addressing}. We retain observations whose fitted conditional density exceeds an empirical quantile within that stratum and take their convex hull. Relative to the convex hull of all observations, the density-trimmed hull classifies more policy-generated values as outside $\calH$. This can improve estimation stability by reducing reliance on outcome-regression and density-ratio estimates in sparse regions, at the cost of wider partial-identification bounds for the same target $\thetad$.

We expand the resulting polytopes outward by radius $\rho$ in the $\mathbf V$-metric to satisfy Assumption~\ref{assump: boundary reg} with $\epsilon_{\max}=\rho$. Outward smoothing may add treatment--covariate values outside the conditional support. The expanded region may then fail Definition~\ref{def: positivity region}, so the identification and inference results in Sections~\ref{sec: partial id} and~\ref{sec: est inf} are not established for that region. For a containment-preserving inward alternative, see Supplementary Material~\ref{append: impl smooth}.

For the cross-fitted estimators in Section~\ref{subsec: estimators}, we construct $\calH$ from each training fold and treat it as given when evaluating the estimating equations on held-out observations. Supplementary Corollary~\ref{cor: split sample} extends coverage to the resulting fold-averaged bounds.

To guide sensitivity analyses, we provide a practical scaling for $\mathbf V$ and a data-informed reference for $L(\cdot)$. We set $\mathbf V$ to the diagonal matrix of inverse sample variances \citep{abadie2006large}. For $L(\cdot)$, we use a pairwise-slope benchmark following \citet{ma2025sensitivity}:

\begin{equation}
\begin{aligned}
     \widehat{L}_{q}(\bfd) = \max_{i}\left\{\text{$q$-th quantile of } \left(\frac{|\widehat{Q}(\bfc_i,\bfd)-\widehat{Q}(\bfc_j,\bfd)|}{\|\bfc_i-\bfc_j\|_{\mathbf{V}}}\right)_{j\neq i}\right\},\label{eq: Lqd}
\end{aligned}
\end{equation}
where $i$ and $j$ index samples used to fit $\widehat Q$ in stratum $\bfd$. For sensitivity analysis, $\widehat L_q$ provides a data-informed reference for the prespecified $L$ grid, not an estimate of the smallest $L$ satisfying Assumption~\ref{assump: weight Lip}.
Whether a given $L$ suffices is generally unverifiable: the observed distribution may not identify $\mu$ at every policy-generated values.
In the numerical studies, we set $q=0.99$ to avoid an overly small empirical reference.
Supplementary Material~\ref{append: impl tuning} gives the remaining tuning details, including the $\epsilon$ grid.

\section{Simulation Study}\label{sec: sim}

We assess finite-sample coverage and interval width under structural and practical positivity violations. The simulations use a bivariate continuous treatment $\bfA=(A_1,A_2)\tr$, a binary covariate $X$, and the policy $\bbmd(\bfA)=(\kappa A_1,A_2)\tr$ for $\kappa\in\{0.7,0.8\}$. DGP1 is bivariate normal truncated to $A_1\geq4$, so the policy sends some values outside the treatment support. DGP2 is bivariate normal and DGP3 is bivariate $t$ with five degrees of freedom. Both have full support, so the positivity problem is practical rather than structural. The heavier tails of DGP3 provide better representation of policy-generated values. The treatment correlation is $0.2$ when $X=0$ and $0.9$ when $X=1$ {(before truncation in DGP1)}, so the one-coordinate shift is less well represented in the latter stratum. We consider two outcome regression functions: an additive linear function $Q_1$ of $A_1$, $A_2$, and $X$, and $Q_2$, which adds treatment--treatment and treatment--covariate interactions.
Supplementary Material~\ref{sec: add sim} gives the complete data-generating processes and parameter values.

We take $n\in\{500,1000,2000,5000\}$ and use 500 replications for each setting. For \texttt{Proposal}, we use separate Super Learner \citep{vanderlaan2007super} ensembles: {for the outcome regression, we use
generalized additive models (GAMs), multivariate adaptive regression
splines (MARS), lasso, and linear
regression with and without two-way interactions; for density-ratio
classification, we use $\ell_1$-penalized logistic regression,
Bayesian additive regression trees (BART), and MARS.}
We construct cross-fitted convex-hull regions as in Section~\ref{sec: practical}, using the inverse-variance choice of $\mathbf V$, the data-informed choice of $L$, and outward smoothing with $\rho=0.05$.

We compare \texttt{Proposal} with three benchmarks. \texttt{Oracle (L+Q)} uses {an outcome-regression plug-in estimator
with} the true outcome regression and the design-specific oracle $L$. {\texttt{Naive MTP} uses a cross-fitted implementation of the
doubly robust one-step approach \citep{diaz2012population}}, evaluating the fitted outcome regression at every policy value, including those outside the estimated region. \texttt{Trimming} {uses a corresponding cross-fitted one-step
estimator for the subpopulation whose policy-generated values
lie in} the true 95\% highest-density region. \texttt{Proposal} and \texttt{Oracle (L+Q)} use the same interior-displaced-projection construction, with multiplier and paired bootstraps, respectively. We report empirical coverage of $\taud$ and mean width {of
nominal 95\% CIs}.

\begin{figure}[htp!]
\centering
\includegraphics[width=\textwidth]{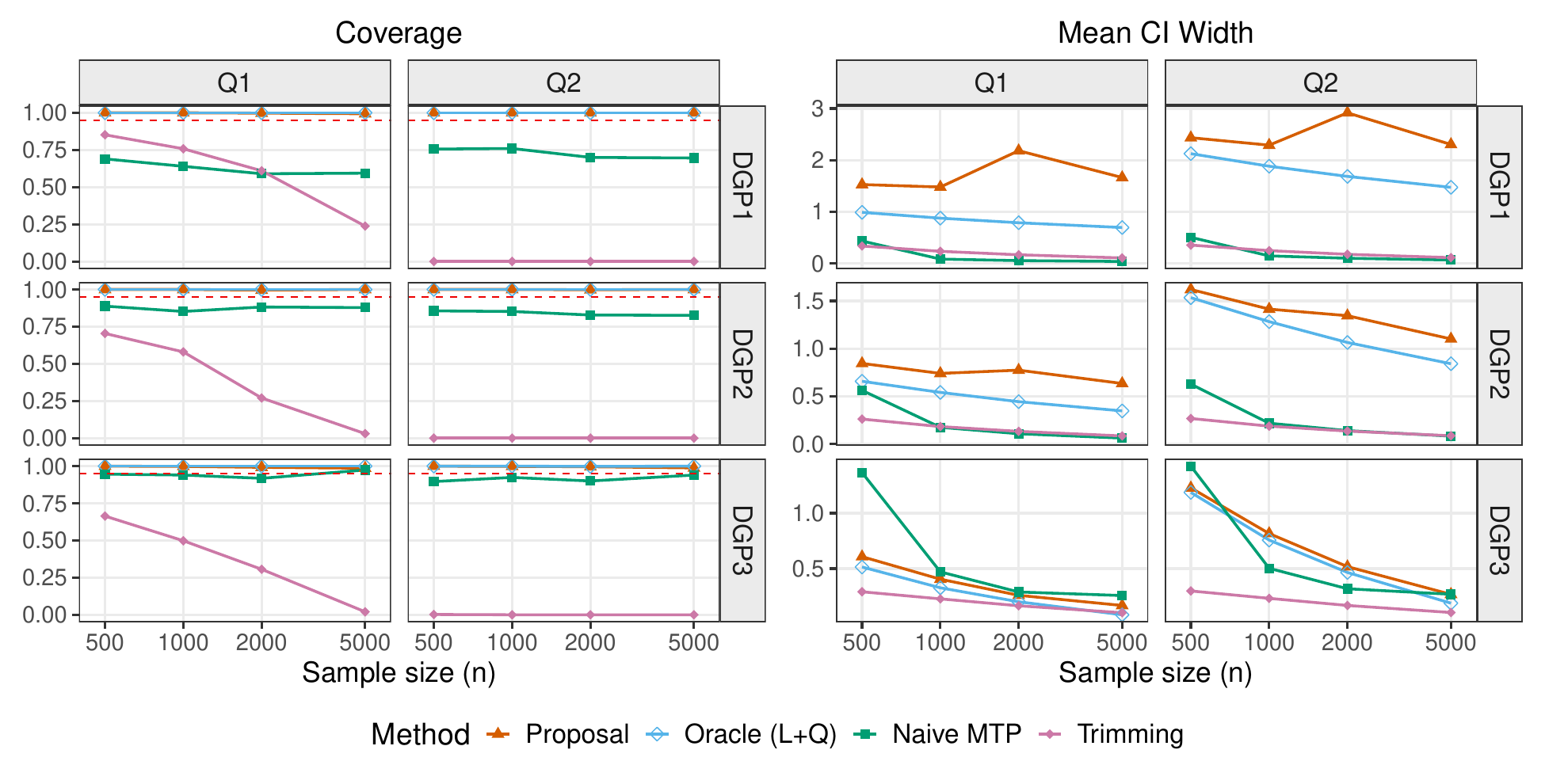}
\caption{Coverage of the causal contrast $\taud$ and mean CI width at $\kappa=0.7$ for \texttt{Proposal}, \texttt{Oracle (L+Q)}, \texttt{Naive MTP} and \texttt{Trimming}. Rows are exposure distributions, columns outcome models. The dashed line in the coverage panels is the nominal level $1-\alpha=0.95$.}
\label{fig: cov width}
\end{figure}

Figure~\ref{fig: cov width} presents coverage and mean CI width at $\kappa=0.7$. \texttt{Oracle (L+Q)} and \texttt{Proposal} attain or exceed 95\% coverage in every reported setting, including $Q_2$. The coverage of \texttt{Naive MTP} follows the proportion of policy values inside $\calH$, which is lowest under DGP1 and highest under DGP3 (Supplementary Figure~\ref{fig: prop inside}). Its coverage ranges from 0.59 under DGP1 to 0.97 under DGP3 and does not improve steadily with $n$. \texttt{Trimming} covers its own subpopulation target between 0.91 and 0.97 but increasingly undercovers $\taud$ as the sample grows. {Mean coverage of $\taud$ across scenarios falls} from 0.37 at $n=500$ to 0.05 at $n=5000$, because its intervals concentrate around a different target.

The width panels show that \texttt{Naive MTP} is generally narrower since it treats $\taud$ as point identified. Unstable weights reverse this ordering in some heavy-tailed settings. \texttt{Proposal} remains close to \texttt{Oracle (L+Q)}. Under DGP1, the width of \texttt{Proposal} approaches a nonzero identification limit as $n$ grows. Under DGP2 and DGP3, the data-adaptive region $\calH$ expands with $n$ and the identification component shrinks, most rapidly under DGP3.

We also examine sensitivity to specifying $L$ (Supplementary Material~\ref{subsec: sensitivity tuning}). Smaller values narrow the intervals, but coverage falls to 0.97 at one half of the oracle value and 0.93 at one quarter.

\section{Real Data Application: CHAMACOS}\label{sec: real data}

We now apply our proposed approach to {bound the causal effects of hypothetical reductions in} pesticide exposure on maternal hypertension risk among $259$ CHAMACOS participants in California's Salinas Valley \citep{eskenazi2003chamacos,rudolph2026everything}. These participants had pesticide exposure measurements at two or more visits {across the five study waves from 2010--2013 through 2022--2024. We set each participant's first observation period as the first of these waves for which at least one year of exposure data was available. We numbered subsequent study waves in order as periods two, three, and so on, giving up to five observation periods per participant.}
For the multivariate exposure, we {used each participant's third-period values of} two-year summaries of agricultural pesticide use within 1 km of {their residence} for seven pesticide classes: organophosphates, pyrethroids, carbamates, neonicotinoids, manganese-containing fungicides, glyphosate, and paraquat. {During preprocessing, the binary hypertension outcome was coded as positive in all periods following a positive record. We used each participant's fifth-period outcome after preprocessing, using the fourth or, if necessary, third-period outcome when it was missing.} We adjusted for {two covariates considered potential common causes of pesticide exposure and hypertension:} continuous maternal {age at index birth} and baseline maternal education (categorized as sixth grade or below, seventh to twelfth grade, or high-school graduate or above). We interpret the results under consistency and no unmeasured confounding conditional on these variables (Assumption~\ref{assump: basic}), though residual confounding is likely. {Supplementary Material~\ref{append: impl chamacos} gives preprocessing details and additional sensitivity analyses.}\footnote{Code for the simulations and CHAMACOS analysis is available at \url{https://github.com/taehyeonkoo/MTPbounds}. The CHAMACOS data are not public but may be requested from the study investigators under an approved data use agreement; see \url{https://cerch.berkeley.edu}.}

The first five pesticide classes have pairwise correlations between 0.70 and 0.88, so reducing one while holding the others fixed can generate joint exposure combinations that are sparsely represented even when each exposure is adequately represented marginally (Supplementary Figure~\ref{fig: CHAMACOS corr t3}). We consider proportional reductions from 0\% to 20\%, applied either to all seven classes jointly, which largely preserves the mixture profile, or to neonicotinoids alone, which disrupts the strongest correlations. \citet{rudolph2026everything} redefined the poorly supported latter policy to leave the exposure vector unchanged when a reduction would require extrapolation. We retain both policies and bound their mean outcomes. {For each policy $\bbmd$, we target the hypertension risk difference $\taud=\E[Y(\bfAd)]-\E[Y]$, comparing the risk under the exposure reduction with that under the observed exposures.}

{For our proposed method (\texttt{Proposal}), we estimate the bound endpoints for $\taud$ using the one-step estimators in Section~\ref{subsec: estimators}. Given the small sample, we implement these estimators without cross-fitting so that each nuisance function can be fitted on the full sample using discrete Super Learner (see Supplementary Material~\ref{append: impl chamacos} for the candidate learners). We report a sensitivity analysis with cross-fitting in Supplementary Material~\ref{append: impl cv1}. }

Within each education stratum, $\calH$ is the full-sample convex hull of the seven pesticide exposures and maternal age, outward-smoothed as in Section~\ref{subsec: boundary smoothing} with $\rho=0.05$ in the $\mathbf V$-scaled coordinates. Including maternal age makes this a joint exposure--covariate region rather than an exposure-only region. We use the full sample {because the stratum-specific samples are small} and because this construction places every observed value inside $\calH$. The cross-fitted alternative gives intervals several times wider at every $L$ (Supplementary Material~\ref{append: impl cv1}). Outward smoothing may add values outside the conditional support, {and the full-sample construction of $\calH$ does not use the sample splitting assumed in Supplementary Corollary~\ref{cor: split sample}, so our theoretical guarantees do not directly cover this implementation. However, simulations in Supplementary Material~\ref{subsec: full sample regions} provide empirical evidence of valid CI coverage under these analytic choices.}

We use the inverse-variance metric $\mathbf V$ from Section~\ref{subsec: practical algorithm} and report $L\in\{0,0.15,0.3,0.6\}$. At $L=0$, {the lower and upper endpoint estimates coincide}. We report this only as a descriptive benchmark, not as a valid bound at $L=0$. Because anchors may change maternal age as well as exposures, the benchmark is not a treatment policy and differs from the supported-shift redefinition of \citet{rudolph2026everything}. The benchmark $\widehat L_{0.99}$ of \eqref{eq: Lqd} ranges from 0.13 to 0.15 across strata, motivating $L=0.15$ {for our primary analysis, with additional sensitivity analyses at $L=0.3$ and $0.6$.} For the binary outcome, $L=0.15$ allows conditional means to differ by at most 15 percentage points per unit of standardized Euclidean distance, corresponding to a one-standard-deviation change in one coordinate with the others fixed. {The less restrictive $L=0.6$ allows 60 percentage points per unit.}

\begin{figure}[htp!]
    \centering
    \includegraphics[width=\textwidth]{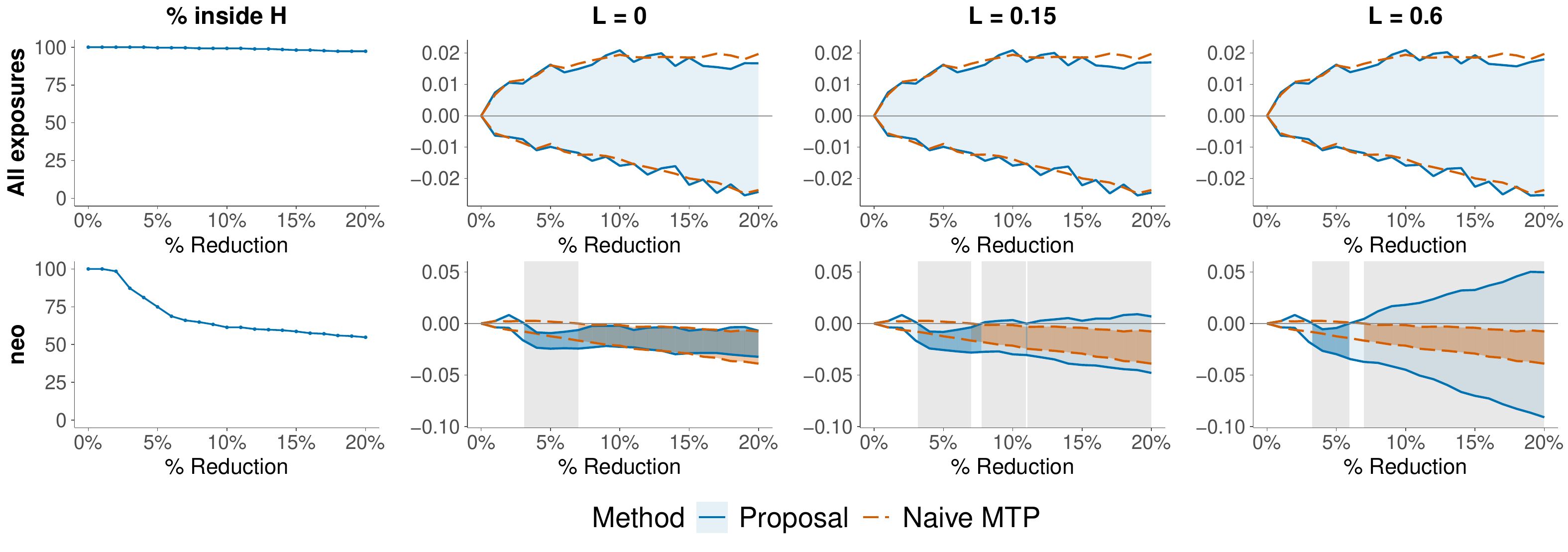}
    \caption{CHAMACOS results for joint (top) and neonicotinoid-only (bottom) reductions. Left panels show the percentage of policy values inside the data-adaptive region. Remaining panels compare the \texttt{Proposal} and \texttt{Naive MTP} intervals at selected $L$ values. Gray shading marks disagreement about excluding zero; darker fill shows where {each band} excludes zero.}
    \label{fig: CHAMACOS Lsens}
\end{figure}

{We compare \texttt{Proposal} with \texttt{Naive MTP}, the doubly robust one-step estimator described in Section~\ref{sec: sim}. Its nuisance functions use the same learner libraries and full-sample fitting as \texttt{Proposal}. We report 95\% CIs separately for each exposure reduction.
For \texttt{Proposal}, we use the multiplier-bootstrap CI in \eqref{eq: unif CI}.}
The left column of Figure~\ref{fig: CHAMACOS Lsens} reports the proportion of policy values inside the region. At least 97\% remain inside under the joint policy through a 20\% reduction. Under the neonicotinoid-only policy, a quarter lie outside by a 5\% reduction and nearly half by 20\%. These diagnostics show that adequate marginal representation does not ensure that policy-generated combinations lie inside the joint treatment--covariate region.

Under the joint policy, nearly every policy-generated value lies in $\calH$, and the proposed and \texttt{Naive MTP} analyses give similar results. Across the reported $L$ values and reductions, their {interval widths} differ by {less than 0.006}, and neither excludes zero. The estimated contrast is at most half a percentage point. Even at a 20\% reduction, partial-identification width accounts for less than 5\% of the proposed interval's width, so most uncertainty reflects sampling variability. Both intervals widen smoothly as their fitted policy density ratios become more variable with the shift (Supplementary Material~\ref{append: impl ratios}).

When many policy values leave $\calH$, as under the neonicotinoid-only policy, the methods can lead to different conclusions. At $L=0.15$, the proposed interval excludes zero for reductions of 4\% to 7\% {and at 11\%}, whereas the \texttt{Naive MTP} interval excludes zero from an 8\% reduction onward. Beyond 5\%, \texttt{Naive MTP} evaluates the fitted outcome regression at 25\% to 50\% of policy values outside $\calH$. Its conclusion that reductions in neonicotinoid exposure were associated with lower hypertension risk depends on predictions at those values. Once the outside conditional means are allowed to vary at the stated Lipschitz rate, the proposed interval contains zero {from a 12\% reduction onward}, and the analysis no longer supports a directional conclusion. This {agrees} qualitatively with \citet{rudolph2026everything}, whose intervals contain zero; however, the estimand differs as they redefined the policy to retain adequately supported shifts.

\section{Conclusion and Discussion}\label{sec: conclusion}

MTPs accommodate continuous and multivariate exposures, but policy-generated values may be unsupported or too sparsely represented for stable estimation. Keeping the policy and estimand unchanged, we bound the contribution of values outside a region of adequate positivity under a Lipschitz sensitivity restriction. Nearest-point projection gives the shortest interval in the one-anchor class, but its boundary-valued anchors {can make} the endpoints non-pathwise differentiable. Our interior-displaced projection moves these anchors into the region, restoring pathwise differentiability and enabling influence function-based inference {under the stated conditions.
In the simulations shown in Figure~\ref{fig: cov width}, our CIs attained at least nominal coverage of $\taud$, whereas CIs from the standard MTP estimator assuming positivity had lower coverage when more policy-generated values fell outside $\calH$. In CHAMACOS, the two approaches gave similar results when nearly all policy-generated values lay within $\calH$. At larger reductions with poor support, the standard analysis suggested lower hypertension risk, whereas our CIs remained compatible with no change in risk.}

Three extensions are especially relevant. {One is to tighten the bounds by using monotonicity to restrict the direction of change in the conditional mean as one exposure increases and Lipschitz continuity to limit the magnitude of that change.}
A second is to account for uncertainty in an estimated $\calH$, treating the region as an additional nuisance parameter. A possible route is to recast its effect on the bound endpoints as a lower-dimensional inferential problem that may permit parametric-rate inference even when the region estimator converges more slowly. Another is to extend the framework to longitudinal MTPs by applying the inside--outside decomposition recursively to sequential outcome regressions, with a positivity region and Lipschitz restriction at each treatment time \citep{diaz2023nonparametric}.

\clearpage
\setcounter{equation}{0}
\renewcommand{\theequation}{S\arabic{equation}}
\providecommand{\theHequation}{}
\renewcommand{\theHequation}{supp.\arabic{equation}}
\setcounter{figure}{0}
\renewcommand{\thefigure}{S\arabic{figure}}
\providecommand{\theHfigure}{}
\renewcommand{\theHfigure}{supp.\arabic{figure}}
\setcounter{table}{0}
\renewcommand{\thetable}{S\arabic{table}}
\providecommand{\theHtable}{}
\renewcommand{\theHtable}{supp.\arabic{table}}
\setcounter{algorithm}{0}
\renewcommand{\thealgorithm}{S\arabic{algorithm}}
\providecommand{\theHalgorithm}{}
\renewcommand{\theHalgorithm}{supp.\arabic{algorithm}}
\setcounter{Theorem}{0}
\renewcommand{\theTheorem}{S\arabic{Theorem}}
\providecommand{\theHTheorem}{}
\renewcommand{\theHTheorem}{supp.\arabic{Theorem}}
\setcounter{Corollary}{0}
\renewcommand{\theCorollary}{S\arabic{Corollary}}
\providecommand{\theHCorollary}{}
\renewcommand{\theHCorollary}{supp.\arabic{Corollary}}
\setcounter{Proposition}{0}
\renewcommand{\theProposition}{S\arabic{Proposition}}
\providecommand{\theHProposition}{}
\renewcommand{\theHProposition}{supp.\arabic{Proposition}}
\setcounter{Lemma}{0}
\renewcommand{\theLemma}{S\arabic{Lemma}}
\providecommand{\theHLemma}{}
\renewcommand{\theHLemma}{supp.\arabic{Lemma}}
\setcounter{Example}{0}
\renewcommand{\theExample}{S\arabic{Example}}
\providecommand{\theHExample}{}
\renewcommand{\theHExample}{supp.\arabic{Example}}
\setcounter{Remark}{0}
\renewcommand{\theRemark}{S\arabic{Remark}}
\providecommand{\theHRemark}{}
\renewcommand{\theHRemark}{supp.\arabic{Remark}}
\setcounter{Definition}{0}
\renewcommand{\theDefinition}{S\arabic{Definition}}
\providecommand{\theHDefinition}{}
\renewcommand{\theHDefinition}{supp.\arabic{Definition}}
\setcounter{Assumption}{0}
\renewcommand{\theAssumption}{S\arabic{Assumption}}
\providecommand{\theHAssumption}{}
\renewcommand{\theHAssumption}{supp.\arabic{Assumption}}
\begin{center}
{\bf \Large Supplementary Material for ``Causal Effects of Modified Treatment
Policies under Positivity Violations: Partial
Identification and Inference''}
\end{center}

\appendix
{These supplements are organized as follows:
\begin{itemize}
    \item Section~\ref{append: algorithm} presents the algorithm for estimation and inference with fixed inputs and gives implementation details.
    \item Building on the main text, Section~\ref{append: methods theory} presents additional methods and theory on alternative formulations of Lipschitz continuity, anchor choices, bound widths, and inference with fixed and data-adaptive regions.
    \item Section~\ref{append: sim} describes the simulation designs and reports additional results on region construction, sample splitting, and tuning parameters.
    \item Section~\ref{append: app} describes preprocessing for the CHAMACOS analysis and reports sensitivity analyses and density-ratio diagnostics.
    \item Section~\ref{sec: proofs} collects the proofs of the main and supplementary results.
\end{itemize}}

\section{Algorithm for Estimation and Inference}\label{append: algorithm}\label{append: impl nuisance}\label{append: impl var}\label{append: impl boot}
The algorithm below collects the estimation and inference steps of Section~\ref{sec: est inf} for fixed inputs.
Section~\ref{append: impl details} describes the numerical exceptions used when computing the residual corrections. Within the algorithm, a superscript $(j)$ denotes evaluation at $\epsilon_j$.
\begin{algorithm}[htp!]
\caption{Estimation and Uniform Inference over $\epsilon$ for $\thetad$ with Fixed Inputs}\label{alg: est tau(H)}
\begin{algorithmic}[1]
\Require Data $\{(Y_i,\bfA_i, \bfX_i)\}_{i=1}^n$; MTP $\bbmd$; fixed region of adequate positivity $\calH$; Lipschitz constants $L(\cdot)$; scale matrix $\mathbf{V}$; train/test folds $\{(\mathcal{T}_k,\mathcal{I}_k)\}_{k=1}^K$; displacement grid $\{\epsilon_1,\ldots,\epsilon_J\}$; significance level $\alpha\in(0,1)$; number of bootstrap draws $M$;
\Ensure CI $\widehat{\rm CI}_{\alpha}$;
\For{$j \gets 1,\ldots,J$}
    \For{$k \gets 1,\ldots,K$}
        \State Estimate the nuisances $\widehat{Q}_k,\widehat{r}^{{\rm in}}_k,\widehat{r}^{(j)}_k$ on $\mathcal{T}_{k}$;
        \State For each $i\in\mathcal{I}_k$, evaluate $\widehat{\Psi}_{*,i}^{(j)}=\thetahat_i^{\rm in}+\thetahat_{*,i}^{(j)}$ from \eqref{eq: theta k}, $*\in\{l,u\}$;
        \State Construct $\thetahat_k^{\rm in},\thetahat_{l,k}^{(j)},\thetahat_{u,k}^{(j)}$ and $(\sigmahat_{l,k}^{(j)})^2,(\sigmahat_{u,k}^{(j)})^2$ as in \eqref{eq: sigma k};
    \EndFor
    \State Aggregate $\widehat{{\rm PI}}_{l}^{(j)}$, $\widehat{{\rm PI}}_{u}^{(j)}$, $(\sigmahat_l^{(j)})^{2}$, and $(\sigmahat_u^{(j)})^{2}$ as in \eqref{eq: aggregation of theta and sigma};
\EndFor
\For{$m \gets 1,\ldots,M$}
    \State Draw i.i.d.\ multipliers $\{\xi_i^{[m]}\}_{i=1}^n\sim^{i.i.d}(0,1)$;
    \State For each $j$, form $\widehat{T}_{l,j}^{[m]},\widehat{T}_{u,j}^{[m]}$ and $\widehat{D}^{[m]}$ as in Section~\ref{subsec: estimators};
\EndFor
\State Set $\widehat{q}_{1-\alpha}$ to the $1-\alpha$ empirical quantile of $\{\widehat{D}^{[m]}\}_{m=1}^M$;
\State Construct $\widehat{\rm CI}_{\alpha}$ as in \eqref{eq: unif CI};
\end{algorithmic}
\end{algorithm}
\FloatBarrier

\paragraph{Implementation details.}
Density-ratio classification uses the observed and generated samples described in Section~\ref{subsec: estimators}. For each ratio, we stack all observed training values (label 0) with the corresponding generated values (label 1), without rebalancing the classes. If $\widehat p_k(\bfw)$ is the fitted probability of label 1, the ratio estimate is $\widehat r_k(\bfw)=\widehat p_k(\bfw)/\{1-\widehat p_k(\bfw)\}$; See Section 5.3 of \citet{diaz2023nonparametric} for details of the density ratio learning with classification procedure.

Each observation's observed and generated rows remain in the same fold {during cross-validation for classifier selection within the training sample}. If {the training sample contains no policy-generated values inside or no policy-generated values outside} $\calH$, {we set the corresponding density-ratio estimate to zero}. Section~\ref{append: impl details} describes two departures from the theoretical procedure: a fixed clipping floor can prevent ratio consistency where the true anchor ratio is zero, and omission of a residual correction can depend on the evaluation fold.

In the following, we show the exact formulas of the terms aggregated in \eqref{eq: aggregation of theta and sigma}. For $i\in\mathcal I_k$,
\begin{equation}\label{eq: theta k}
\begin{aligned}
     \thetahat_{i}^{\rm in} &=\left(Y_i-\widehat{Q}_k(\bfW_i)\right)\widehat{r}^{{\rm in}}_k(\bfW_i)\,\I(\bfW_i\in\calH)+\widehat{Q}_k(\bfWd_i)\,\I(\bfWd_i\in\calH),\\
     \thetahat_{l,i}^{(\epsilon)} &= \{Y_i-\widehat{Q}_k(\bfW_i)\}\widehat{r}_k^{(\epsilon)}(\bfW_i)\I(\bfW_i\in\calH)\\     &\quad+\big[\widehat{Q}_k(\tilde{\Pi}^{\epsilon}(\bfWd_i))- L(\bfX_{D,i})\|\bfWd_i-\tilde{\Pi}^{\epsilon}(\bfWd_i)\|_{\mathbf{V}}\big]\I(\bfWd_i\notin\calH),\\
     \thetahat_{u,i}^{(\epsilon)} &= \{Y_i-\widehat{Q}_k(\bfW_i)\}\widehat{r}_k^{(\epsilon)}(\bfW_i)\I(\bfW_i\in\calH)\\     &\quad+\big[\widehat{Q}_k(\tilde{\Pi}^{\epsilon}(\bfWd_i))+ L(\bfX_{D,i})\|\bfWd_i-\tilde{\Pi}^{\epsilon}(\bfWd_i)\|_{\mathbf{V}}\big]\I(\bfWd_i\notin\calH),
\end{aligned}
    \end{equation}
so we have $\widehat\Psi_{*,i}^{(\epsilon)}=\thetahat_i^{\rm in}+\thetahat_{*,i}^{(\epsilon)}$. Let $\thetahat_k^{\rm in}$ and $\thetahat_{*,k}^{(\epsilon)}$ denote the averages of $\thetahat_i^{\rm in}$ and $\thetahat_{*,i}^{(\epsilon)}$ over $i\in\mathcal I_k$, respectively. The empirical variance of $\widehat\Psi_{*,i}^{(\epsilon)}$ within fold $k$ is
\begin{align}
    &\left(\sigmahat_{*,k}^{(\epsilon)}\right)^2 =   \frac{1}{|\mathcal{I}_k|-1}\sum_{i\in\mathcal{I}_k}\left[\widehat{\Psi}_{*,i}^{(\epsilon)}-\left(\thetahat_{k}^{\rm in}+\thetahat_{*,k}^{(\epsilon)}\right)\right]^2,\quad\text{for}\quad \text{$* = l$ and $u$.} \label{eq: sigma k}
\end{align}

We average these within-fold variances equally rather than weighting them by fold size as in \eqref{eq: aggregation of theta and sigma}. The two averages coincide for equal-sized folds and have the same limit when $|\mathcal I_k|/n\to1/K$.

\section{Additional Methods and Theory}\label{append: methods theory}
We provide additional methods and theory that expand on the main text, including results beyond those presented there.

\subsection{Alternative Lipschitz Assumptions and Anchor Choices}\label{append: notation}\label{append: add meth}\label{append: source anchor}
{This section considers alternatives to two features of the main analysis: how Lipschitz continuity is formulated and the choice of anchor.}

{Assumption~\ref{assump: weight Lip} in the main paper uses a single Lipschitz constant within each stratum and measures distance in a weighted Euclidean norm.} An alternative assigns a separate constant to each continuous coordinate.

\begin{Assumption}[Coordinate-wise Lipschitz continuity] \label{assump: coord Lip}
    For given discrete value $\bfd$, a function $\mu(\cdot,\bfd)$ is coordinate-wise Lipschitz continuous with a vector of nonnegative constants $\mathbf{L}(\bfd)=(L_1(\bfd),\dots,L_p(\bfd))\tr$, with respect to the continuous $\bfc$, for every $\bfc\in\calH(\bfd)$ and every $\bfc'$:
    \begin{equation*}
        \left|\mu(\bfc',\bfd)-\mu(\bfc,\bfd)\right|\leq \sum_{j=1}^pL_j(\bfd)|c_j-c_j'|.
    \end{equation*}
    When no discrete argument is present, $\mu(\cdot)$
satisfies the same condition with a fixed vector of nonnegative constants $\mathbf{L}=(L_1,\dots,L_p)\tr$.
\end{Assumption}

Under Assumption~\ref{assump: coord Lip}, $L_j(\mathbf d)$ bounds the change in the conditional mean potential outcome per unit change in coordinate $j$ when the other coordinates are held fixed. When several coordinates change, their contributions to the bound add rather than combine through the weighted Euclidean distance.
The two assumptions imply each other up to a change of constants. Suppose Assumption~\ref{assump: coord Lip} holds. Suppose $\mathbf{V}$ is a diagonal matrix with positive diagonal elements as in Assumption~\ref{assump: weight Lip}. By the Cauchy--Schwarz inequality,
\begin{equation*}
    \sum_{j=1}^p L_j(\bfd)\,|c_j-c_j'| \;\le\; \Bigl(\textstyle\sum_{j=1}^p L_j(\bfd)^2/V_{jj}\Bigr)^{1/2}\,\|\bfc-\bfc'\|_{\mathbf{V}},
\end{equation*}
so Assumption~\ref{assump: weight Lip} holds with $L(\bfd)=\{\mathbf{L}(\bfd)\tr\mathbf{V}^{-1}\mathbf{L}(\bfd)\}^{1/2}$. Conversely, if Assumption~\ref{assump: weight Lip} holds, then $\|\bfc-\bfc'\|_{\mathbf V}\leq\sum_{j=1}^pV_{jj}^{1/2}|c_j-c_j'|$ gives Assumption~\ref{assump: coord Lip} with $L_j(\bfd)=L(\bfd)V_{jj}^{1/2}$ for the same pairs of points.

Under Assumption~\ref{assump: coord Lip}, the partial identification results of Proposition~\ref{prop: partial gen} with anchor map $g$ carry over after replacing the extrapolation penalty $L(\bfd)\|\bfw-g(\bfw)\|_{\mathbf V}$ with $\sum_j L_j(\bfd)|w_j-g_j(\bfw)|$.

When $L_j(\bfd)>0$ for every $j$, a bound-minimizing anchor is a metric projection in the weighted $\ell_1$ norm. We use the weighted Euclidean norm because projection onto a closed convex region is unique, whereas weighted $\ell_1$ projection can be set-valued. This uniqueness supports the interior-displacement and change-of-variables arguments in Section~\ref{subsec: reg proj}. In practice, coordinate-wise constants elicited from subject-matter knowledge can be converted to the Euclidean form through the display above; the two penalties agree when the vector of absolute coordinate differences is proportional to $\mathbf{V}^{-1}\mathbf{L}(\bfd)$.
\paragraph{A source-anchor benchmark.}
{Here we retain Assumption~\ref{assump: weight Lip} but consider an alternative to projection: we anchor each policy-generated value $\bfWd$ outside $\calH$} at its observed source value $\bfW$. This gives
$ \thetad \in [\thetainH+\thetal^{({\rm src})},\ \thetainH+\thetau^{({\rm src})}]$, where $\thetainH$ is identified as \eqref{eq: id of theta in} and
\begin{equation}\label{eq: theta source}
    \begin{aligned}
        &\thetal^{({\rm src})} = \E\left[\left(Y-L(\bfX_D)\|\bfW-\bfWd\|_{\mathbf V}\right)\I(\bfWd\notin\calH)\right],\\
        &\thetau^{({\rm src})}=\E\left[\left(Y+L(\bfX_D)\|\bfW-\bfWd\|_{\mathbf V}\right)\I(\bfWd\notin\calH)\right].
    \end{aligned}
\end{equation}
The source-anchor interval does not require an invertible policy. It is valid when every observed source value lies in $\calH$, because Assumption~\ref{assump: weight Lip} then covers each pair $(\bfW,\bfWd)$. {In this case, the hard projection gives an interval no wider than the source-anchor interval because it selects the nearest point in $\calH$ (Proposition~\ref{prop: partial gen}). When both the source and its policy-generated value lie outside $\calH$, the source may be closer to the policy-generated value than the hard projection is. Although Assumption~\ref{assump: weight Lip} does not cover this pair, we can still use the shorter distance in the bound if the same Lipschitz inequality holds between the two values.} No general ordering holds against the interior-displaced interval of Section~\ref{subsec: reg proj} either.

{To illustrate this distinction, we compare the two anchors under DGP1 from Section~\ref{sec: sim} (Table~\ref{tab: source benchmark}). Within each stratum of $X$, we define $\calH$ as the set of exposure pairs with $A_1$ at or above the stratum-specific first percentile of $A_1$. We use the globally Lipschitz outcome model $Q_1$, with $\mathbf V=\mathbf I$ and $L=\sqrt{0.61}$, so the source-anchor bound remains valid even when both values fall below this threshold. When a reduction takes $A_1$ from above the threshold to below it, the hard projection gives the narrower bound. When $A_1$ is already below the threshold before reduction, the source lies between the threshold and the reduced value, so the source anchor gives the narrower bound.}

\begin{table}[htp!]
\centering
{\begin{tabular}{lrrr}
\hline
$A_1$ before reduction & Reduction in $A_1$ & Hard projection & Source anchor \\
\hline
At or above the threshold & 20\% & 0.563 & 1.465 \\
                          & 30\% & 0.845 & 2.413 \\
Below the threshold       & 20\% & 1.419 & 1.284 \\
                          & 30\% & 2.061 & 1.926 \\
\hline
\end{tabular}}
\caption{{Mean Lipschitz-bound widths among exposure pairs for which the reduced $A_1$ falls below the stratum-specific threshold in DGP1, by $A_1$ before reduction.}}
\label{tab: source benchmark}
\end{table}

\subsection{The Width Gap Between Sharp and Projection Intervals}\label{append: sharpness}

We compare the hard-projection interval ${\rm PI}^{\Pi}$ with the sharp interval ${\rm PI}^{\star}$ based on the Lipschitz envelopes in \eqref{eq: envelopes main} \citep{mcshane1934extension,whitney1934analytic}. We consider finite endpoints; for binary $Y$, each pointwise interval is intersected with $[0,1]$ before taking expectations.

Throughout, we work under Assumption~\ref{assump: weight Lip} and hold fixed the distribution of $\bfW=(\bfA\tr,\bfX\tr)\tr$ and the policy $\bbmd$. We write $\bfw=(\bfc,\bfd)$ and abbreviate $\Pi=\Pi_{\calH}$.

The two intervals coincide when the continuous space is one-dimensional and $\calH(\bfd)$ is an interval. In higher dimensions, however, the projection interval can be wider when the change in $Q$ along the boundary outweighs the additional distance penalty.

\begin{figure}[htbp]
\centering
\includegraphics[width=\linewidth]{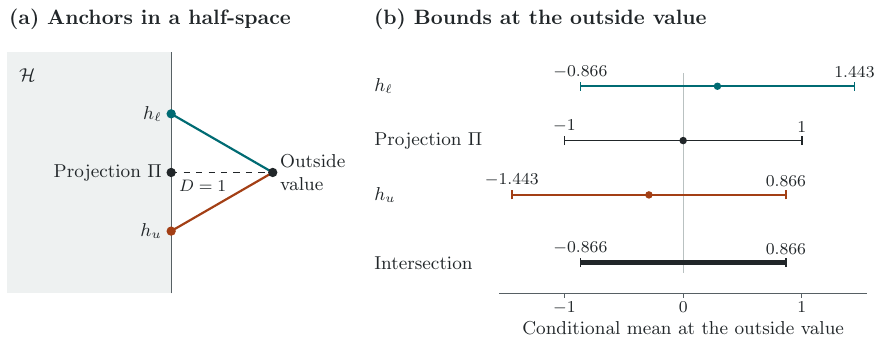}
\caption{Pointwise bounds in a half-space with $D=L(\bfd)=1$, $a=0.5$, and $Q=0$ at the projection. Panel (a) shows the projection and the boundary anchors $h_\ell$ and $h_u$ that attain the lower and upper endpoints, respectively. Panel (b) shows their anchor-specific intervals, centered at the corresponding values of $Q$ (dots), and the intersection over all anchors.}
\label{fig: halfspace anchors}
\end{figure}

To see when this happens, let $\calH(\bfd)$ be a half-space, take $\mathbf V=\mathbf I$, and consider an outside value $\bfw$ at distance $D$ from the boundary. By Assumption~\ref{assump: weight Lip}, the boundary point on the segment from any interior anchor to $\bfw$ gives an interval contained in that anchor's interval. If $Q$ is linear along the boundary with tangential slope norm $a<L(\bfd)$, minimizing $at+L(\bfd)(D^2+t^2)^{1/2}$ over signed displacement $t$ along the slope from the projection gives $t^\star=-aD/\{L(\bfd)^2-a^2\}^{1/2}$ for the upper endpoint. Before any truncation to the outcome range, the pointwise widths and anchor distances at $\bfw$ satisfy
\begin{equation}\label{eq: width ratio}
    \frac{\text{sharp width}}{\text{projection width}}=\left\{1-\left(\frac{a}{L(\bfd)}\right)^{2}\right\}^{1/2},
    \quad
    \frac{\|\bfw-\bfh^{\star}\|_{\mathbf V}}{\|\bfw-\Pi(\bfw)\|_{\mathbf V}}=\left\{1-\left(\frac{a}{L(\bfd)}\right)^{2}\right\}^{-1/2},
\end{equation}
where $\bfh^\star$ is the boundary anchor attaining the upper endpoint. At $a/L(\bfd)=0.5$, the projection interval is about $15\%$ wider than the sharp pointwise interval, and the extremal anchor is about $15\%$ farther from $\bfw$ than the projection; see Figure~\ref{fig: halfspace anchors}.

Table~\ref{tab: sharp pop} compares policy-mean interval widths across the designs of Section~\ref{sec: sim}. {We use $Q_1$ because it satisfies the Lipschitz condition globally with the oracle constant given in Section~\ref{append: impl details}. For each design, we generated 200,000 Monte Carlo draws. Within each stratum, we took the convex hull of the $99\%$ of draws with the highest conditional density. For each sampled policy value outside the region, we optimized the lower and upper bounds over all anchors in the polygon before averaging the resulting widths.}
\begin{table}[htp!]
\centering
\begin{tabular}{llcccccc}
\toprule
 & & \multicolumn{6}{c}{Reduction in both exposures} \\
\cmidrule(lr){3-8}
Design & & $5\%$ & $10\%$ & $15\%$ & $20\%$ & $25\%$ & $30\%$ \\
\midrule
\multirow{2}{*}{DGP1} & Exterior fraction & 0.042 & 0.089 & 0.153 & 0.238 & 0.344 & 0.470 \\
 & {Width ratio} & {0.727} & {0.765} & {0.777} & {0.785} & {0.795} & {0.810} \\
\midrule
\multirow{2}{*}{DGP2} & Exterior fraction & 0.008 & 0.011 & 0.016 & 0.028 & 0.049 & 0.089 \\
 & {Width ratio} & {0.735} & {0.839} & {0.883} & {0.914} & {0.936} & {0.949} \\
\midrule
\multirow{2}{*}{DGP3} & Exterior fraction & 0.009 & 0.008 & 0.008 & 0.008 & 0.009 & 0.012 \\
 & {Width ratio} & {0.803} & {0.814} & {0.835} & {0.854} & {0.881} & {0.910} \\
\bottomrule
\end{tabular}
\caption{{Estimated ratio of sharp to projection interval width under $Q_1$ and} multiplicative reductions of both exposures, by design. The exterior fraction {is} $\Pr(\bfWd\notin\calH)$.}
\label{tab: sharp pop}
\end{table}

The projection interval is between {$5.4\%$} and {$37.5\%$} wider {than the sharp interval}. About $8.9\%$ of policy values fell outside the region under both the $10\%$ reduction in DGP1 and the $30\%$ reduction in DGP2, yet the sharp-to-projection width ratios were $0.765$ and $0.949$, respectively. In DGP1, moving along the truncation boundary changes $A_2$ while leaving $A_1$ fixed, so a more distant anchor can tighten the bound. In DGP2, moving along the boundary near most projections trades an increase in one exposure for a decrease in the other. These changes largely offset in $Q_1$, leaving less to gain from a more distant anchor. These comparisons concern identification width, not the lengths of confidence intervals.

\subsection{Hard Projection and the Displacement Trade-off}\label{append: hard nondiff}
We examine why the hard-projection endpoints can lack influence functions and how moving the anchors into the interior affects bound width and estimation precision.
We work in the rescaled coordinates $\mathbf V^{1/2}\bfc$, in which the $\mathbf V$-weighted distance is Euclidean.

Hard projection maps exterior policy values to the boundary of $\calH$, where the observed distribution assigns zero probability. This rules out the usual density-ratio representation but leaves open whether the endpoints have an influence function. The following proposition gives conditions under which neither endpoint has one.

\begin{Proposition}[No influence function for the hard projection]\label{prop: hard nondiff}
    {Suppose Assumption~\ref{assump: basic} holds, $\calH$ is as in Definition~\ref{def: positivity region}, $Y$ is bounded, and $\mu$ satisfies Assumption~\ref{assump: weight Lip} with the common margin $\eta>0$ of Proposition~\ref{prop: efficiency}.} Suppose the observed distribution assigns zero probability to the boundary of $\calH(\bfX_D)$, i.e., $\prob\{\bfC\in\partial\calH(\bfX_D)\}=0$, and, for some $\delta_0,\sigma^2_{\min}>0$, $\Var(Y\mid\bfW=\bfw)\geq\sigma^2_{\min}$ whenever $\operatorname{dist}(\bfc,\partial\calH(\bfd))<\delta_0$. If $\E[L(\bfX_D)\|\bfWd-\Pi_{\calH}(\bfWd)\|_{\mathbf V}]<\infty$ and $\prob(\bfWd\notin\calH)>0$, then neither hard-projection endpoint in \eqref{eq: partial theta out} is pathwise differentiable under the maintained model.
\end{Proposition}

See Section~\ref{pf: hard nondiff} for the proof.

The interior-displaced projection in Section~\ref{subsec: reg proj} of the main text moves the anchors off the boundary. The resulting endpoints have influence functions under the conditions of Theorem~\ref{thm: EIF}, allowing us to conduct inference, but the bounds can be wider. The following proposition quantifies the trade-off between bound width and influence-function variance as the displacement decreases. Write ${\rm PI}_*^{(\Pi)}$ and ${\rm PI}_*^{(\epsilon)}$, $*\in\{l,u\}$, for the endpoints of Proposition~\ref{prop: partial gen} with $g=\Pi_{\calH}$ and $g=\tilde\Pi_{\calH}^{\epsilon}$, and let $\Psi_*^{(\epsilon)}=\psi^{\rm in}+\psi_*^{(\epsilon)}$ denote the uncentered influence-function representations of Section~\ref{subsec: eif} for the latter.

\begin{Proposition}[Cost of interior displacement]\label{prop: eps tradeoff}
    Fix $\epsilon\in(0,\epsilon_{\max})$ and let $R=\|\bfWd-\Pi_{\calH}(\bfWd)\|$ and $\Delta(\epsilon)=\E[L(\bfX_D)\,\phi^{\epsilon}(R)\,\I(\bfWd\notin\calH)]$.
    \begin{enumerate}
        \item[(i)] Under the conditions of Proposition~\ref{prop: partial gen} and Assumption~\ref{assump: boundary reg}, suppose in addition that $\E[L(\bfX_D)\I(\bfWd\notin\calH)]<\infty$. Then
        \[
        {\rm PI}_l^{(\Pi)}-2\Delta(\epsilon)\ \leq\ {\rm PI}_l^{(\epsilon)}\ \leq\ {\rm PI}_l^{(\Pi)},\quad
        {\rm PI}_u^{(\Pi)}\ \leq\ {\rm PI}_u^{(\epsilon)}\ \leq\ {\rm PI}_u^{(\Pi)}+2\Delta(\epsilon),
        \]
        and $\Delta(\epsilon)\leq \epsilon\,\E[L(\bfX_D)\I(\bfWd\notin\calH)]$ because $\phi^{\epsilon}(r)<\epsilon$.
        \item[(ii)] Suppose also that Theorem~\ref{thm: EIF} applies to $g=\tilde\Pi_{\calH}^{\epsilon}$. On the interior layer $\mathcal L_{\epsilon}(\bfd)=\{\bfc\in{\rm int}\,\calH(\bfd):{\rm dist}(\bfc,\partial\calH(\bfd))<\epsilon\}$, suppose the conditional density of $\bfC$ given $\bfX_D=\bfd$ is bounded above by $\bar f(\bfd)$, with $0<c_{\calH}=\sum_{\bfd}\prob(\bfX_D=\bfd)\,\bar f(\bfd)\,|\partial\calH(\bfd)|_{p-1}<\infty$, where $|\cdot|_{p-1}$ denotes the $(p-1)$-dimensional surface area. Finally, suppose $\Var(Y\mid\bfW)\geq\sigma^2_{\min}>0$ almost surely on $\{r^{(\epsilon)}(\bfW)>0\}$. Then, for $*\in\{l,u\}$,
        \[
            \Var\left(\Psi_*^{(\epsilon)}\right)\ \geq\ \frac{\sigma^2_{\min}\,\prob(\bfWd\notin\calH)^2}{c_{\calH}\,\epsilon}.
        \]
    \end{enumerate}
\end{Proposition}

See Section~\ref{pf: eps tradeoff} for the proof.

Interior displacement increases the width of the hard-projection interval by exactly $2\Delta(\epsilon)$, which is at most of order $\epsilon$. As $\epsilon$ decreases, this excess width vanishes, but the same exterior policy mass becomes concentrated in a thinner interior layer. If the exterior probability is positive and the bounds in part (ii) hold with the same constants across displacement levels, the influence-function variance for each endpoint is bounded below by a positive constant times $1/\epsilon$ and can grow faster.

\subsection{Fixed-Displacement Inference and Bounded Outcomes}\label{append: est inf}\label{append: IM interval}\label{append: impl bounded}
For a single displacement level $\epsilon$, the Imbens--Manski interval combines uncertainty about both endpoints \citep{imbens2004confidence}:
\begin{align}
   {\rm CI}_{\alpha}^{(\epsilon)} = \left[\widehat{{\rm PI}}_l^{(\epsilon)}-C_{\alpha}\cdot\sigmahat_l^{(\epsilon)}/\sqrt{n},\ \widehat{{\rm PI}}_u^{(\epsilon)}+C_{\alpha}\cdot\sigmahat_u^{(\epsilon)}/\sqrt{n}\right],\label{eq: CI theta}
\end{align}
where $C_{\alpha}$ satisfies
\begin{align*}
    \Phi\left(C_{\alpha}+\sqrt{n}\left(\widehat{{\rm PI}}_u^{(\epsilon)}-\widehat{{\rm PI}}_l^{(\epsilon)}\right)/\max\left\{\sigmahat_l^{(\epsilon)},\sigmahat_u^{(\epsilon)}\right\}\right)-\Phi\left(-C_{\alpha}\right) = 1-\alpha.
\end{align*}
If the estimated bound interval has zero width, $C_\alpha=z_{\alpha/2}$, where $z_q$ denotes the upper $q$ quantile of the standard normal distribution. For a wide interval, $C_\alpha\approx z_\alpha$.
Combining the joint asymptotic normality established in Theorem~\ref{thm: asy-normal} with the CI construction of \citet{imbens2004confidence}, we obtain the following coverage guarantee.
\begin{Proposition}\label{prop: coverage}
Under the assumptions of Theorem~\ref{thm: asy-normal}, suppose the marginal asymptotic variances are positive. For fixed $\epsilon$, ${\rm CI}_{\alpha}^{(\epsilon)}$ in \eqref{eq: CI theta} satisfies $ \liminf_{n\to\infty} \prob(\thetad\in{\rm CI}_{\alpha}^{(\epsilon)})\geq 1-\alpha.$
\end{Proposition}

\begin{proof}
For positive bound width, the result follows from Theorem~\ref{thm: asy-normal}, variance consistency, and the calibration of \citet{imbens2004confidence}. At zero width, the nonnegative Lipschitz penalty vanishes almost surely, so the two endpoint contributions in \eqref{eq: theta k} coincide and the interval reduces to the usual two-sided Wald interval.
\end{proof}
We establish asymptotic coverage for each fixed distribution satisfying the assumptions of the proposition, but we do not establish uniform coverage over this class of distributions \citep[see][]{stoye2009more}.
\paragraph{Bounded outcomes.}
When the outcome is bounded, the parameter space supplies additional information at no inferential cost. For binary $Y$, $\thetad\in[0,1]$ and $\taud\in[-1,1]$, so the estimated bound endpoints and every reported CI may be intersected with the corresponding range. Intersecting the population bounds and the CI with the known target range preserves validity of the bounds and coverage of the target. The set-coverage guarantee of Corollary~\ref{cor: uniform set coverage} then concerns the population bound interval intersected with the same range. Truncating instead inside the expectations in \eqref{eq: partial theta out} yields narrower valid bounds, but introduces a kink in the endpoint functionals whose influence-function analysis requires an additional condition (e.g., margin condition). We therefore apply the range restriction only after computing the bounds and CI.

\subsection{Region Construction and Inference with Data-Adaptive Regions}\label{append: region}\label{append: impl region}\label{append: impl smooth}\label{append: impl smoothprop}\label{append: impl tuning}\label{append: impl adaptive}\label{append: impl split}
We construct the convex hull of the continuous treatment--covariate values separately within each stratum $\bfd$:
\begin{equation}\label{eq: full sample hull}
\Hhat_{\Delta} =\left\{(\bfc,\bfd): \bfc = \sum_{i:\bfX_{D,i}=\bfd}\lambda_i\bfC_i,\quad \lambda_i\geq 0,\quad \sum_{i:\bfX_{D,i}=\bfd} \lambda_i = 1\right\}.
\end{equation}
Combining discrete levels in one hull would create artificial interpolation between strata.

The full hull can stretch across sparsely populated tails. A more conservative alternative begins with the highest-density region (HDR) \citep{hyndman1996computing,bao2025addressing}
\begin{equation}\label{def: HDR}
    \calH_{\alpha}(\bfd)=\left\{\bfc : f(\bfc\mid\bfd) \geq q_{\alpha}\right\},
\end{equation}
where $q_{\alpha}$ satisfies $\int_{f(\bfc\mid\bfd) \geq q_{\alpha}} f(\bfc\mid\bfd) d\bfc = 1-\alpha$. To select observations for the hull using the HDR in \eqref{def: HDR}, we fit the density within each stratum and set $\widehat t$ to the empirical $\alpha$-quantile of the fitted densities. Taking the convex hull of the retained observations gives
\begin{equation}\label{def: HDR hull}
    \widehat{\calH}_{\rm dens}(\bfd) = \mathrm{Conv}\bigl\{\bfc_i : i\in\mathcal{I}(\bfd),\ \widehat{f}(\bfc_i\mid\bfd)\ge \widehat{t}\,\bigr\}.
\end{equation}
Here $\mathcal I(\bfd)$ indexes the observations used to construct the region in stratum $\bfd$. The implementation retains ties at the cutoff. If fewer than $p+2$ observations remain, it uses all training observations in that stratum. As in Section~\ref{subsec: pos region}, this changes the region used for extrapolation, not the target population.

\begin{figure}[htp!]
   \centering
   \includegraphics[width = .8\textwidth]{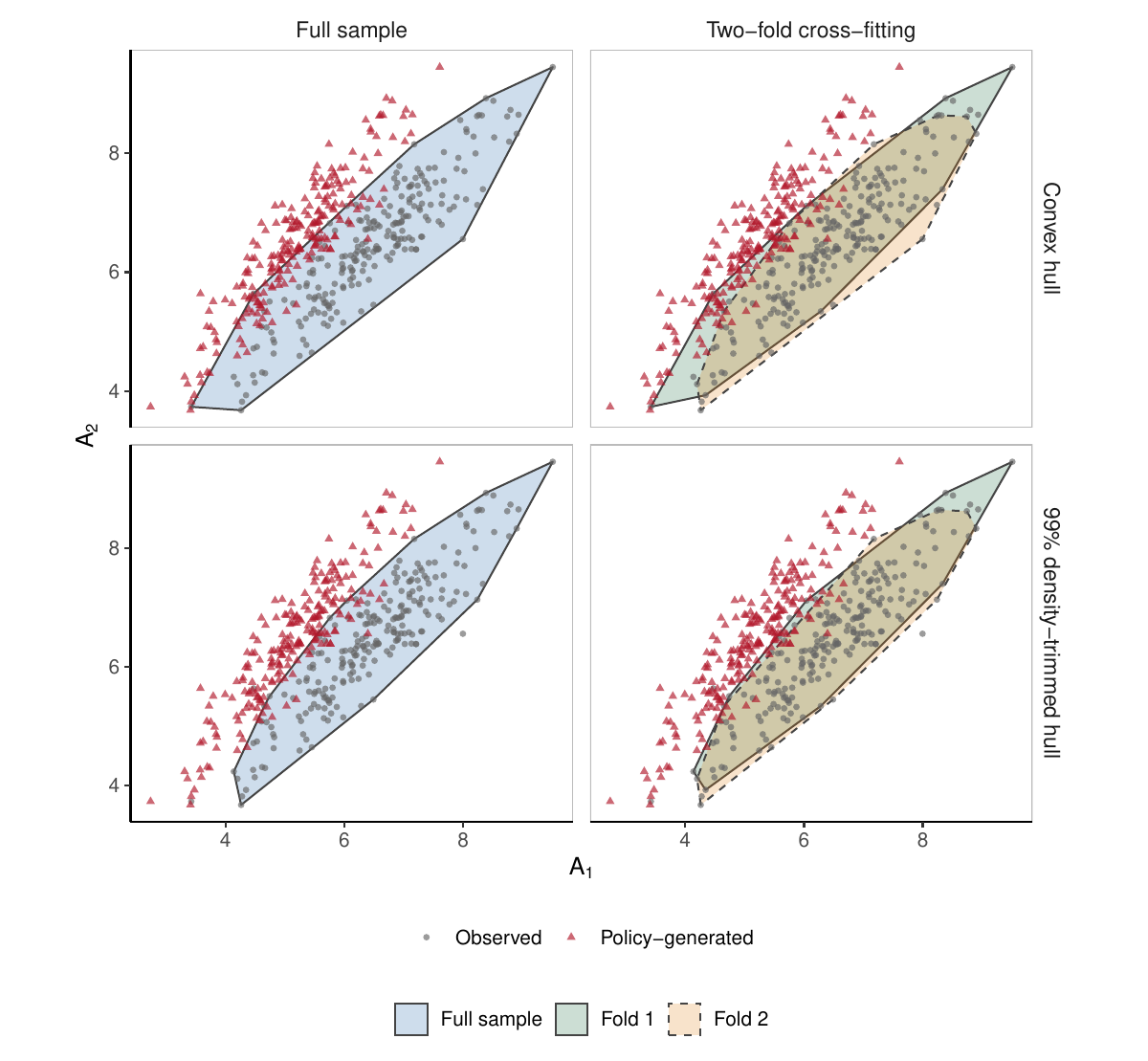}
   \caption{Estimated regions of adequate positivity: the full convex hull (top) and 99\% trimmed hull (bottom), constructed from the full sample (left) or by two-fold sample splitting (right).}
   \label{fig: hull hdr eg}
\end{figure}

Figure~\ref{fig: hull hdr eg} compares the full convex hull in \eqref{eq: full sample hull} with the 99\% density-trimmed hull in \eqref{def: HDR hull} under DGP2 (Section~\ref{append: impl designs}).

Definition~\ref{def: positivity region} in the main text requires the region in each stratum to lie within the conditional support. In practice, this support is unknown, and either sample hull can bridge a gap if the support is nonconvex. The following lemma gives sufficient conditions for both hulls to satisfy the definition.
\begin{Lemma}[Sample hulls satisfy Definition~\ref{def: positivity region} under convex support]\label{lem: hull admissible}
\rm Suppose that in stratum $\bfd$ the conditional support of $\bfC$ is convex with nonempty interior and the conditional distribution has a $p$-dimensional density positive on that interior. If the stratum is observed, the convex hull of any nonempty subset of its observed $\bfC$ values lies in that interior almost surely and satisfies Definition~\ref{def: positivity region}. If these conditions hold in every stratum and $\supp(\bfX_D)$ is finite, all strata are observed with probability tending to one as the sample size grows.
\end{Lemma}
\begin{proof}
The boundary of a convex set with nonempty interior has zero $p$-dimensional volume, so a law admitting a $p$-dimensional density assigns it probability zero, and every observation lies in the interior of the support almost surely. The interior of a convex set is convex, so the hull of the observations is contained in it, and a hull of finitely many points is nonempty, closed, and convex.
With finitely many strata, the probability that any stratum is empty tends to zero, since each has positive probability.
\end{proof}

\begin{figure}[htp!]
    \centering
    \includegraphics[width=0.4\linewidth]{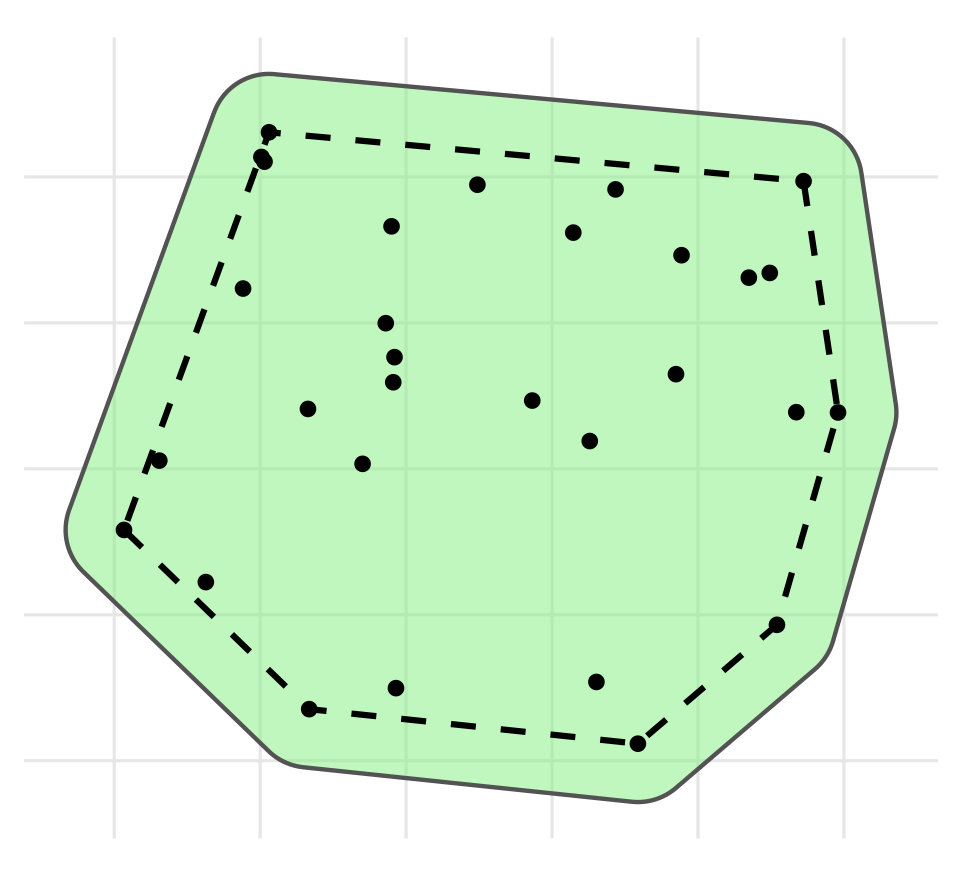}
    \caption{Boundary smoothing. The dashed line is the preliminary convex hull $\calH$ and the filled set is its radius-$\rho$ expansion $\calH_\rho$.}
    \label{fig: smooth conv hull}
\end{figure}

This containment result applies to the preliminary hulls. To obtain the unique outward normal needed for the interior-displaced projection $\tilde{\Pi}$, we round their corners by outward expansion. In the rescaled coordinates where $\mathbf V=\mathbf I$, expand each $\calH(\bfd)$ by a ball of radius $\rho$,
\begin{equation}\label{eq: H rho}
    \calH_{\rho}(\bfd)=\{\bfh+\bfu:\bfh\in\calH(\bfd),\ \|\bfu\|\leq \rho\},
\end{equation}
and let $\calH_\rho$ be the union of the resulting regions. The outer parallel body of a convex set is convex, but in dimensions $p\geq2$ it differs from the polytope obtained by displacing each facet outward by $\rho$: that polytope is larger near the corners, where the parallel body is rounded. We use the larger polytope only to rule out points: a point belongs to the rounded region if and only if its distance to the original hull is at most $\rho$.
Figure~\ref{fig: smooth conv hull} illustrates the construction.

When each preliminary hull has inradius greater than $\rho$, we can consider an inward construction. Specifically, it erodes each $\calH(\bfd)$ by the same radius before expanding it, so the smoothed region is the opening $\{\calH(\bfd)\ominus\rho B\}\oplus\rho B$, which is contained in $\calH(\bfd)$ and whose boundary has reach at least $\rho$: every point less than $\rho$ from the boundary has a unique nearest boundary point. Because the region lies inside the preliminary hull, it satisfies Definition~\ref{def: positivity region} whenever the hull does, and every interior-displaced anchor lies inside the hull. Section~\ref{subsec: sensitivity pos} compares the two constructions numerically.

The next result characterizes the boundary of the expanded region.
\begin{Proposition}\label{prop: reg proj inj}
    For any nonempty proper closed convex set $\calH\subsetneq\mathbb R^p$ and $\rho>0$, $\partial\calH_\rho$ is continuously differentiable and has reach at least $\rho$. Consequently, for every $\epsilon\in(0,\rho)$, the interior-displaced projection onto $\calH_\rho$ satisfies Lemma~\ref{lem: reg proj cov}. In particular, it is injective on $\calH_\rho^c$ and its image is $\mathcal L_\epsilon=\{\bfw\in{\rm int}(\calH_\rho):0<{\rm dist}(\bfw,\partial\calH_\rho)<\epsilon\}$.
\end{Proposition}

See Section~\ref{pf: reg proj inj} for the proof.

The expansion radius $\rho$ gives a simple way to choose the displacement grid without computing the reach of the expanded hull. At each grid point $0<\epsilon<\rho$, distinct policy values outside the hull map to distinct anchors in its interior.

The default simulation grid uses $J=5$ values logarithmically spaced from $0.002\rho$ to $0.95\rho$. For small $\epsilon$, the exponential displacement $\phi^\epsilon(r)$ can round to $\epsilon$ for distinct exterior distances even though the exact map is injective. We report density-ratio diagnostics in Section~\ref{append: impl ratios} rather than removing unstable grid values after inspection. Section~\ref{subsec: sensitivity tuning} examines sensitivity to $q$ and $L$.

\paragraph{Inference with data-adaptive regions.}
Let $\Hhat_{\mathcal I}$ be a region constructed from observations indexed by $\mathcal I$. Substituting the full-sample region into the fixed-region functional gives the data-adaptive components \citep{vanderlaan2015targeted,hubbard2016statistical,montoya2022estimators}
\begin{equation}\label{def: whole sample thetahat}
\thetain(\Hhat_{1:n}) = \left.\E\left[Y(\bfAd)\I(\bfWd\in\calH)\right]\right\vert_{\calH = \Hhat_{1:n}}, \quad\thetaout(\Hhat_{1:n}) = \thetad - \thetain(\Hhat_{1:n}).
\end{equation}
With $K$-fold sample splitting, construct the region on $\mathcal T_k=\{1,\ldots,n\}\setminus\mathcal I_k$ and define
\begin{equation}\label{def: split sample thetahat}
\thetain(\Hhat_{\rm cv}) = \sum_{k=1}^K\frac{|\mathcal I_k|}{n} \left(\left.\E\left[Y(\bfAd)\I(\bfWd\in\calH)\right]\right\rvert_{\calH = \Hhat_{\mathcal{T}_k}}\right), \quad \thetaout(\Hhat_{\rm cv}) = \thetad - \thetain(\Hhat_{\rm cv}).
\end{equation}
The corresponding contrast components are defined analogously.
When the region and the sensitivity inputs are rebuilt on each training fold, the fixed-region theory applies within a fold but not to the fold average. Corollary~\ref{cor: split sample} gives unconditional coverage of the policy mean using the fold-averaged bounds.
{\begin{Corollary}[Inference with cross-validated regions]\label{cor: split sample}
{\normalcolor Use the cross-fitted estimators of Section~\ref{subsec: estimators}, with folds assigned independently of the data and $|\mathcal I_k|/n\to1/K$. For each fold $k$, construct $(\calH_k,L_k,\mathbf V_k)$ using only the training sample $\mathcal T_k$.}
{\normalcolor For $*\in\{l,u\}$, let $\mathrm{PI}^{(j)}_{*,k}$ denote the population endpoint from Proposition~\ref{prop: partial gen} for fold $k$ at displacement $\epsilon_j$, and set $\bar{\mathrm{PI}}^{(j)}_*=\sum_k (|\mathcal I_k|/n)\,\mathrm{PI}^{(j)}_{*,k}$.} Suppose that
\begin{enumerate}
\item[{\normalcolor(i)}] {\normalcolor the conditions of Corollary~\ref{cor: uniform set coverage} hold uniformly over a fixed class $\mathfrak C$, and with probability tending to one, $(\calH_k,L_k,\mathbf V_k)\in\mathfrak C$ for every fold $k$. The nuisance rates and moment bounds in Assumption~\ref{assump: reg nuisance} hold for each fold's selected inputs;}
\item[{\normalcolor(ii)}] {\normalcolor the fold-specific vectors $\boldsymbol\Psi_k$ of uncentered endpoint representations across the grid satisfy $\max_k\E_P\!\left[\|\boldsymbol\Psi_k-\boldsymbol\Psi_\infty\|_2^2\right]\xrightarrow{p}0$ for a fixed $\boldsymbol\Psi_\infty$ with finite, nonsingular covariance.}
\end{enumerate}
{\normalcolor Then $\thetad\in[\max_j\bar{\rm PI}^{(j)}_{l},\min_j\bar{\rm PI}^{(j)}_{u}]$ with probability tending to one, and
$\liminf_{n\to\infty}\prob\big(\thetad\in\widehat{\rm CI}_{\alpha}\big)\geq 1-\alpha$,
where the probability is unconditional.}
\end{Corollary}}

See Section~\ref{pf: split sample} for the proof.

The $K$-fold case is the cross-fitted construction of the main text, in which the region and $(L,\mathbf V)$ are built on each training fold. It is the cross-validated data-adaptive construction of \citet{hubbard2016statistical} and \citet{vanderlaan2015targeted}, with the difference that the target $\thetad$ is fixed and only the bounds are data-adaptive. Condition (ii) asks the fold-specific influence-function representations to converge in $L_2$, but at no rate.

\section{Simulation Details and Additional Results}\label{append: sim}
This section specifies the designs of Section~\ref{sec: sim} and reports the results omitted there.

\subsection{Simulation designs}\label{append: impl designs}\label{sec: add sim}
The distributions and outcome functions used in Section~\ref{sec: sim} are specified below; Section~\ref{append: impl details} gives the Lipschitz constants used by the oracle methods.

\begin{figure}[htp!]
    \centering
    \includegraphics[width=.78\linewidth]{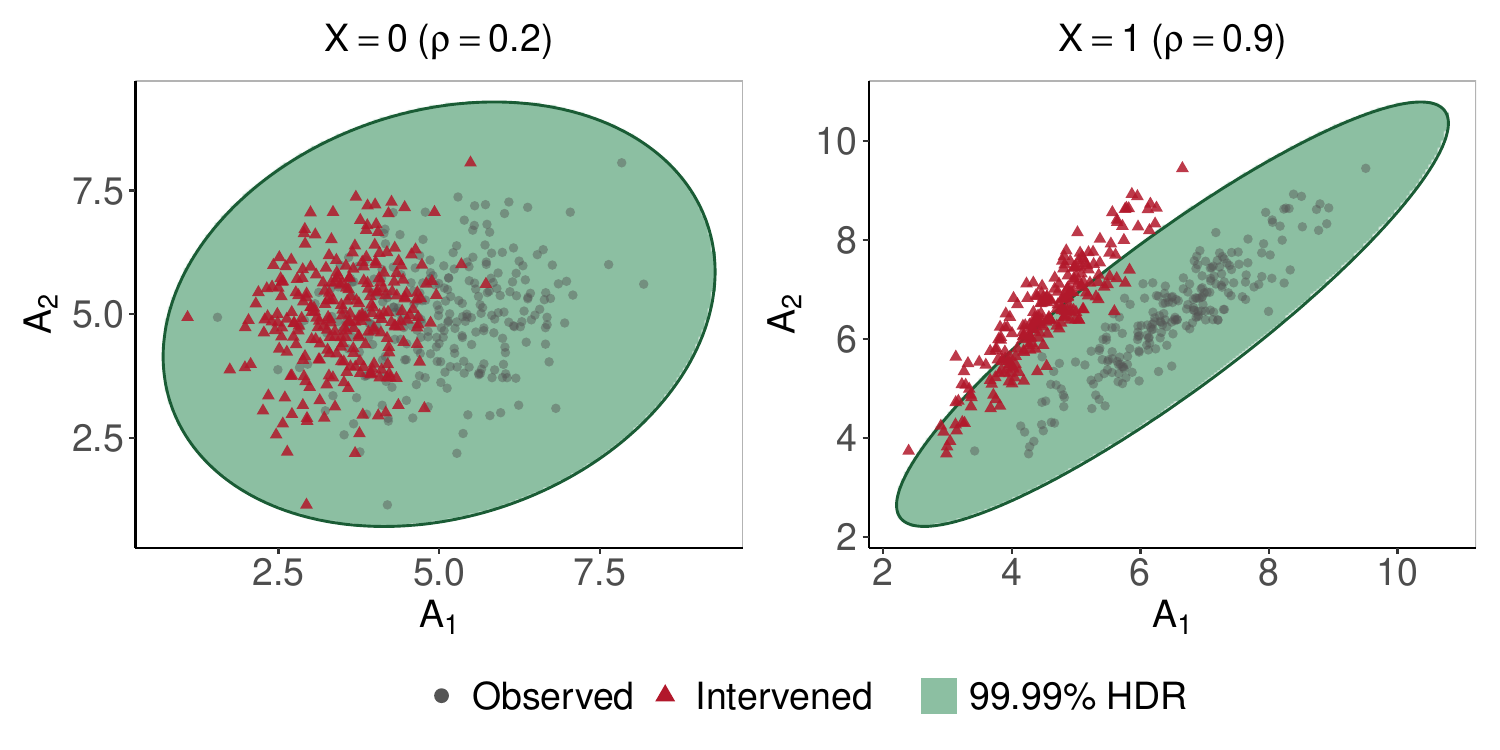}
    \caption{Observed treatments (gray circles) and policy-generated treatments (red triangles) after a 30\% reduction in $A_1$, within each stratum of a binary covariate $X$. The panel header gives the conditional correlation. The shaded set contains all but $0.01\%$ of the conditional treatment probability.}
    \label{fig: hdr illus}
\end{figure}

We generate $X\sim\mathrm{Bernoulli}(0.5)$. Conditional on $X=x$, the treatment follows
\begin{align*}
&\text{DGP1 (truncated normal):}\quad
     \bfA_i \mid X_i=x_i\sim \mathcal{TN}\bigl({\bm \mu}(x_i),\,{\bm \Sigma}(x_i);[4,\infty) \times \mathbb{R}\bigr),\\
    &\text{DGP2 (normal):}\quad
      \bfA_i \mid X_i=x_i\sim \mathcal{N}\bigl({\bm \mu}(x_i),\,{\bm \Sigma}(x_i)\bigr),\\
      &\text{DGP3 ($t$-distribution):}\quad
      \bfA_i \mid X_i=x_i\sim t_5\bigl({\bm \mu}(x_i),\,{\bm \Sigma}(x_i)\bigr),
\end{align*}
where ${\bm \mu}(x)=(5+1.5x)\mathbf 1_2$ and ${\bm \Sigma}(x)$ has unit variances and correlation $0.2+0.7x$. The $t_5$ distribution is parameterized by its covariance matrix ${\bm \Sigma}(x)$, corresponding to scale matrix $3{\bm \Sigma}(x)/5$. Figure~\ref{fig: hdr illus} illustrates the effect of this correlation on practical positivity under DGP2.

The outcome is $Y=Q(\bfA,X)+\varepsilon$, with $\varepsilon\sim\mathcal N(0,0.5^{2})$ independent of $(\bfA,X)$, under two regressions centered at $\bfA=(5,5)\tr$:
\begin{equation}\label{eq: outcome model}
\begin{aligned}
        Q_1(\bfA, X) &= 0.5X + 0.6(A_1 - 5) + 0.5(A_2 - 5),\\
        Q_2(\bfA, X) &= 0.5X + (0.6+0.6X)(A_1 - 5)
        + (0.5+0.2X)(A_2 - 5)\\
        &\quad + 0.1\,(A_1 - 5)(A_2 - 5).
\end{aligned}
\end{equation}
\begin{figure}[htp!]
\centering
\includegraphics[width=\textwidth]{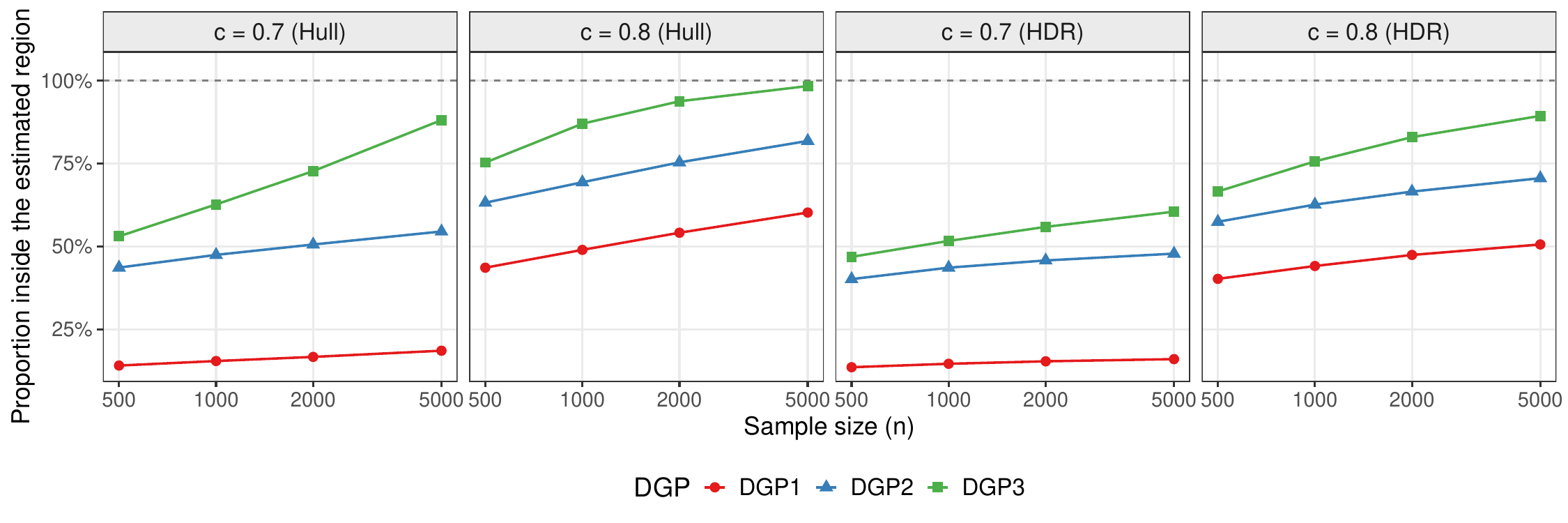}\caption{Proportion of policy values inside the estimated region of adequate positivity, by exposure distribution, shift, sample size, and region construction. The plotted proportion does not involve the outcome model.}
\label{fig: prop inside}
\end{figure}

\subsection{Implementation details}\label{append: impl details}

To choose $L$ for the oracle methods, we bound the slope of the outcome regression over a specified range of treatment values. When a policy value and its anchor both lie in this range, the maximum gradient norm bounds the change in the regression per unit distance between them.

The linear model $Q_1$ is globally Lipschitz, with $L(0)=L(1)\approx0.78$ under $\mathbf V=\mathbf I$. The interaction in $Q_2$ rules out a finite global constant on an unbounded domain. Over the box $[-4,4]^2$ in treatments centered at $(5,5)$, that is $[1,9]^2$, the gradient norm of $Q_2$ is maximized at the corner $(4,4)$, giving constants $\sqrt{1.81}\approx1.35$ for $X=0$ and $\sqrt{3.77}\approx1.94$ for $X=1$. The oracle methods use the rounded working constants $1.345$ and $1.942$. At either reduction, approximately $0.3\%$ of policy values fall outside this box under DGP1 and DGP2 and $0.9\%$ under DGP3; for observed values the corresponding proportions are $0.5\%$ and $1.2\%$. These proportions describe the treatments, not the displaced anchors, and do not establish the global Lipschitz condition required by Theorem~\ref{thm: asy-normal}.

Estimated density ratios are formed from clipped classification probabilities, truncated to $[0.001,0.999]$, so a ratio cannot exceed $999$. {If either the training or the evaluation fold has no policy values inside the region, we omit the inside residual correction. We omit the outside correction when the evaluation fold has no exterior policy values. Even when exterior evaluation values are present,} the outside augmentation is set to zero and the event is recorded {if fewer than five training anchors are available for exterior policy values}. The plug-in bound contribution remains.

For density-ratio classification {in the DGP3 rerun}, candidates with fitting or convergence failures or invalid {training-sample} predictions are excluded. Selection uses cross-validated log-loss with probabilities clipped to $[10^{-15},1-10^{-15}]$. An analysis is flagged as failed if no usable classifier is available {or if the selected classifier fails to return valid probabilities on the evaluation sample}. Clipping {limits extreme fitted weights but may introduce} bias, and omitting the augmentation can affect both bias and variance.

We remark that DGP3 does not satisfy the finite-variance condition in Theorem~\ref{thm: EIF} with the exponential interior displacement used here. These simulations nevertheless provide a finite-sample assessment of the clipped implementation.

Unless stated otherwise, simulations use $\rho=0.05$ in the rescaled coordinates, five cross-fitting folds, 200 multiplier draws, and five logarithmically spaced $\epsilon$ values from $10^{-4}$ to $0.95\rho$. The grid floor is a fixed fraction of $\rho$, equal to $10^{-4}$ when $\rho=0.05$. Section~\ref{subsec: sensitivity tuning} examines these tuning choices.

We use the learner libraries listed in Section~\ref{sec: sim}. Both Super Learners are discrete, selecting the learner with the smallest cross-validated risk. The outcome library uses continuous prediction rules for Lipschitz calibration. The density-ratio learners are not subject to this restriction; the penalized logistic regression leaves treatment coordinates unpenalized.

The \texttt{Oracle (L+Q)} procedure uses the oracle constant, which is the global Lipschitz constant under $Q_1$ and the effective-range constant under $Q_2$, together with the true outcome regression. It estimates the endpoints by plug-in because {$Q$ is known. Its paired bootstrap resamples each observation's plug-in contributions together with its observed outcome.} \texttt{Oracle (L)} uses the oracle constant but estimates $Q$ {and uses the same one-step estimator and multiplier bootstrap as \texttt{Proposal}.} Both oracle procedures use $\mathbf V=\mathbf I$ and the same displacement grid as {\texttt{Proposal}}.

\subsection{Sensitivity to region construction}\label{subsec: sensitivity pos}\label{append: impl trimmed}
\begin{figure}[htp!]
\centering
\includegraphics[width=\textwidth]{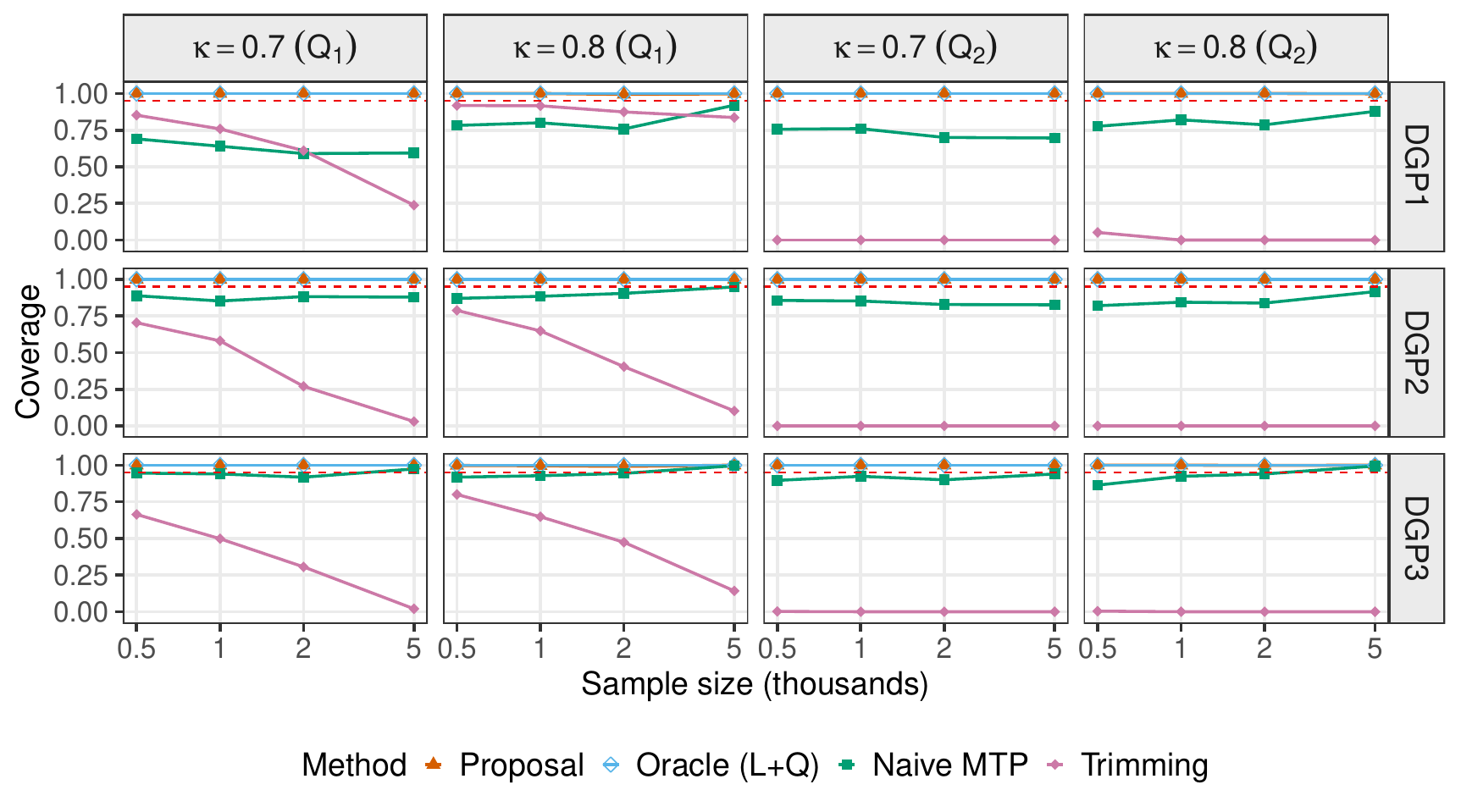}\caption{Empirical coverage with the 99\% trimmed hull and cross-fitting. Methods are as in Figure~\ref{fig: cov width}. Rows are exposure distributions; columns are combinations of the outcome model and $\kappa$.}
\label{fig: cov hdr cv1}
\end{figure}

\begin{figure}[htp!]
\centering
\includegraphics[width=\textwidth]{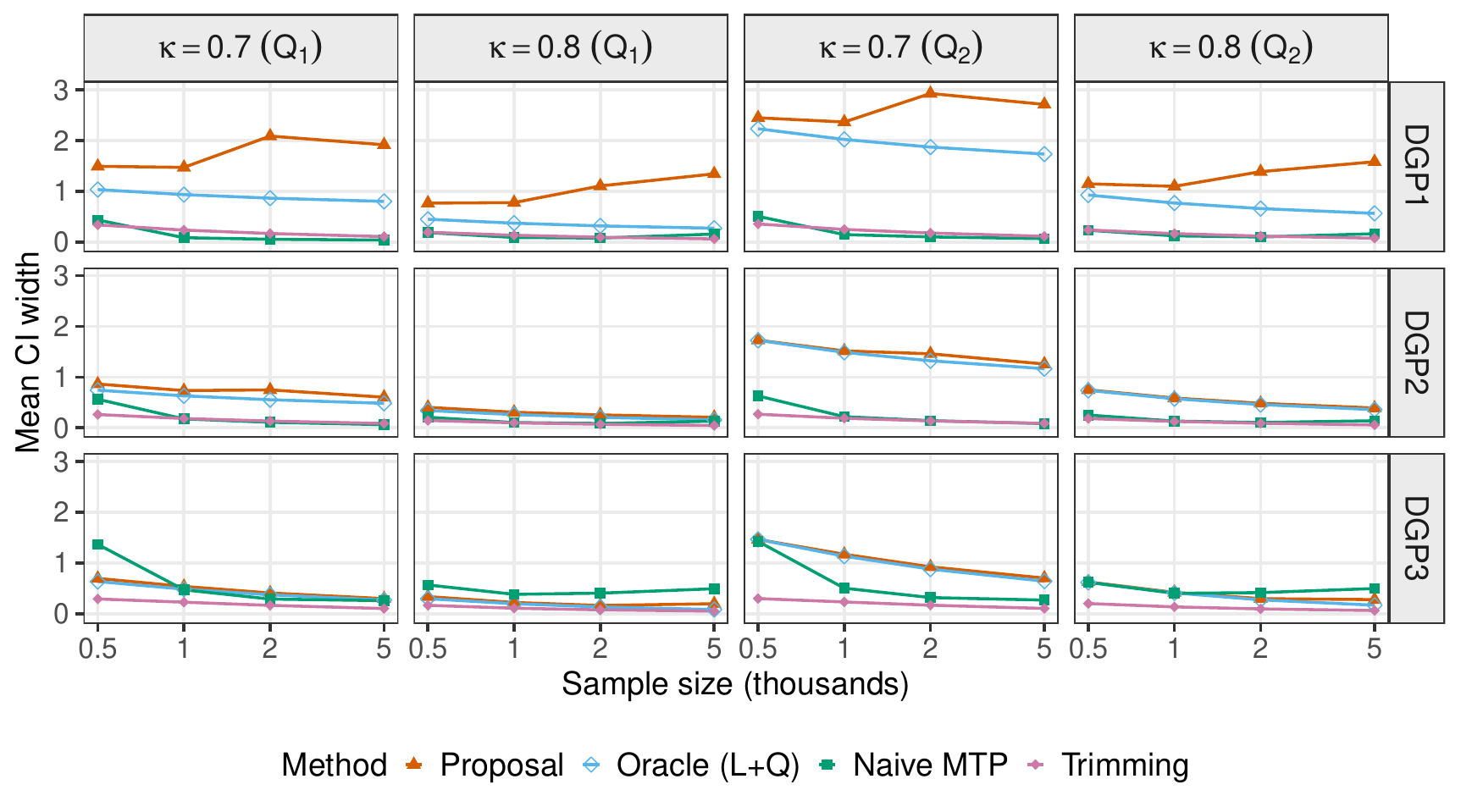}\caption{Mean confidence-interval width with the 99\% trimmed hull and cross-fitting. Rows are exposure distributions; columns are combinations of the outcome model and $\kappa$.}
\label{fig: wid hdr cv1}
\end{figure}

Figures~\ref{fig: cov hdr cv1} and \ref{fig: wid hdr cv1} report coverage and mean CI width for the 99\% trimmed hull with cross-fitting. The coverage conclusions agree with the full-hull results. Across sample sizes, outcome models, and both shifts, median width ratios relative to the full hull are 1.01, 1.07, and 1.30 under DGP1, DGP2, and DGP3. Under DGP1 with $n=1000$ and $\kappa=0.8$, one full-hull intersection per outcome model was empty. Counting its length as zero rather than using the signed endpoint difference gives mean widths of $0.8501$ and $1.1244$ instead of $0.8486$ and $1.1238$ for $Q_1$ and $Q_2$; coverage and the reported median ratio are unchanged. At $n=5000$ and $\kappa=0.7$, the largest $\widehat r^{\rm in}$ falls from $73$ to $29$ under DGP1 and from $572$ to $39$ under DGP2; under DGP3 both constructions reach the clipping bound. Trimming can therefore stabilize the ratios while widening the bounds by placing more policy values outside the region.

\paragraph{Inward versus outward smoothing.}
Table~\ref{tab: inward outward} compares the two constructions in the main configuration of Section~\ref{sec: sim} (convex hull, cross-fitted, $\kappa=0.7$, estimated $L$) over the 24 design-by-outcome-by-$n$ cells. Coverage is unchanged to within Monte Carlo error and the inward intervals are wider by a median of 7\%. Under DGP1 at $n\leq2000$ they are narrower. In this design, outward smoothing can place anchors outside the conditional support, where the observed density vanishes.

\begin{table}[htp!]
\centering
\begin{tabular}{lccc}
\hline
 & Width ratio, range & Min.\ coverage, outward & Min.\ coverage, inward \\
\hline
DGP1 & 0.87--1.13 & 0.998 & 0.998 \\
DGP2 & 1.03--1.11 & 0.996 & 1.000 \\
DGP3 & 1.05--1.15 & 0.990 & 0.988 \\
\hline
\end{tabular}
\caption{Outward versus inward smoothing in a matched experiment using the earlier classifier implementation for both constructions. Width ratio is mean inward width divided by mean outward width across the eight cells of each design. Minimum coverage is over the same cells.}
\label{tab: inward outward}
\end{table}

\FloatBarrier
\subsection{Sensitivity to the Lipschitz constant and tuning parameters}\label{subsec: sensitivity tuning}\label{append: impl adversarial}

We examined sensitivity to the supplied Lipschitz constant, its calibration quantile, and the displacement grid in three earlier full-sample experiments without cross-fitting. The first used the final projection geometry with an earlier classifier; the other two predated both the geometry and classifier revisions.

All three experiments used the exposure distributions and outcome models of Section~\ref{sec: sim} at $n=2000$ and $\kappa=0.7$. Coverage and width are means over these six combinations, and ``below'' counts how many of those six combinations fall under $0.95$.

The first sweep reduces the supplied constant relative to the oracle reference (the global constant for $Q_1$ and the working box constant for $Q_2$). Coverage falls monotonically with this reduction. It is $0.990$ at the reference constant, then $0.985$, $0.973$ and $0.926$ at three quarters, one half and one quarter of it. The number of combinations below nominal rises from none, to none, to one, to four. Mean width falls from $1.82$ to $1.32$ over the same sequence.

The second sweep lowers the quantile $q$ in \eqref{eq: Lqd}. Coverage is $0.992$ at $q=0.99$ and $0.991$ at $q=0.9$, $0.75$ and $0.5$, while the mean calibrated constant falls from $1.13$ to $1.05$ for $X=0$ and from $1.68$ to $1.59$ for $X=1$. In these configurations, reducing $q$ changed the calibrated constant little, motivating the direct scaling in the first sweep.

The third sweep varies the lower endpoint of the $\epsilon$ grid as a fraction of $\rho$, using ten grid values throughout so that the floor is not confounded with grid density. For \texttt{Proposal}, coverage is $0.992$, $0.991$ and $0.992$ at floors of $0.02$, $0.05$ and $0.2$ times $\rho$, respectively; the corresponding mean widths are $1.785$, $1.786$ and $1.847$. Lowering the grid floor slightly narrowed the CIs in these settings. This should not be read as an argument for taking $\epsilon$ as small as possible. Under its additional conditions, Proposition~\ref{prop: eps tradeoff} gives a lower bound of order $\epsilon^{-1}$ for the influence-function variance, while the endpoint discrepancy is bounded by order $\epsilon$.

\FloatBarrier
\subsection{Full-sample regions of adequate positivity}\label{subsec: full sample regions}
\begin{figure}[htbp!]
\centering
\includegraphics[width=.9\textwidth]{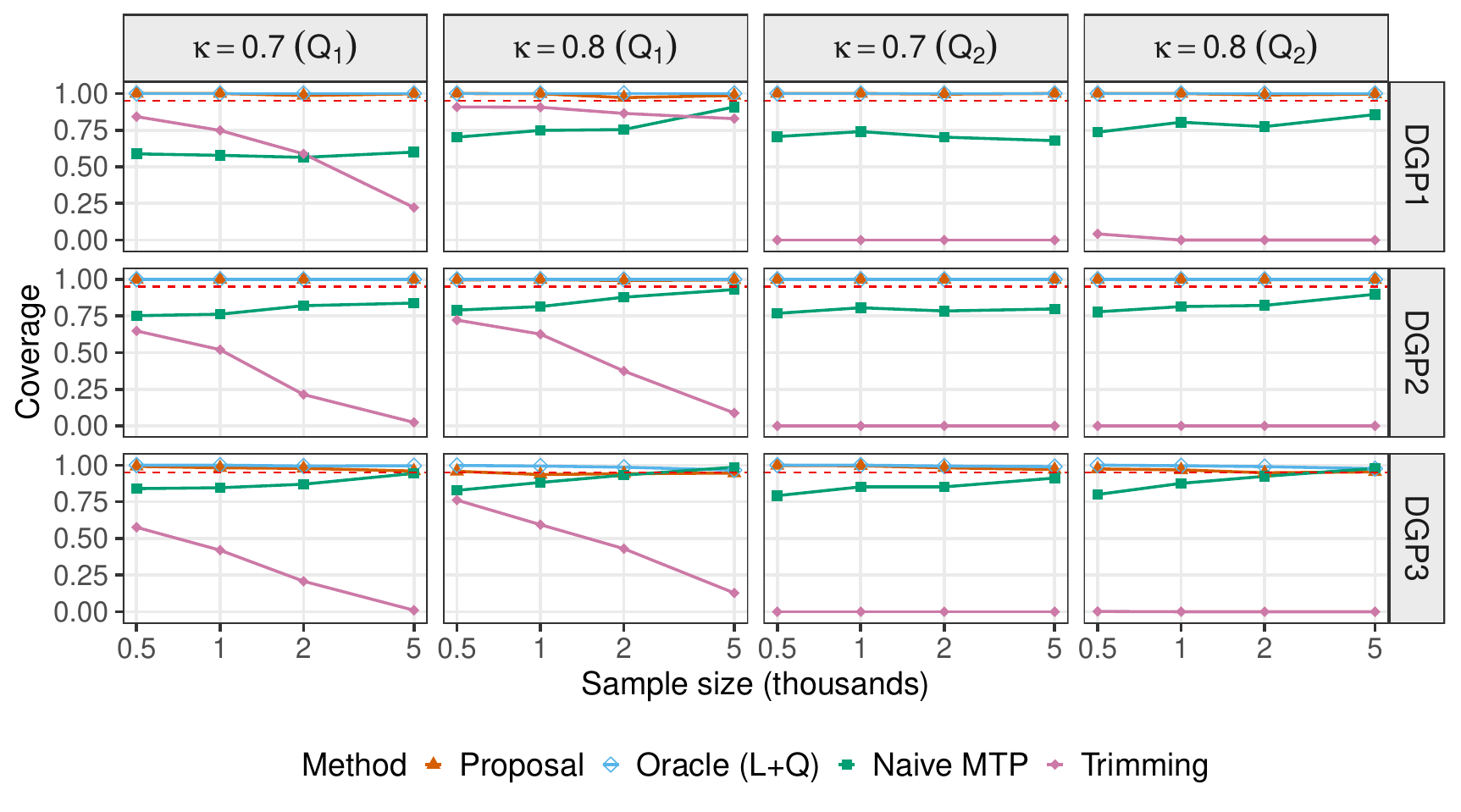}\caption{Empirical coverage with the convex hull and without cross-fitting. Rows are exposure distributions; columns are combinations of the outcome model and $\kappa$. The dashed line marks 95\%.}
\label{fig: cov hull cv0}
\end{figure}

\begin{figure}[htbp!]
\centering
\includegraphics[width=.9\textwidth]{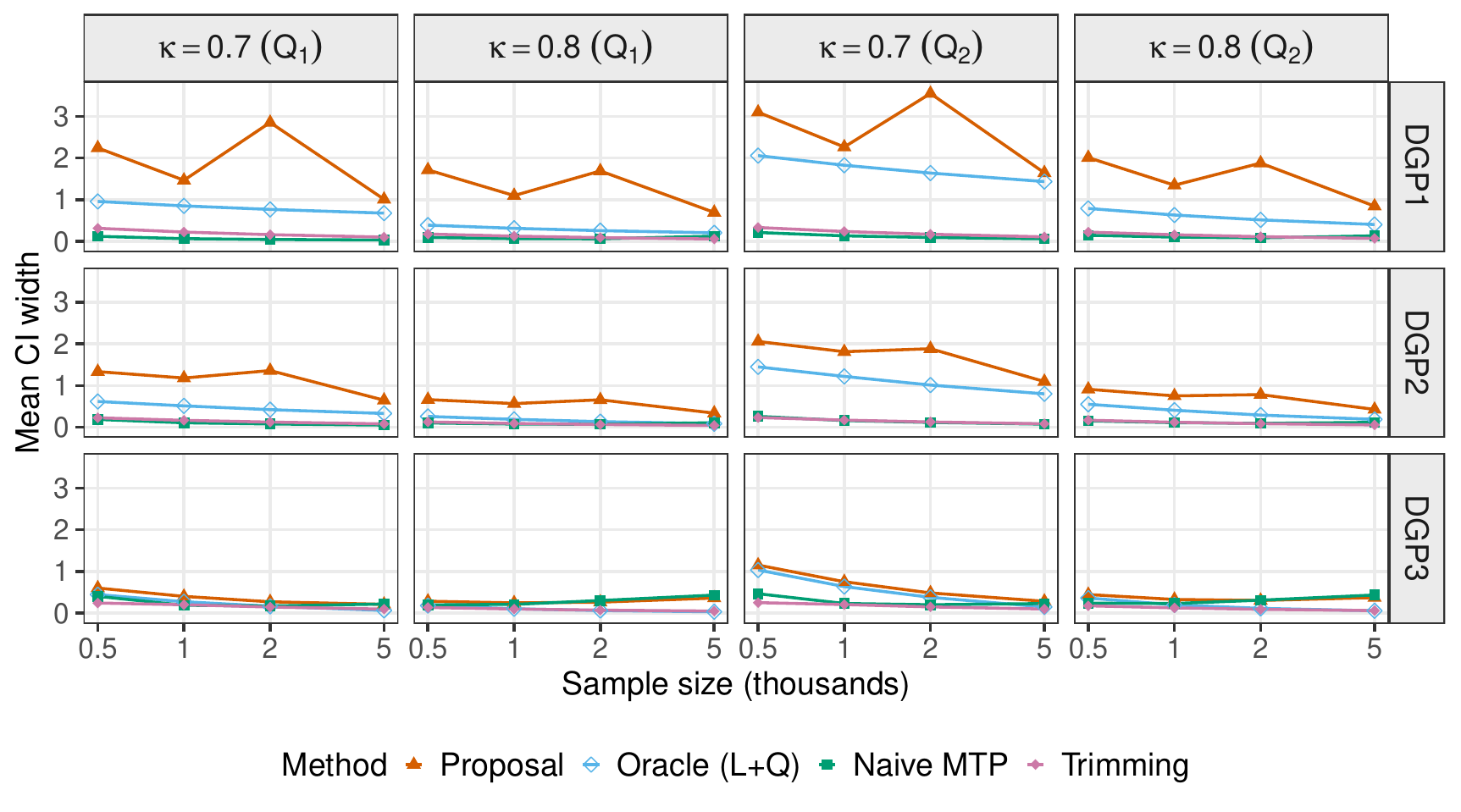}\caption{Mean CI width with the convex hull and without cross-fitting.}
\label{fig: wid hull cv0}
\end{figure}

Figures~\ref{fig: cov hull cv0} and~\ref{fig: wid hull cv0} report empirical coverage and mean CI width, respectively, for the convex hull when both region construction and nuisance estimation use the full sample, without cross-fitting. The coverage conclusions of Section~\ref{sec: sim} are unchanged. The lowest coverages of the full hull, $0.936$ to $0.948$ in four cells under DGP3 with $\kappa=0.8$, lie within about two Monte Carlo standard errors of nominal ($0.010$ at 500 replications).

With the 99\% trimmed hull, empirical coverage of the proposed CI was at least 98.2\% in every setting. Trimming had little effect on median relative width under DGP1, but tended to narrow the CIs under DGP2 and widen them under DGP3. Under DGP3, the full hull includes sparse tail observations that the trimmed hull removes. Both full-sample analyses lie outside the fixed-region theory (Sections~\ref{sec: partial id} and~\ref{sec: est inf} of the main text).

\section{Application Details and Additional Results}\label{append: app}\label{sec: add real}
This section supplements the analysis of Section~\ref{sec: real data} with preprocessing details, sensitivity analyses, and density-ratio diagnostics.

{\sloppy
The CHAMACOS cohort has been supported by the EPA (RD83451301), NIH-NIEHS (P01ES009605, R01ES017054, R01ES021369, R01ES026994, R24ES028529, UH3ES030631, U24ES028529-06), NIH-NIDA (R01DA035300), and NIH-NIA (R01AG069090).\par}

\subsection{Estimation details for the application}\label{append: impl chamacos}
The primary analysis fits nuisance functions on the full sample as described in Section~\ref{sec: real data}; cross-validation selects learners but does not cross-fit the endpoint estimators. The outcome-regression library includes GAMs, MARS, and logistic regression with and without two-way interactions. Density-ratio classification uses $\ell_1$-penalized logistic regression and BART, with exposure coefficients left unpenalized in the former.

For four participants, we carried all seven exposures forward from period two (three participants) or period one (one participant). We capped each exposure class at its 95th percentile among the 303 participants with nonmissing third-period values after carry-forward.
Both policies reduce exposures on this capped scale. Besides the adjustment covariates (maternal age and education), complete-case selection required information on baseline high blood pressure, baseline diabetes, and whether the participant was born in the United States.

Nine outcomes were carried forward from the fourth observation period and twenty from the third, where exposure and outcome are contemporaneous. The processed outcome captures cumulative recorded hypertension, not new onset after exposure: 39 of the 259 participants had a positive record by their aligned third period. Restricting the analysis to the 230 participants with a nonmissing processed outcome in their own fifth observation period leaves the qualitative conclusions unchanged. Under the joint policy, neither method's interval excludes zero at any reduction. Under the neonicotinoid policy, allowing conditional mean outcomes outside $\calH$ to vary can change the conclusion suggested by \texttt{Naive MTP}. Intersecting the reported risk-difference CIs with $[-1,1]$ as in Section~\ref{append: impl bounded} leaves them unchanged.

\begin{figure}[htp!]
    \centering
    \includegraphics[width=0.5\linewidth]{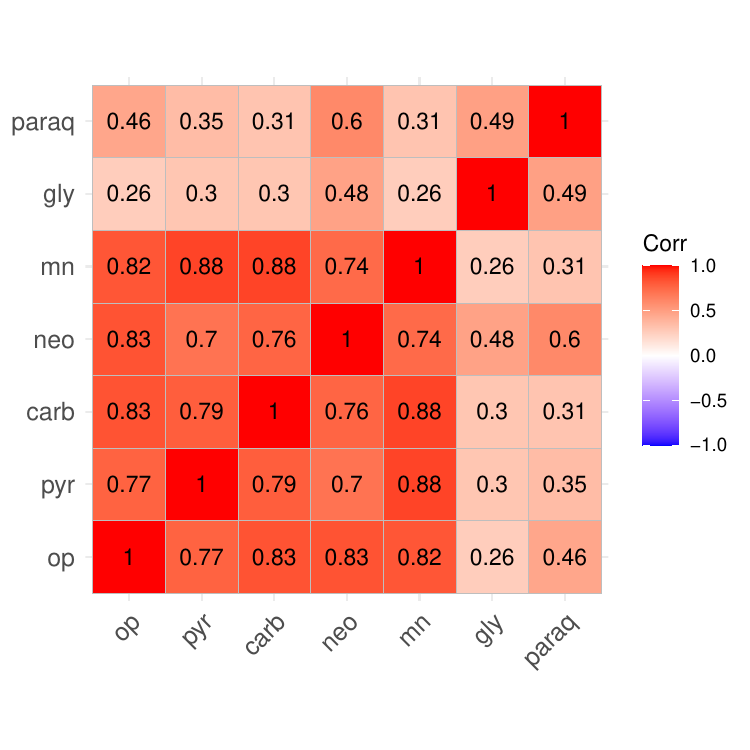}
    \caption{Pairwise correlations between the seven pesticide exposure classes in each participant's third observation period.}
    \label{fig: CHAMACOS corr t3}
\end{figure}

Figure~\ref{fig: CHAMACOS corr t3} shows the pairwise correlations among the seven exposure classes in each participant's third observation period.

\subsection{Sensitivity of the inside proportions}\label{append: impl levels}
Table~\ref{tab: policy spectrum} reports the inside proportions for ten candidate policies. These are a proportional reduction of each single class, of the five strongly correlated classes, of glyphosate and paraquat together, and of all seven classes. Only the last keeps nearly every policy value inside the estimated region. At larger reductions, every restricted policy leaves the region, illustrating that positivity is a property of the joint treatment--covariate geometry rather than of any single margin. Even the two weakly correlated classes reduced together leave up to two in five values outside. Trimming the sparsest observations before taking the hull barely moves these levels. At trimming level $0.01$ it drops four observations and changes each entry by at most $1.5$ percentage points, with the neonicotinoid entry changing from $74.9\%$ to $74.5\%$ at a $5\%$ reduction and remaining unchanged at $10\%$ and $20\%$.

\begin{table}[htp!]
\centering
\begin{tabular}{lcccc}
\hline
Policy & 1\% & 5\% & 10\% & 20\% \\
\hline
All seven classes        & 100\% & 99.6\% & 99.2\% & 97.3\% \\
First five classes       & 100\% & 87.3\% & 79.2\% & 72.6\% \\
Glyphosate and paraquat  & 100\% & 86.5\% & 71.4\% & 61.4\% \\
Organophosphates         & 100\% & 81.9\% & 68.3\% & 58.7\% \\
Pyrethroids              & 100\% & 79.9\% & 67.6\% & 56.4\% \\
Carbamates               & 100\% & 80.7\% & 70.3\% & 58.7\% \\
Neonicotinoids           & 100\% & 74.9\% & 61.4\% & 54.8\% \\
Mn fungicides            & 100\% & 78.0\% & 69.5\% & 56.4\% \\
Glyphosate               & 100\% & 85.3\% & 74.1\% & 59.5\% \\
Paraquat                 & 100\% & 93.8\% & 79.5\% & 71.0\% \\
\hline
\end{tabular}
\caption{Proportion of policy values inside the region, by policy and reduction. Full-sample convex hull with the $\rho$-expansion of Section~\ref{append: impl smooth}, $\rho=0.05$ in standardized units.}
\label{tab: policy spectrum}
\end{table}

\subsection{Sensitivity to region construction}\label{append: impl cv1}\label{append: impl trimmed cham}
\paragraph{Cross-fitted regions.}
We rebuild the region and compute $\mathbf V$ on each training fold, following the sample-splitting construction in Corollary~\ref{cor: split sample}. Outward smoothing may nevertheless extend the region beyond the conditional support, so we report this analysis as an empirical diagnostic rather than claim that all conditions of the corollary hold. The five validation folds contained 51 or 52 participants; the within-fold variances were averaged equally as described in Section~\ref{append: impl var}. With eight coordinates treated as continuous (including maternal age recorded in whole years) and 38 to 99 training observations per stratum, 71\% of held-out values fall outside their training-fold region before any intervention, so most of the target is bounded at every reduction and the bound interval widens with $L$. Relative to the full-sample intervals, the cross-fitted intervals at a 20\% reduction are wider by factors of 2.5, 4.5, 6.6 and 10.6 for the joint policy and 4.1, 4.2, 4.5 and 4.6 for the neonicotinoid policy at $L=0,0.15,0.3,0.6$. The proposed cross-fitted intervals contain zero at every reduction. We retain the full-sample construction for the primary analysis. The wider cross-fitted intervals illustrate the dimensional limitation noted in the introduction of the main text.

\paragraph{Trimmed regions.}
Table~\ref{tab: trimmed cham} repeats the analysis at $L=0.15$ with the 99\% trimmed hull of Section~\ref{append: impl region} in place of the full hull. At a 0\% reduction the identity policy has a risk difference of zero regardless of the region. The full-hull CI collapses to zero; the trimmed-hull procedure bounds the contribution from the 1.5\% of observed values outside its region and returns a CI of width 0.0266. The joint-policy intervals then contain zero at every reduction, as under the full hull, and from 4\% on the neo-policy intervals are 1.25 to 2.05 times as wide as under the full hull, which eliminates the exclusions of zero in the 200-draw analysis at reductions of 4\% to 7\% and 11\% under the full hull.

Two of the full-hull conclusions were sensitive to the number of multiplier draws. Recalibrating the same neonicotinoid intervals at $L=0.15$ with 20,000 draws instead of 200 changed the CI at a 3\% reduction from including zero to excluding it, and the CI at an 11\% reduction from excluding zero to including it. The CIs at reductions of 4\% through 7\% excluded zero under both calibrations.

\begin{table}[htp!]
\centering
\begin{tabular}{lcccc}
\toprule
 & \multicolumn{2}{c}{Joint policy} & \multicolumn{2}{c}{Neo policy} \\
\cmidrule(lr){2-3}\cmidrule(lr){4-5}
Reduction & Full hull & Trimmed hull & Full hull & Trimmed hull \\
\midrule
0\%  & 0 & 0.0266 & 0 & 0.0266 \\
5\%  & 0.0262 & 0.0381 & 0.0175$^\ast$ & 0.0347 \\
10\% & 0.0368 & 0.0382 & 0.0332 & 0.0454 \\
15\% & 0.0409 & 0.0488 & 0.0429 & 0.0598 \\
20\% & 0.0415 & 0.0489 & 0.0548 & 0.0688 \\
\bottomrule
\end{tabular}
\caption{Width of the proposed confidence interval for the contrast at $L=0.15$, simultaneous over the displacement grid, under the full-sample convex hull and the 99\% trimmed hull, both with the $\rho$-expansion. An asterisk marks an interval that excludes zero.}
\label{tab: trimmed cham}
\end{table}

\subsection{Density-ratio behavior in the application}\label{append: impl ratios}
BART is selected for most density-ratio fits. Under the joint policy, the standard deviation of the fitted inside ratio rises from 0.10 at a 1\% reduction to 0.36 at 20\%, and the intervals generally widen. Where no policy value lies outside $\calH$, the two methods' classifiers use the same rows and return the same fit.

Interior displacement concentrates anchors in a thin layer inside the boundary of $\calH_\rho$. Outward expansion can place this layer where the observed distribution has little or no density, which \eqref{eq: H rho} does not rule out. We therefore examine the fitted anchor ratios: at the smallest displacement level, their maxima are 1.9 under the neonicotinoid policy and 0.3 under the joint policy. At $L=0.15$, fewer than five exterior training anchors triggered omission of the outside residual correction at one or more displacement levels in 11 of the 42 primary policy--reduction settings (Section~\ref{append: impl details}). These empirical diagnostics do not establish the density condition of Assumption~\ref{assump: anchor reg}.

\FloatBarrier
\section{Proofs}\label{sec: proofs}
In this section, we provide proofs of the results in the main text and the supplementary material.

\subsection{Proof of Proposition~\ref{prop: partial gen}}\label{pf: partial gen}
\begin{proof}
In view of the rescaling introduced after Assumption~\ref{assump: weight Lip}, take $\mathbf V=\mathbf I$ and write $\|\cdot\|$ for $\|\cdot\|_{\mathbf V}$.
{In the coordinates of \eqref{eq: mu}, the second equality in Assumption~\ref{assump: basic}(ii) gives}
\begin{equation}\label{eq: causal conditional mean}
\E[Y(\bfAd)\mid\bfA,\bfX]=\mu(\bfCd,\bfX_D)\quad\text{almost surely}.
\end{equation}
{The first equality gives} $Q(\bfW)=\mu(\bfC,\bfX_D)$ almost surely. The continuous-version argument of Section~\ref{subsec: decomposition} extends this agreement to $\calH$. Multiplying \eqref{eq: causal conditional mean} by $\I(\bfWd\in\calH)$ and taking expectations therefore proves \eqref{eq: id of theta in}. Likewise,
\[
    \thetaoutH=\E\left[\mu(\bfWd)\I(\bfWd\notin\calH)\right].
\]
On the exterior event, $g(\bfWd)$ belongs to $\calH$ and preserves $\bfX_D$. {The same agreement at the anchors gives} $\mu\{g(\bfWd)\}=Q\{g(\bfWd)\}$. Assumption~\ref{assump: weight Lip} then yields, pointwise on this event,
\[
Q\{g(\bfWd)\}-L(\bfX_D)\|\bfWd-g(\bfWd)\|
\leq \mu(\bfWd)
\leq Q\{g(\bfWd)\}+L(\bfX_D)\|\bfWd-g(\bfWd)\|.
\]
Multiplying by $\I(\bfWd\notin\calH)$ and taking expectations proves that $\thetaoutH\in[\theta_l^g,\theta_u^g]$.
Combining this interval for $\thetaoutH$ with $\thetad=\thetainH+\thetaoutH$ proves that $\thetad\in[{\rm PI}_l^g,{\rm PI}_u^g]$.
The length of the above PI interval is
\begin{equation*}
    2\E\left[L(\bfX_D)\|\bfWd-g(\bfWd)\|\I(\bfWd\notin\mathcal{H})\right].
\end{equation*}
For every anchor map $g$ and every exterior point, the definition of the metric projection gives
\[
\|\bfWd-g(\bfWd)\|\geq\|\bfWd-\Pi_{\calH}(\bfWd)\|.
\]
Because $L(\bfX_D)\geq0$, multiplying by $2L(\bfX_D)\I(\bfWd\notin\calH)$ and taking expectations shows that no anchor map in the class of Proposition~\ref{prop: partial gen} has smaller width than $\Pi_{\calH}$. The projection belongs to that class and attains the lower bound, proving the minimization claim.
\end{proof}

\subsection{Proof of Theorem~\ref{thm: EIF}}\label{pf: EIF}
\begin{proof}
We hold $\bbmd$, $\calH$, $g$, $L$, and $\mathbf V$ fixed. We first express the inside and outside evaluations using the observed density, then differentiate the bound endpoints.
For expressions multiplied by indicators, extend $g$ by the identity on $\calH$ and $r^{\rm in}$ by zero outside $\calH$; these extensions do not change either endpoint.

Write ${\rm PI}_l^g = \Theta_{\rm in} + \Theta_{l}^{\rm out}$, where
\begin{align*}
    \Theta_{\rm in} &= \E[Q(\bfWd)\I(\bfWd\in\calH)],\\
    \Theta_{l}^{\rm out} &= \E\left[\left\{Q(g(\bfWd)) - L(\bfX_D)\|\bfWd - g(\bfWd)\|_{\mathbf{V}}\right\}\I(\bfWd\notin\calH)\right].
\end{align*}
The upper endpoint replaces $-L(\bfX_D)$ by $+L(\bfX_D)$.

\medskip
\noindent\textbf{Density representation of the inside portion.}
The open cells in Assumption~\ref{assump: smooth inv} cover the conditional treatment distribution up to a set of probability zero. Applying change of variables on each cell and summing gives \citep[e.g.,][]{haneuse2013estimation,diaz2023nonparametric}:
\begin{equation}\label{eq: fd}
     f^{\bbmd}(\bfa\mid \bfx)
  =
  \sum_{j=1}^{J(\bfx)}
    \I\left\{\bfa\in\bbmd_j(\operatorname{int}\mathcal A_j(\bfx),\bfx)\right\}
    f\left(\bbmb_j(\bfa,\bfx)\mid \bfx\right)
    \left|\det J_{\bbmb_j}(\bfa,\bfx)\right|,
\end{equation}
where $ J_{\bbmb_j}(\cdot,\bfx)$ is the Jacobian matrix of $\bbmb_j(\cdot,\bfx) $ with respect to $\bfa$.
Each summand is defined to be zero outside the image of its cell interior.

Using \eqref{eq: fd}, the inside portion is
\begin{align*}
    \Theta_{\rm in} &= \int Q(\bfa',\bfx)\I((\bfa',\bfx)\in\calH) f^{\bbmd}(\bfa'|\bfx)\, d\bfa'\, dP_{\bfX}(\bfx)\\
    &= \int Q(\bfa',\bfx)\I((\bfa',\bfx)\in\calH)\, r^{\rm in}(\bfa',\bfx)\, f(\bfa'|\bfx)\, d\bfa'\, dP_{\bfX}(\bfx)= \E\left[r^{\rm in}(\bfW) Q(\bfW)\I(\bfW\in\calH)\right],
\end{align*}
where the second equality follows from the conditional-density assumption of Section~\ref{sec: est inf} and the definition of $r^{\rm in}$; the last line relabels $\bfa'$ as the observed treatment.

\medskip
\noindent\textbf{Density representation of the outside portion.}
Write $\Theta_{l}^{\rm out} = T_Q - T_L$, where
\begin{align*}
    T_Q &= \E\left[Q(g(\bfWd))\I(\bfWd\notin\calH)\right],\quad
    T_L = \E\left[L(\bfX_D)\|\bfWd-g(\bfWd)\|_{\mathbf{V}}\I(\bfWd\notin\calH)\right].
\end{align*}
The integrand in $T_L$ is a fixed function of $\bfW$.

For $T_Q$, fix $\bfx_d$ and let $\{\mathcal D_{j}(\bfx_d)\}_j$ be the finite partition of Assumption~\ref{assump: anchor reg}. On each cell, let $h_j$ map the observed continuous coordinates to those of the assigned anchor:
\[
    h_j(\bfa,\bfx_c;\bfx_d)
    =\{g(\bbmd(\bfa,\bfx),\bfx_c;\bfx_d)\}_{C}.
\]
The area formula and the one-to-one property on each cell give the induced subdensity
\[
f^g(\bfc'\mid\bfx_d)
=\sum_j \I\{\bfc'\in h_j(\mathcal D_j(\bfx_d))\}
 f\{h_j^{-1}(\bfc';\bfx_d)\mid\bfx_d\}
 \left|\det J_{h_j^{-1}}(\bfc';\bfx_d)\right|.
\]
Each $h_j^{-1}$ is extended arbitrarily outside $h_j(\mathcal D_j)$, which does not affect the expression because of the corresponding indicator. The displayed subdensity integrates to $\prob(\bfWd\notin\calH\mid\bfX_D=\bfx_d)$. Therefore,
\begin{align*}
    T_Q &= \sum_{\bfx_d}\int Q(\bfa',\bfx_c',\bfx_d)\, f^g(\bfa',\bfx_c'|\bfx_d)\, d\bfa'\, d\bfx_c'\, P(\bfX_D=\bfx_d)= \E\left[r^{g}(\bfW) Q(\bfW)\right],
\end{align*}
where the last equality follows from the definition of $r^g$, since $f^g$ vanishes outside $\calH$ and the observed conditional density is positive almost everywhere on $\calH$. Consequently, $r^g(\bfW)=r^g(\bfW)\I(\bfW\in\calH)$ almost surely.

\medskip
\noindent\textbf{Endpoint representation.}
Define the fixed map and cost
\[
T(\bfW)=
\begin{cases}
\bfWd,&\bfWd\in\calH,\\
g(\bfWd),&\bfWd\notin\calH,
\end{cases}
\quad
K(\bfW)=L(\bfX_D)\|\bfWd-g(\bfWd)\|_{\mathbf V}\I(\bfWd\notin\calH).
\]
Write $Ph=\E_P[h(O)]$ and set
\[
\rho(\bfW)=\{r^{\rm in}(\bfW)+r^g(\bfW)\}\I(\bfW\in\calH),
\quad R=Y-Q(\bfW),\quad F_{l,u}(\bfW)=Q(T(\bfW))\mp K(\bfW).
\]
The density representations above give $P[h(T(\bfW))]=P[\rho(\bfW)h(\bfW)]$ for every nonnegative measurable $h$, and hence for every integrable $h$ under the law of $T(\bfW)$. In particular, $P\rho=1$ and
\[
\theta_*(P):={\rm PI}_*^g(P)=PF_*,\quad
\Psi_{*,P}=\rho R+F_*,\quad *\in\{l,u\}.
\]
The integrability conditions of Proposition~\ref{prop: partial gen} imply $F_l,F_u\in L_1(P)$. Since $\Psi_{*,P}\in L_2(P)$, the identity $\rho R=\Psi_{*,P}-F_*$ first gives integrability of $\rho R$. Thus $\E[\Psi_{*,P}\mid\bfW]=F_*$, and conditional Jensen's inequality gives $F_*\in L_2(P)$ and then $\rho R\in L_2(P)$. Also $P\Psi_{*,P}=\theta_*(P)$.

\medskip
\noindent\textbf{Pathwise derivative.}
Let $\{P_t:t\}$ be a regular parametric submodel through $P=P_0$, with score $s\in L_2^0(P)$, where $L_2^0(P)$ denotes the space of mean-zero, square-integrable functions under $P$. Following the smooth-submodel calculation for MTP means in Supplementary Section~3 of \citet{diaz2023nonparametric}, we assume that the joint and conditional laws admit score derivatives. We also assume that differentiation at $t=0$ can be interchanged with the conditional integrals defining $Q_t$ and the outer expectation defining $\theta_*(P_t)$, applying the product rule to its $t$-dependent integrand; see also \citet[Section~3.4]{kennedy2024semiparametric}. We remark that these analytical regularity conditions are premises of the calculation, not consequences of the $L_2(P)$ conditions alone.

Write $s_{\bfW}(\bfw)=\E_P[s(O)\mid\bfW=\bfw]$ and $s_Y(O)=s(O)-s_{\bfW}(\bfW)$ for the marginal and conditional scores. Differentiating the conditional mean gives, for almost every $\bfw$ under the law of $T(\bfW)$,
\begin{align*}
\dot Q(\bfw):=\left.\frac{d}{dt}Q_t(\bfw)\right|_{t=0}
&=\E_P[(Y-Q(\bfw))s_Y(O)\mid\bfW=\bfw]\\
&=\E_P[R s(O)\mid\bfW=\bfw].
\end{align*}
The second equality uses $\E_P[R\mid\bfW]=0$. Since $\rho R,F_*\in L_2(P)$, Cauchy--Schwarz gives $P[|\rho R s|]+P[|F_*s|]<\infty$. The density identity therefore yields
\[
P[\dot Q(T(\bfW))]
=P[\rho(\bfW)\dot Q(\bfW)]
=P[\rho R s].
\]
Finally, $T$ and $K$ are fixed along the submodel, so differentiating the outer expectation gives
\begin{align*}
\left.\frac{d}{dt}\theta_*(P_t)\right|_{t=0}
&=P[\dot Q(T(\bfW))]+P[F_*s_{\bfW}]\\
&=P[(\rho R+F_*)s]\\
&=P[(\Psi_{*,P}-\theta_*(P))s],
\end{align*}
where $P[s]=0$ and $F_*$ is a function of $\bfW$. Under the stated analytical regularity conditions, $\Psi_{*,P}-\theta_*(P)$ is therefore an influence function. The same calculation applies to submodels within the maintained Lipschitz model, without requiring a nonbinding Lipschitz restriction. Expanding $\rho R+F_*$ recovers exactly $\psi^{\rm in}+\psi_*^g$ in Theorem~\ref{thm: EIF}.
\end{proof}

\subsection{Proof of Theorem~\ref{thm: asy-normal}}\label{pf: asy-normal}
\begin{proof}
The fold assignments are independent of the observed values, with fixed $K$ and $n_k/n\to\pi_k>0$ for each $k$.
Fix $g$ and write $\Psi_*^{g}=\psi^{\rm in}+\psi_*^{g}$ for $*\in\{l,u\}$, the uncentered representations in Theorem~\ref{thm: EIF}, so that ${\rm PI}_*^{g}=\E[\Psi_*^{g}]$. The cross-fitted one-step estimator is $\widehat{\rm PI}_*^{g}=n^{-1}\sum_{k=1}^K\sum_{i\in\mathcal I_k}\widehat\psi_{*,i}^{\rm PI}$, where $\widehat\psi_{*,i}^{\rm PI}$ evaluates $\Psi_*^{g}$ at the fold-$k$ nuisance estimators $\widehat\eta_k$. Standard cross-fitting algebra gives, for each $*$,
\[
\sqrt n\big(\widehat{\rm PI}_*^{g}-{\rm PI}_*^{g}\big)
=\frac1{\sqrt n}\sum_{i=1}^n\big(\Psi_*^{g}(O_i)-{\rm PI}_*^{g}\big)
+\sqrt n\,R_{n,*}
+o_p(1),
\]
where $R_{n,*}$ collects the conditional bias and the empirical-process remainder. We show that each is $o_p(n^{-1/2})$. Throughout, condition on a training sample and suppress its fold index. The finite plug-in expectations in Assumption~\ref{assump: reg nuisance} mean that
\[
P\left[|\widehat Q(\bfWd)|\I(\bfWd\in\calH)
+|\widehat Q(g(\bfWd))|\I(\bfWd\notin\calH)\right]<\infty.
\]
Bounded fitted outcome regressions satisfy this condition. The fixed Lipschitz term is integrable by Proposition~\ref{prop: partial gen}. The two density-ratio identities in Theorem~\ref{thm: EIF} therefore give the exact conditional bias
\[
P\widehat\Psi_*^{g}- {\rm PI}_*^{g}
=P\left[(Q-\widehat Q)
\left\{(\widehat r^{\rm in}-r^{\rm in})\I(\bfW\in\calH)
+(\widehat r^{g}-r^{g})\I(\bfW\in\calH)\right\}\right].
\]
The sign of the fixed Lipschitz term does not enter this expression. By Cauchy--Schwarz, its absolute value is bounded by
\[
\|\widehat Q-Q\|_2
\left(\|\widehat r^{\rm in}-r^{\rm in}\|_2
+\|\widehat r^{g}-r^{g}\|_2\right)
=o_p(n^{-1/2})
\]
under Assumption~\ref{assump: reg nuisance}(ii). Write $P_{n,k}f=n_k^{-1}\sum_{i\in\mathcal I_k}f(O_i)$ for the validation-fold average. The remaining empirical-process term is a foldwise average of $(P_{n,k}-P)(\widehat\Psi_{*,k}^{g}-\Psi_*^{g})$. Conditional on $\mathcal T_k$, the variance of $\sqrt n(n_k/n)(P_{n,k}-P)(\widehat\Psi_{*,k}^{g}-\Psi_*^{g})$ is at most $(n_k/n)\|\widehat\Psi_{*,k}^{g}-\Psi_*^{g}\|_2^2=o_p(1)$ by Assumption~\ref{assump: reg nuisance}(iii). Since $K$ is fixed, summing over folds gives an $o_p(1)$ contribution to the scaled expansion \citep{chernozhukov2018double,kennedy2024semiparametric}. The leading term is an average of mean-zero i.i.d.\ summands with finite second moment. Applying the multivariate Lindeberg CLT via the Cram\'er--Wold theorem yields
\[
\sqrt n\begin{pmatrix}\widehat{\rm PI}_l^{g}-{\rm PI}_l^{g}\\[2pt] \widehat{\rm PI}_u^{g}-{\rm PI}_u^{g}\end{pmatrix}\dto\mathcal N(\mathbf 0_2,\Sigma),
\]
where $\Sigma=E[(\boldsymbol\Psi-\boldsymbol\theta)(\boldsymbol\Psi-\boldsymbol\theta)\tr]$. Finally, Assumption~\ref{assump: reg nuisance}(iii), the foldwise law of large numbers, and consistency of $\widehat{\boldsymbol\theta}$ imply
\[
\widehat\Sigma:=\frac1n\sum_{i=1}^n
(\widehat{\boldsymbol\Psi}_i-\widehat{\boldsymbol\theta})
(\widehat{\boldsymbol\Psi}_i-\widehat{\boldsymbol\theta})\tr
\pto
E[(\boldsymbol\Psi-\boldsymbol\theta)(\boldsymbol\Psi-\boldsymbol\theta)\tr].
\]
This proves consistency of $\widehat\Sigma$ and its diagonal entries. The estimator in \eqref{eq: aggregation of theta and sigma} centers within each fold rather than at $\widehat{\boldsymbol\theta}$. Because the fold-$k$ mean differs from $\widehat{\boldsymbol\theta}$ by $O_p(n_k^{-1/2})$ and $K$ is fixed, the fold-size-weighted difference contributes $O_p(n^{-1})$ and the fold-size-weighted variances share the limit above.
\end{proof}

\subsection{Proof of Proposition~\ref{prop: efficiency}}\label{pf: efficiency}
To establish efficiency, we show that the Lipschitz constraint leaves the tangent space unchanged under the conditions of Proposition~\ref{prop: efficiency}. For a real-valued function $h$ on the observed support and a fixed stratum $\bfd$, define the Lipschitz seminorm
\begin{equation*}
    |h(\cdot,\bfd)|_{\rm Lip}
    =\sup_{\substack{\bfc\in\calH(\bfd)\\
                     \bfc'\in\supp(\bfC\mid\bfX_D=\bfd)\setminus\{\bfc\}}}
      \frac{|h(\bfc',\bfd)-h(\bfc,\bfd)|}{\|\bfc'-\bfc\|_{\mathbf V}}.
\end{equation*}
Let $\mathcal P_L$ denote the model of observed laws admitting a version of $Q$ with $|Q(\cdot,\bfd)|_{\rm Lip}\leq L(\bfd)$ in every stratum with positive probability. Any such version of $Q$ can be extended beyond the observed support by the lower envelope in \eqref{eq: envelopes main}, without violating Assumption~\ref{assump: weight Lip}. The following lemma characterizes its tangent space.

\begin{Lemma}[Tangent space under the Lipschitz restriction]\label{lem: tangent}
\rm Let $P\in\mathcal P_L$ and write
$R=Y-Q(\bfW)$ and $\sigma^2(\bfw)=\Var(Y\mid\bfW=\bfw)$. Suppose there are a version $Q$ and a common $\eta>0$ such that, in every stratum with positive probability,
\begin{enumerate}
\item[(i)] $|Q(\cdot,\bfd)|_{\rm Lip}\leq L(\bfd)-\eta$;
\item[(ii)] $\sigma^2(\bfw)\geq\sigma^2_{\min}>0$ for $P_{\bfW}$-almost every $\bfw$;
\item[(iii)] $\lim_{M\to\infty}\ \operatorname*{ess\,sup}_{\bfw}\ \E[R^2\I(|R|>M)\mid\bfW=\bfw]=0$.
\end{enumerate}
Then the closed linear span of the tangent set of $\mathcal P_L$ at $P$ is $L_2^0(P)$.
\end{Lemma}

See Section~\ref{pf: tangent} for the proof of Lemma~\ref{lem: tangent}.
Condition (i) leaves room to perturb $Q$ in either direction without exceeding the Lipschitz bound. Condition (ii) keeps the conditional variance of $Y$ bounded away from zero, while (iii) ensures that sufficiently large residuals contribute arbitrarily little to that variance, uniformly in $\bfw$.
The lemma shows that every mean-zero, square-integrable function can be approximated in $L_2(P)$ by linear combinations of scores from submodels that obey the Lipschitz constraint. Thus the constraint does not reduce the tangent space at $P$.

\begin{proof}[Proof of Proposition~\ref{prop: efficiency}]
Under Assumption~\ref{assump: basic}, the restriction of $\mu$ to the conditional observed support is a version of $Q$. The common margin $\eta$ in Proposition~\ref{prop: efficiency} therefore gives condition (i) of Lemma~\ref{lem: tangent}.
The conditional-variance and uniform-integrability assumptions give conditions (ii) and (iii).
Thus Lemma~\ref{lem: tangent} shows that the tangent space of the observed-data model $\mathcal P_L$ at $P$ is
$L_2^0(P)$.

For each $*\in\{l,u\}$, Theorem~\ref{thm: EIF} gives the influence function
$\varphi_*^g=\Psi_*^g-{\rm PI}_*^g\in L_2^0(P)$. The efficient influence function is the
orthogonal projection of any influence function onto the model's tangent space
\citep[e.g.,][]{bickel1993efficient}. Since this space is $L_2^0(P)$, the projection of
$\varphi_*^g$ is $\varphi_*^g$ itself. Hence the influence function is efficient.
\end{proof}

\subsection{Proof of Corollary~\ref{cor: uniform set coverage}}\label{pf: uniform set coverage}
\begin{proof}
(i) For each $j$, the proof of Theorem~\ref{thm: asy-normal} gives the asymptotically linear expansion $\sqrt n(\widehat{\rm PI}_*^{(j)}-{\rm PI}_*^{(j)})=n^{-1/2}\sum_i(\Psi_*^{(j)}(O_i)-{\rm PI}_*^{(j)})+o_p(1)$. Stacking over the finite index set $\{(j,*):j\le J,\,*\in\{l,u\}\}$ and applying the Lindeberg CLT with the Cram\'er--Wold theorem to the $2J$-vector of mean-zero influence functions gives joint convergence to $\mathcal N(\mathbf 0,\Sigma_J)$, with $\Sigma_J$ nonsingular by assumption.

(ii) Conditionally on the data, each $\widehat T_{*,j}^{[m]}$ is a multiplier-bootstrap average of the estimated, centered, and studentized influence functions with i.i.d.\ mean-zero, unit-variance multipliers $\xi_i^{[m]}$. Let $\widehat Z_i$ denote the stacked, centered, and studentized estimated representations, and let $\Gamma_J$ be the covariance matrix of the studentized Gaussian limit in~(i). Assumption~\ref{assump: reg nuisance}(iii) and $\widehat\sigma_*^{(j)}\pto\sigma_*^{(j)}$ give
\[
\frac1n\sum_{i=1}^n\widehat Z_i\widehat Z_i\tr\pto\Gamma_J,
\quad
\frac1{\sqrt n}\max_{1\le i\le n}\|\widehat Z_i\|\pto0.
\]
By the conditional multiplier CLT, the bootstrap vector has the same studentized Gaussian limit $(G_{l,j},G_{u,j})_{j\le J}$ as in~(i). Applying the fixed-grid argument of \citet[Appendix~A.9]{susmann2025non}, with $M_n\to\infty$, gives $\widehat q_{1-\alpha}\pto q_{1-\alpha}$, where $q_{1-\alpha}$ is the $1-\alpha$ quantile of $D=\max_j\max\{G_{l,j},-G_{u,j}\}$.

(iii) Let $T_{*,j}=\sqrt n(\widehat{\rm PI}_*^{(j)}-{\rm PI}_*^{(j)})/\widehat\sigma_*^{(j)}$. By~(i)--(ii) and continuity of the distribution of $D$, the event $\max_j\max\{T_{l,j},-T_{u,j}\}\le\widehat q_{1-\alpha}$ has probability tending to $1-\alpha$. On this event, each calibrated interval contains its population bound interval, so their intersection $\widehat{\rm CI}_\alpha$ contains $[\max_j{\rm PI}_l^{(j)},\min_j{\rm PI}_u^{(j)}]$. Since $\thetad$ belongs to every population bound interval, both coverage claims follow.
\end{proof}

\subsection{Proof of Proposition~\ref{prop: hard nondiff}}\label{pf: hard nondiff}
\begin{proof}
Let $\mathbf Z=\Pi_{\calH}(\bfWd)$ and $D=L(\bfX_D)\|\bfWd-\mathbf Z\|_{\mathbf V}$. Inside $\calH$, $\mathbf Z=\bfWd$ and $D=0$. Thus the endpoints in \eqref{eq: partial theta out} are $\E[\mu(\mathbf Z)]\mp\E[D]$, using the continuous version of the conditional mean on $\calH$. The anchor mean is bounded because $Y$ is bounded, and the penalty has finite mean by assumption. We vary only the conditional law of $Y$ given $\bfW$, so the distributions of $\mathbf Z$ and $D$ stay fixed.

On the exterior event, $\Pi_{\calH}(\bfWd)$ lies in
\[
\mathcal B=\bigcup_{\bfd\in\supp(\bfX_D)}\partial\calH(\bfd)\times\{\bfd\}.
\]
{By assumption, $P_{\bfW}(\mathcal B)=0$.}

To perturb the conditional mean near the boundary while leaving it unchanged elsewhere, define, for $k>1/\delta_0$,
\[
b_k(\bfc,\bfd)=
\left\{1-k\,\operatorname{dist}\bigl(\bfc,\partial\calH(\bfd)\bigr)\right\}_+
\]
and set $s_k=0$ wherever $b_k=0$. Elsewhere, writing $\sigma^2(\bfw)=\Var(Y\mid\bfW=\bfw)$, let
\[
s_k(y,\bfw)=\{y-Q(\bfw)\}\frac{b_k(\bfw)}{\sigma^2(\bfw)}.
\]
The variance lower bound and boundedness of $Y$ make $s_k$ bounded. Since $\E[s_k\mid\bfW]=0$, the submodel $dP_{t,k}=(1+t s_k)\,dP$ leaves the law of $\bfW$ unchanged for sufficiently small $|t|$. Its conditional mean is $Q+t b_k$ almost surely, since $\E[Ys_k\mid\bfW]=b_k(\bfW)$. The function $\mu+t b_k$ agrees with this mean almost surely and satisfies Assumption~\ref{assump: weight Lip} for $|t|k<\eta$, because $b_k$ is $k$-Lipschitz in the rescaled coordinates. It therefore gives the continuous version on $\calH$, including at anchors of observed probability zero.

Along this submodel the penalty is unchanged, so the derivative of either endpoint is $\E[b_k(\mathbf Z)]$. As $b_k=1$ on the boundary and $b_k\geq0$ elsewhere,
\[
\E[b_k(\mathbf Z)]
=\prob(\bfWd\notin\calH)+\E[b_k(\bfWd)\I(\bfWd\in\calH)]
\geq\prob(\bfWd\notin\calH)>0
\]
for every $k$, where the exterior probability $\prob(\bfWd\notin\calH)$ is fixed because neither the policy nor the distribution of $\bfW$ changes with $k$.

If $\varphi\in L_2(P)$ represented the derivative of either endpoint, then $\E[\varphi s_k]=\E[b_k(\mathbf Z)]$. At any point a distance $r>0$ from its stratum's boundary, $b_k$ is zero once $1/k<r$. Since $P_{\bfW}(\mathcal B)=0$, $b_k(\bfW)\to0$ almost surely. As $0\leq b_k\leq1$, the dominated convergence theorem gives $\E[b_k(\bfW)^2]\to0$, so $\|s_k\|_2^2\leq \E[b_k(\bfW)^2]/\sigma^2_{\min}\to0$. The Cauchy--Schwarz inequality gives
$|\E[\varphi s_k]|\leq\|\varphi\|_2\|s_k\|_2\to0$, a contradiction. Thus neither endpoint has an $L_2(P)$ influence function.
\end{proof}

\subsection{Proof of Proposition~\ref{prop: eps tradeoff}}\label{pf: eps tradeoff}\label{pf: reg proj cov}

\begin{proof}[Proof of Proposition~\ref{prop: eps tradeoff}]
Work in the rescaled coordinates, and on the exterior event $\{\bfWd\notin\calH\}$ abbreviate $\bfa_0=\Pi_{\calH}(\bfWd)$, $\bfa_{\epsilon}=\tilde\Pi_{\calH}^{\epsilon}(\bfWd)$, and $\phi=\phi^{\epsilon}(R)$.

\medskip\noindent\textbf{Part (i).}
By \eqref{eq: proj notation} and \eqref{eq: reg proj}, $\bfa_{\epsilon}=\bfa_0-\phi\,\mathbf n$ and $\bfWd=\bfa_{\epsilon}+(R+\phi)\,\mathbf n$ for the outward unit normal $\mathbf n$, so $\|\bfWd-\bfa_{\epsilon}\|=R+\phi$ and $\|\bfa_0-\bfa_{\epsilon}\|=\phi$. Both anchors lie in $\calH$, where $Q$ coincides with the continuous version of $\mu$, and Assumption~\ref{assump: weight Lip} gives $|\mu(\bfa_{\epsilon})-\mu(\bfa_0)|\leq L(\bfX_D)\phi$. The lower-endpoint integrands of \eqref{eq: partial theta out} for $g=\Pi_{\calH}$ and $g=\tilde\Pi_{\calH}^{\epsilon}$ therefore differ, pointwise on the exterior event, by
\[
    \begin{aligned}
    &\left\{\mu(\bfa_0)-L(\bfX_D)R\right\}
      -\left\{\mu(\bfa_{\epsilon})-L(\bfX_D)(R+\phi)\right\}=\mu(\bfa_0)-\mu(\bfa_{\epsilon})+L(\bfX_D)\phi,
    \end{aligned}
\]
which belongs to $\left[0,\ 2L(\bfX_D)\phi\right]$.
Taking expectations gives ${\rm PI}_l^{(\Pi)}-2\Delta(\epsilon)\leq{\rm PI}_l^{(\epsilon)}\leq{\rm PI}_l^{(\Pi)}$. The upper-endpoint integrands differ by $\mu(\bfa_{\epsilon})-\mu(\bfa_0)+L(\bfX_D)\phi\in[0,\ 2L(\bfX_D)\phi]$, giving the second display. Finally, the range condition $\phi^{\epsilon}(r)<\epsilon$ gives $\Delta(\epsilon)\leq\epsilon\,\E[L(\bfX_D)\I(\bfWd\notin\calH)]$.

\medskip\noindent\textbf{Part (ii).}
The uncentered representation $\Psi_*^{(\epsilon)}$ equals $(Y-Q(\bfW))\{r^{\rm in}(\bfW)\I(\bfW\in\calH)+r^{(\epsilon)}(\bfW)\}$ plus a function of $(\bfA,\bfX)$ alone. Conditioning on $(\bfA,\bfX)$,
\[
    \Var\left(\Psi_*^{(\epsilon)}\right)
    \geq \E\left[\Var(Y\mid\bfW)\left\{r^{\rm in}(\bfW)\I(\bfW\in\calH)+r^{(\epsilon)}(\bfW)\right\}^2\right]
    \geq \sigma^2_{\min}\,\E\left[r^{(\epsilon)}(\bfW)^2\right],
\]
using that both density ratios are nonnegative and the variance floor holds wherever $r^{(\epsilon)}>0$. To bound this second moment from below, we use the following lemma to identify the interior layer containing the displaced anchors.

\begin{Lemma}[Change of variables for interior-displaced projection]\label{lem: reg proj cov}
Fix a stratum $\bfX_D=\bfd$ and let $\calH(\bfd)\subseteq\mathbb R^p$ be nonempty, closed, and convex. Suppose Assumption~\ref{assump: boundary reg} holds and fix $\epsilon\in(0,\epsilon_{\max})$. Suppose $\phi^\epsilon$ is continuously differentiable with positive derivative and maps $[0,\infty)$ onto $[0,\epsilon)$. The continuous component of $\tilde\Pi_{\calH}^{\epsilon}$ is a bijection from $\mathbb R^p\setminus\calH(\bfd)$ onto the interior layer $\mathcal L_\epsilon(\bfd)$ in Proposition~\ref{prop: eps tradeoff}(ii). The map and its inverse are locally Lipschitz with nonzero Jacobian determinants almost everywhere. Thus, if the exterior policy values have a density within the stratum, so do their displaced anchors.
\end{Lemma}

See Section~\ref{pf: reg proj cov lemma} for the proof. Other displacement functions satisfying the lemma's conditions may replace the exponential form in Section~\ref{subsec: reg proj} of the main text.

By Lemma~\ref{lem: reg proj cov}, every displaced anchor in stratum $\bfd$ lies in $\mathcal L_\epsilon(\bfd)$. Thus the anchor density, and hence $r^{(\epsilon)}$, vanishes almost everywhere outside this layer. Since the anchor density in each stratum integrates to the conditional probability that $\bfWd\notin\calH$, averaging over strata gives $\E[r^{(\epsilon)}(\bfW)]=\prob(\bfWd\notin\calH)$. The Cauchy--Schwarz inequality gives
\[
    \begin{aligned}
    \prob(\bfWd\notin\calH)^2
    &=\left(\E\left[r^{(\epsilon)}(\bfW)\,\I\{r^{(\epsilon)}(\bfW)>0\}\right]\right)^2\leq \E\left[r^{(\epsilon)}(\bfW)^2\right]\,
      \prob\left(r^{(\epsilon)}(\bfW)>0\right).
    \end{aligned}
\]
It remains to bound the support probability. For $0<t<\epsilon$, the level set $\{\bfc\in\calH(\bfd):{\rm dist}(\bfc,\partial\calH(\bfd))=t\}$ is the boundary of the inner parallel body $\calH_{-t}(\bfd)=\{\bfc\in\calH(\bfd):{\rm dist}(\bfc,\partial\calH(\bfd))\geq t\}$, which is convex and contained in $\calH(\bfd)$. Surface area is monotone under inclusion among closed convex sets \citep[e.g.,][]{schneider2014convex}, so $|\partial\calH_{-t}(\bfd)|_{p-1}\leq|\partial\calH(\bfd)|_{p-1}$. The distance function to $\partial\calH(\bfd)$ is $1$-Lipschitz with unit gradient almost everywhere on $\{{\rm dist}>0\}$, so the coarea formula yields
\[
    {\rm vol}\left(\mathcal L_{\epsilon}(\bfd)\right)
    =\int_0^{\epsilon}\left|\partial\calH_{-t}(\bfd)\right|_{p-1}\,dt
    \leq \epsilon\,\left|\partial\calH(\bfd)\right|_{p-1}.
\]
Hence
\[
    \begin{aligned}
    \prob\left(r^{(\epsilon)}(\bfW)>0\right)
    &\leq \sum_{\bfd}\prob(\bfX_D=\bfd)\,\bar f(\bfd)\,
      {\rm vol}\left(\mathcal L_{\epsilon}(\bfd)\right)\leq c_{\calH}\,\epsilon.
    \end{aligned}
\]
Combining the three displays proves part (ii).
\end{proof}

\subsection{Proof of Corollary~\ref{cor: split sample}}\label{pf: split sample}
\begin{proof}
Fix $j$ and $*\in\{l,u\}$, and write $P_{n,k}$ for the empirical measure of $\mathcal I_k$ and $\Psi^{(j)}_{*,k}$ for the corresponding component of $\boldsymbol\Psi_k$. Conditionally on $\mathcal T_k$ the inputs and nuisance functions of fold $k$ are fixed and $\mathcal I_k$ is an independent sample from $P$, so the cross-fitting argument in the proof of Theorem~\ref{thm: asy-normal}, applied at that fold's selected inputs, gives
\[
\thetahat^{\rm in}_k+\thetahat^{(j)}_{*,k}-{\rm PI}^{(j)}_{*,k}=(P_{n,k}-P)\Psi^{(j)}_{*,k}+o_P(n^{-1/2}),
\]
where Condition (i) controls the nuisance product and the conditional variance of the score error at the selected inputs. Each fold's inputs belong to $\mathfrak C$ with probability tending to one, so the expansion holds unconditionally.

Weighting the previous display by $|\mathcal I_k|/n$ gives
\[
\widehat{\rm PI}^{(j)}_{*}-\bar{\rm PI}^{(j)}_{*}=(P_n-P)\Psi^{(j)}_{*,\infty}+\sum_{k=1}^K\frac{|\mathcal I_k|}{n}(P_{n,k}-P)\big(\Psi^{(j)}_{*,k}-\Psi^{(j)}_{*,\infty}\big)+o_P(n^{-1/2}).
\]

Each term of the sum has conditional mean zero given $\mathcal T_k$ and conditional variance at most $\|\Psi^{(j)}_{*,k}-\Psi^{(j)}_{*,\infty}\|^2_{L_2(P)}/|\mathcal I_k|$, which is $o_P(n^{-1})$ by (ii), so Chebyshev's inequality conditionally on $\mathcal T_k$ makes the sum $o_P(n^{-1/2})$.
The stacked vector $\sqrt n(\widehat{\rm PI}^{(j)}_{*}-\bar{\rm PI}^{(j)}_{*})_{j,*}$ is therefore asymptotically normal with the covariance of $\boldsymbol\Psi_\infty$, by the central limit theorem for $(P_n-P)\boldsymbol\Psi_\infty$. The aggregated variance estimators and the multiplier bootstrap quantile are consistent by the same conditioning argument together with (ii), so the coverage conclusions of Corollary~\ref{cor: uniform set coverage} hold with $\bar{\rm PI}^{(j)}_{*}$ in place of ${\rm PI}^{(j)}_{*}$, and $\prob([\max_j\bar{\rm PI}^{(j)}_{l},\min_j\bar{\rm PI}^{(j)}_{u}]\subseteq\widehat{\rm CI}_\alpha)\geq1-\alpha$ in the limit.

On the event in (i), Proposition~\ref{prop: partial gen} applies with each fold's $(\calH_k,L_k,\mathbf V_k)$, so $\thetad\in[{\rm PI}^{(j)}_{l,k},{\rm PI}^{(j)}_{u,k}]$ for all $j$ and $k$, hence $\thetad$ lies in the fold averages and in their intersection over $j$. Steps 4 and 5 together give the claim.
\end{proof}

\subsection{Proof of Proposition~\ref{prop: reg proj inj}}\label{pf: reg proj inj}
\begingroup
To show that the interior-displaced projection maps the exterior of $\calH_\rho$ bijectively onto its interior layer, we apply Lemma~\ref{lem: reg proj cov}. The following lemma verifies the boundary regularity required by that result.
\begin{Lemma}\label{lem: smooth boundary}
      For any nonempty proper closed convex set $\calH\subsetneq\mathbb R^p$ and $\rho>0$:
      \begin{itemize}
          \item[\textcolor{black}{\rm(i)}] {\color{black}$\calH_{\rho}$ in \eqref{eq: H rho} has a $C^{1,1}$ boundary, meaning that near each boundary point it can be represented as the graph of a function with Lipschitz gradient. The outward unit normal is $\mathbf{n}(\mathbf{p})=(\mathbf{p}-\Pi_{\calH}(\mathbf{p}))/\rho$ at each $\mathbf{p}\in\partial\calH_{\rho}$.}
          \item[\rm(ii)] ${\rm reach}(\partial\calH_\rho)\geq\rho$.
      \end{itemize}
   \end{Lemma}

See Section~\ref{pf: smooth boundary} for the proof of Lemma~\ref{lem: smooth boundary}. Part (i) gives the outward normal to the expanded boundary. By part (ii), every point at distance less than $\rho$ from that boundary has a unique nearest boundary point.
\par\endgroup

\begin{proof}[Proof of Proposition~\ref{prop: reg proj inj}]
   Lemma~\ref{lem: smooth boundary}(i) and (ii) show that $\calH_\rho$ satisfies Assumption~\ref{assump: boundary reg} with $\epsilon_{\max}=\rho$. Since $\calH_\rho$ is also closed and convex, Lemma~\ref{lem: reg proj cov} shows that, for every $0<\epsilon<\rho$, the interior-displaced projection maps the exterior of $\calH_\rho$ bijectively onto $\mathcal L_\epsilon$.
\end{proof}

\subsection{Proof of Lemma~\ref{lem: tangent}}\label{pf: tangent}
\begingroup\color{black}
{\color{black}Fix a version of $Q$ satisfying condition (i) of Lemma~\ref{lem: tangent}. This version agrees with the continuous identified version on $\calH$, but its values at individual supported points outside $\calH$ may not be unique.}

\begin{proof}
{\color{black}Every score of a regular submodel in $\mathcal P_L$ belongs to $L_2^0(P)$. It therefore suffices to show that linear combinations of these scores are dense in $L_2^0(P)$.}

For bounded $s$ with $\E[s]=0$ and $\|s\|_\infty\leq B$, let $dP_t=(1+ts)\,dP$ for
$|t|<1/B$. Each $P_t$ is a probability measure and $t\mapsto P_t$ is a regular parametric submodel with
score $s$ at $t=0$. Since $1+ts>0$, each $P_t$ has the same support as $P$, so the seminorm in the
definition of $\mathcal P_L$ is taken over a fixed domain along every such submodel. It remains to identify
which $s$ keep $P_t$ inside $\mathcal P_L$.

{\color{black}\emph{Step 1: how $Q$ changes along the submodel.}} Write $s_{\bfW}(\bfw)=\E[s\mid\bfW=\bfw]$ and
$s_Y=s-s_{\bfW}$, so that $\E[s_Y\mid\bfW]=0$ and $L_2^0(P)=\mathcal T_{\bfW}\oplus\mathcal T_Y$
orthogonally, where $\mathcal T_{\bfW}=\{h(\bfW):\E h=0\}$ and $\mathcal T_Y=\{g:\E[g\mid\bfW]=0\}$.
Since $\E[Ys\mid\bfW]=a+s_{\bfW}Q$ with $a:=\E[R s_Y\mid\bfW]$,
{\color{black}
\[
  \begin{aligned}
  Q_t(\bfw)
    &=\frac{\E_P[Y\{1+t s(O)\}\mid\bfW=\bfw]}{\E_P[1+t s(O)\mid\bfW=\bfw]}\\
    &=\frac{Q(\bfw)+t\,\E_P[Ys\mid\bfW=\bfw]}{1+t\,s_{\bfW}(\bfw)}
     =Q(\bfw)+\frac{t\,a(\bfw)}{1+t\,s_{\bfW}(\bfw)} .
  \end{aligned}
\]
}
This identity holds exactly $P_{\bfW}$-almost everywhere. We choose versions with $Q_t=Q$ when $a\equiv0$, and $Q_t=Q+ta$
when $s_{\bfW}\equiv0$.

\emph{Step 2: directions that leave $Q$ unchanged.} If $a\equiv0$ then $Q_t=Q$ for every admissible $t$,
so $|Q_t(\cdot,\bfd)|_{\rm Lip}=|Q(\cdot,\bfd)|_{\rm Lip}\leq L(\bfd)$ and $P_t\in\mathcal P_L$. {\color{black}Thus $Q_t=Q$ for two classes of scores:
every bounded $s_{\bfW}\in\mathcal T_{\bfW}$, and every bounded $s_Y\in\mathcal T_Y$ with
$\E[Rs_Y\mid\bfW]=0$.}

\emph{Step 3: moving $Q$.} Let bounded $s=s_Y\in\mathcal T_Y$ be given. Define
$A:=\sup_{\bfd}|a(\cdot,\bfd)|_{\rm Lip}$ and suppose $A<\infty$. {\color{black}Since $s_{\bfW}=\E_P[s_Y\mid\bfW]=0$, Step 1 gives $Q_t=Q+ta$. By subadditivity and the bound $|Q(\cdot,\bfd)|_{\rm Lip}\leq L(\bfd)-\eta$ in condition (i) of Lemma~\ref{lem: tangent},}
\[
  |Q_t(\cdot,\bfd)|_{\rm Lip}\ \leq\ |Q(\cdot,\bfd)|_{\rm Lip}+|t|\,|a(\cdot,\bfd)|_{\rm Lip}
  \ \leq\ L(\bfd)-\eta+|t|A ,
\]
which is at most $L(\bfd)$ for $|t|\leq\eta/A$. {\color{black}Thus $P_t$ remains in $\mathcal P_L$ for all sufficiently small positive and negative $t$, so its score $s$ belongs to the tangent set at $P$.} Without a positive margin, this argument does not
guarantee that both signs of $t$ remain admissible.

\emph{Step 4: realizing a prescribed $a$ by a bounded score.} {\color{black}Step 3 determined how a given score changes $Q$; here we start with a desired change in $Q$ and construct a score that produces it.} Let $\tau_M(u)=\max(-M,\min(M,u))$ and put
$\xi_M=\tau_M(R)-\E[\tau_M(R)\mid\bfW]$, so $|\xi_M|\leq2M$ and $\E[\xi_M\mid\bfW]=0$. Since
$\E[R\mid\bfW]=0$,
\[
  c_M(\bfw):=\E[R\xi_M\mid\bfW=\bfw]=\E[R\tau_M(R)\mid\bfW=\bfw].
\]
The integrand $u\tau_M(u)$ is nonnegative and increases to $u^2$, so $c_M\leq\sigma^2$ and
\[
  \sigma^2-c_M=\E\big[(|R|-M)|R|\,\I(|R|>M)\mid\bfW\big]\leq\E\big[R^2\I(|R|>M)\mid\bfW\big].
\]
By (iii) choose $M_0$ with $\operatorname*{ess\,sup}_{\bfw}(\sigma^2-c_{M_0})\leq\sigma^2_{\min}/2$;
then $c_{M_0}\geq\sigma^2_{\min}/2$ uniformly by (ii). For bounded $a$, set
$$s_Y=a(\bfW)\,\xi_{M_0}/c_{M_0}(\bfW).$$ This score is bounded by
$4M_0\|a\|_\infty/\sigma^2_{\min}$, satisfies $\E[s_Y\mid\bfW]=0$, and gives $\E[Rs_Y\mid\bfW]=a$.
{\color{black}Thus, for any bounded $a$, this score gives $Q_t=Q+ta$ by Step 1.}

{\color{black}\emph{Step 5: approximating both score components.} We show that the closed linear span of the tangent set contains both $\mathcal T_{\bfW}$ and $\mathcal T_Y$. It suffices to consider bounded functions: in both $\mathcal T_{\bfW}$ and $\mathcal T_Y$, every element is an $L_2(P)$ limit of bounded elements of the same space.} For $h\in\mathcal T_{\bfW}$ the
truncations $\tau_M(h)-\E[\tau_M(h)]$ lie in $\mathcal T_{\bfW}$, and for $g\in\mathcal T_Y$ the
truncations $\tau_M(g)-\E[\tau_M(g)\mid\bfW]$ lie in $\mathcal T_Y$. Both are bounded and converge in
$L_2(P)$ by dominated convergence. By Step 2 every bounded element of $\mathcal T_{\bfW}$ lies in the
tangent set, so $\mathcal T_{\bfW}$ lies in its closed linear span.

{\color{black}It remains to approximate a bounded $g\in\mathcal T_Y$. Let $a_g=\E[Rg\mid\bfW]$. By condition (iii), for sufficiently large $M$,
\[
  \sigma^2(\bfW)\leq M^2+\operatorname*{ess\,sup}_{\bfw}
  \E[R^2\I(|R|>M)\mid\bfW=\bfw]<\infty
  \quad\text{almost surely}.
\]
Thus $\sigma$ is essentially bounded, and conditional Cauchy--Schwarz gives
\[
  |a_g|\leq \sigma(\bfW)\,\E[g^2\mid\bfW]^{1/2}
  \leq \|\sigma\|_\infty\|g\|_\infty .
\]}
{\color{black}Although $a_g$ is bounded, it may not be Lipschitz. The score constructed in Step 4 with $a=a_g$ induces exactly the same change in $Q$ as $g$. Subtracting it therefore leaves a score that preserves $Q$:}
\[
  g=\underbrace{\Big(g-\frac{a_g}{c_{M_0}}\xi_{M_0}\Big)}_{=:g_\perp}
    +\frac{a_g}{c_{M_0}}\xi_{M_0},
\]
{\color{black}where $g_\perp$ is bounded and satisfies $\E[Rg_\perp\mid\bfW]=a_g-a_g=0$, so $g_\perp$ lies in the
tangent set of $\mathcal P_L$ at $P$ by Step 2.} Given $\zeta>0$, choose a bounded $\tilde a$ with
$\|\tilde a-a_g\|_{L_2(P_{\bfW})}<\zeta$ that is Lipschitz in $\bfc$ within each stratum and vanishes
outside finitely many strata. {\color{black}First, set $a_g$ to zero outside finitely many strata chosen so that
the resulting $L_2(P_{\bfW})$ error is arbitrarily small. Within each retained stratum, approximate $a_g$
arbitrarily well in $L_2$ by a bounded Lipschitz function.} The finitely many Lipschitz constants have a common finite
maximum, as required in Step 3. {\color{black}The score $\tilde a\,\xi_{M_0}/c_{M_0}$ lies in the same tangent set by Steps 3 and 4,
and}
\[
  \Big\|g-\Big(g_\perp+\frac{\tilde a}{c_{M_0}}\xi_{M_0}\Big)\Big\|_{L_2(P)}
  =\Big\|\frac{a_g-\tilde a}{c_{M_0}}\xi_{M_0}\Big\|_{L_2(P)}
  \leq\frac{4M_0}{\sigma^2_{\min}}\,\zeta .
\]
{\color{black}Since $4M_0/\sigma_{\min}^2$ does not depend on $\zeta$, choosing $\zeta$ sufficiently small makes the $L_2(P)$ distance between $g$ and the sum of these two scores arbitrarily small. The closed linear span of the tangent set therefore contains $\mathcal T_Y$ as well as $\mathcal T_{\bfW}$, and hence equals $L_2^0(P)$.}
\end{proof}
\par\endgroup

\subsection{Proof of Lemma~\ref{lem: reg proj cov}}\label{pf: reg proj cov lemma}
\begingroup\color{black}
\begin{proof}
Fix a stratum and work in the rescaled coordinates where $\mathbf V=\mathbf I$. An exterior point $\mathbf p+r\mathbf n(\mathbf p)$ maps to $\mathbf y=\mathbf p-\phi^\epsilon(r)\mathbf n(\mathbf p)$. Since the displacement is less than the reach of $\partial\calH$, $\mathbf p$ is the unique nearest boundary point of $\mathbf y$ \citep[Theorem~4.8(12)]{federer1959curvature}, and $t=\operatorname{dist}(\mathbf y,\partial\calH)=\phi^\epsilon(r)$. Conversely, for any $\mathbf y\in\mathcal L_\epsilon$, its nearest boundary point $\mathbf p$ and depth $t$ give the unique exterior preimage
\[
    \bigl(\tilde\Pi_{\calH}^{\epsilon}\bigr)^{-1}(\mathbf y)=\mathbf p+(\phi^\epsilon)^{-1}(t)\,\frac{\mathbf p-\mathbf y}{t}.
\]
Thus the map is a bijection onto $\mathcal L_\epsilon$.

For an exterior point $\mathbf w$, write $\mathbf p(\mathbf w)=\Pi_{\calH}(\mathbf w)$ and $r(\mathbf w)=\operatorname{dist}(\mathbf w,\calH)$. The interior-displaced projection is
\[
    \tilde\Pi_{\calH}^{\epsilon}(\mathbf w)
    =\mathbf p(\mathbf w)-\phi^\epsilon(r(\mathbf w))\,
      \frac{\mathbf w-\mathbf p(\mathbf w)}{r(\mathbf w)}.
\]
Both $\mathbf p$ and $r$ are Lipschitz. Near any exterior point, $r$ is bounded away from zero, so the unit normal $(\mathbf w-\mathbf p(\mathbf w))/r(\mathbf w)$ is Lipschitz on that neighborhood. Since $\phi^\epsilon$ is continuously differentiable, the displacement distance $\phi^\epsilon(r(\mathbf w))$ is also locally Lipschitz. Thus $\tilde\Pi_{\calH}^{\epsilon}$ is locally Lipschitz on the exterior. For the inverse, positive reach makes the nearest boundary point locally Lipschitz in $\mathbf y$ \citep[Theorem~4.8(8)]{federer1959curvature}, and $(\phi^\epsilon)'>0$ makes $(\phi^\epsilon)^{-1}$ locally Lipschitz in $t$. The depth $t$ is a Lipschitz function of $\mathbf y$ and is bounded away from zero near each point of the interior layer. Thus every term in the formula for $(\tilde\Pi_{\calH}^{\epsilon})^{-1}$ is locally Lipschitz. The map is locally bi-Lipschitz, so its Jacobian and that of its inverse are nonsingular almost everywhere. The change-of-variables formula then gives a density for the displaced anchors whenever the exterior policy values have a density.
\end{proof}
\par\endgroup

\subsection{Proof of Lemma~\ref{lem: smooth boundary}}\label{pf: smooth boundary}
\begingroup\color{black}
\begin{proof}
Fix $\mathbf p\in\partial\calH_\rho$ and let $\mathbf q=\Pi_{\calH}(\mathbf p)$. By \eqref{eq: H rho}, $\|\mathbf p-\mathbf q\|=\rho$ and $\overline B(\mathbf q,\rho)\subseteq\calH_\rho$. Any supporting hyperplane of $\calH_\rho$ at $\mathbf p$ must also support this ball there, so the unique outward unit normal is $\mathbf n(\mathbf p)=(\mathbf p-\Pi_{\calH}(\mathbf p))/\rho$. Since projection onto a closed convex set is nonexpansive, this normal field is Lipschitz. Thus the convex boundary is $C^{1,1}$, proving (i).

For (ii), an exterior point has a unique nearest boundary point by convexity, and a boundary point is its own nearest point. Now let $\mathbf z\in\operatorname{int}(\calH_\rho)$ have distance $0<t<\rho$ from the boundary, and let $\mathbf p$ be any nearest boundary point. By (i), $\mathbf z=\mathbf p-t\mathbf n(\mathbf p)$. Writing $\mathbf q=\Pi_{\calH}(\mathbf p)$ gives
\[
    \mathbf z=\mathbf q+(\rho-t)\mathbf n(\mathbf p).
\]
Since $\rho-t>0$, $\mathbf z$ lies on the same outward normal ray from $\mathbf q$ as $\mathbf p$. The characterization of projection onto a closed convex set therefore gives $\mathbf q=\Pi_{\calH}(\mathbf z)$. Thus $\mathbf z$ determines $\mathbf q$, then $\mathbf n(\mathbf p)=(\mathbf z-\mathbf q)/\|\mathbf z-\mathbf q\|$, and hence $\mathbf p=\mathbf q+\rho\mathbf n(\mathbf p)$. The nearest boundary point is unique, so $\operatorname{reach}(\partial\calH_\rho)\geq\rho$.
\end{proof}
\par\endgroup

\clearpage


\clearpage
\begingroup
\begin{thebibliography}{}

\bibitem[Abadie and Imbens, 2006]{abadie2006large}
Abadie, A. and Imbens, G.~W. (2006).
\newblock Large sample properties of matching estimators for average treatment
  effects.
\newblock {\em Econometrica}, 74(1):235--267.

\bibitem[Antonelli and Zigler, 2024]{antonelli2024causal}
Antonelli, J. and Zigler, C. (2024).
\newblock Causal analysis of air pollution mixtures: Estimands, positivity, and
  extrapolation.
\newblock {\em Am. J. Epidemiol.}, 193(10):1392--1398.

\bibitem[Armstrong and Koles{\'a}r, 2021]{armstrong2021finite}
Armstrong, T.~B. and Koles{\'a}r, M. (2021).
\newblock Finite-sample optimal estimation and inference on average treatment
  effects under unconfoundedness.
\newblock {\em Econometrica}, 89(3):1141--1177.

\bibitem[Balkus et~al., 2026]{balkus2026causal}
Balkus, S.~V., Delaney, S.~W., and Hejazi, N.~S. (2026).
\newblock The causal effects of modified treatment policies under network
  interference.
\newblock {\em J. R. Stat. Soc. Ser. B Stat. Methodol.}, page qkag052.

\bibitem[Bao and Schomaker, 2025]{bao2025addressing}
Bao, H. and Schomaker, M. (2025).
\newblock Addressing positivity violations in continuous interventions through
  data-adaptive strategies.
\newblock arXiv:2502.14566.

\bibitem[Ben-Michael et~al., 2025]{ben2025safe}
Ben-Michael, E., Greiner, D.~J., Imai, K., and Jiang, Z. (2025).
\newblock Safe policy learning through extrapolation: Application to pre-trial
  risk assessment.
\newblock {\em Journal of the American Statistical Association},
  120(551):1386--1399.

\bibitem[Bibaut and {van der Laan}, 2017]{bibaut2017data}
Bibaut, A.~F. and {van der Laan}, M.~J. (2017).
\newblock Data-adaptive smoothing for optimal-rate estimation of possibly
  non-regular parameters.
\newblock arXiv:1706.07408.

\bibitem[Bickel et~al., 1993]{bickel1993efficient}
Bickel, P.~J., Klaassen, C.~A., Ritov, Y., and Wellner, J.~A. (1993).
\newblock {\em Efficient and Adaptive Estimation for Semiparametric Models},
  volume~4 of {\em Johns Hopkins Series in the Mathematical Sciences}.
\newblock Johns Hopkins University Press, Baltimore.

\bibitem[Bobb et~al., 2015]{bobb2015bayesian}
Bobb, J.~F., Valeri, L., Claus~Henn, B., Christiani, D.~C., Wright, R.~O.,
  Mazumdar, M., Godleski, J.~J., and Coull, B.~A. (2015).
\newblock Bayesian kernel machine regression for estimating the health effects
  of multi-pollutant mixtures.
\newblock {\em Biostatistics}, 16(3):493--508.

\bibitem[Branson et~al., 2023]{branson2023causal}
Branson, Z., Kennedy, E.~H., Balakrishnan, S., and Wasserman, L. (2023).
\newblock Causal effect estimation after propensity score trimming with
  continuous treatments.
\newblock arXiv:2309.00706.

\bibitem[Carlin et~al., 2013]{carlin2013unraveling}
Carlin, D.~J., Rider, C.~V., Woychik, R., and Birnbaum, L.~S. (2013).
\newblock Unraveling the health effects of environmental mixtures: an niehs
  priority.
\newblock {\em Environmental health perspectives}, 121(1):a6.

\bibitem[Chernozhukov et~al., 2018]{chernozhukov2018double}
Chernozhukov, V., Chetverikov, D., Demirer, M., Duflo, E., Hansen, C., Newey,
  W., and Robins, J. (2018).
\newblock Double/debiased machine learning for treatment and structural
  parameters.
\newblock {\em Econom. J.}, pages C1--C68.

\bibitem[Chernozhukov et~al., 2013]{chernozhukov2013intersection}
Chernozhukov, V., Lee, S., and Rosen, A.~M. (2013).
\newblock Intersection bounds: Estimation and inference.
\newblock {\em Econometrica}, 81(2):667--737.

\bibitem[Crump et~al., 2009]{crump2009dealing}
Crump, R.~K., Hotz, V.~J., Imbens, G.~W., and Mitnik, O.~A. (2009).
\newblock Dealing with limited overlap in estimation of average treatment
  effects.
\newblock {\em Biometrika}, 96(1):187--199.

\bibitem[D{\'\i}az and Hejazi, 2020]{diaz2020causal}
D{\'\i}az, I. and Hejazi, N.~S. (2020).
\newblock Causal mediation analysis for stochastic interventions.
\newblock {\em J. R. Stat. Soc. Ser. B Stat. Methodol.}, 82(3):661--683.

\bibitem[D{\'\i}az and van~der Laan, 2012]{diaz2012population}
D{\'\i}az, I. and van~der Laan, M. (2012).
\newblock Population intervention causal effects based on stochastic
  interventions.
\newblock {\em Biometrics}, 68(2):541--549.

\bibitem[D{\'\i}az et~al., 2023]{diaz2023nonparametric}
D{\'\i}az, I., Williams, N., Hoffman, K.~L., and Schenck, E.~J. (2023).
\newblock Nonparametric causal effects based on longitudinal modified treatment
  policies.
\newblock {\em J. Amer. Statist. Assoc.}, 118(542):846--857.

\bibitem[D’Amour et~al., 2021]{d2021overlap}
D’Amour, A., Ding, P., Feller, A., Lei, L., and Sekhon, J. (2021).
\newblock Overlap in observational studies with high-dimensional covariates.
\newblock {\em J. Econom.}, 221(2):644--654.

\bibitem[Eskenazi et~al., 2003]{eskenazi2003chamacos}
Eskenazi, B., Bradman, A., Gladstone, E.~A., Jaramillo, S., Birch, K., and
  Holland, N. (2003).
\newblock Chamacos, a longitudinal birth cohort study: lessons from the fields.
\newblock {\em Journal of Children's Health}, 1(1):3--27.

\bibitem[Federer, 1959]{federer1959curvature}
Federer, H. (1959).
\newblock Curvature measures.
\newblock {\em Transactions of the American Mathematical Society},
  93(3):418--491.

\bibitem[Haneuse and Rotnitzky, 2013]{haneuse2013estimation}
Haneuse, S. and Rotnitzky, A. (2013).
\newblock Estimation of the effect of interventions that modify the received
  treatment.
\newblock {\em Stat. Med.}, 32(30):5260--5277.

\bibitem[Hejazi et~al., 2022]{hejazi2022efficient}
Hejazi, N.~S., Benkeser, D., Diaz, I., and van~der Laan, M.~J. (2022).
\newblock Efficient estimation of modified treatment policy effects based on
  the generalized propensity score.
\newblock arXiv:2205.05777.

\bibitem[Hirano and Porter, 2012]{hirano2012impossibility}
Hirano, K. and Porter, J.~R. (2012).
\newblock Impossibility results for nondifferentiable functionals.
\newblock {\em Econometrica}, 80(4):1769--1790.

\bibitem[{Huang} et~al., 2026]{huang2026multivariate}
{Huang}, Z., {Dong}, K., {Lin}, T., and {Antonelli}, J. (2026).
\newblock Multivariate incremental effects for continuous treatments: Studying
  the health effects of environmental mixtures.
\newblock arXiv:2604.23534.

\bibitem[Hubbard et~al., 2016]{hubbard2016statistical}
Hubbard, A.~E., Kherad-Pajouh, S., and van~der Laan, M.~J. (2016).
\newblock Statistical inference for data adaptive target parameters.
\newblock {\em Int. J. Biostat.}, 12(1):3--19.

\bibitem[Hyndman, 1996]{hyndman1996computing}
Hyndman, R.~J. (1996).
\newblock Computing and graphing highest density regions.
\newblock {\em Amer. Statist.}, 50(2):120--126.

\bibitem[Imbens and Manski, 2004]{imbens2004confidence}
Imbens, G.~W. and Manski, C.~F. (2004).
\newblock Confidence intervals for partially identified parameters.
\newblock {\em Econometrica}, 72(6):1845--1857.

\bibitem[Kennedy, 2019]{kennedy2019nonparametric}
Kennedy, E.~H. (2019).
\newblock Nonparametric causal effects based on incremental propensity score
  interventions.
\newblock {\em J. Amer. Statist. Assoc.}, 114(526):645--656.

\bibitem[Kennedy, 2024]{kennedy2024semiparametric}
Kennedy, E.~H. (2024).
\newblock Semiparametric doubly robust targeted double machine learning: A
  review.
\newblock {\em Handbook of Statistical Methods for Precision Medicine}, pages
  207--236.

\bibitem[Khan et~al., 2024]{khan2024off}
Khan, S., Saveski, M., and Ugander, J. (2024).
\newblock Off-policy evaluation beyond overlap: Sharp partial identification
  under smoothness.
\newblock In {\em Proc. 41st Int. Conf. Mach. Learn.}, volume 235 of {\em
  Proceedings of Machine Learning Research}, pages 23734--23757. PMLR.

\bibitem[Levis et~al., 2025]{levis2025covariate}
Levis, A.~W., Bonvini, M., Zeng, Z., Keele, L., and Kennedy, E.~H. (2025).
\newblock Covariate-assisted bounds on causal effects with instrumental
  variables.
\newblock {\em J. R. Stat. Soc. Ser. B Stat. Methodol.}, 87(5):1508--1527.

\bibitem[Ma et~al., 2025]{ma2025sensitivity}
Ma, Y., Huang, Y., and Namkoong, H. (2025).
\newblock A sensitivity approach to causal inference under limited overlap.
\newblock arXiv:2511.22003.

\bibitem[McShane, 1934]{mcshane1934extension}
McShane, E.~J. (1934).
\newblock Extension of range of functions.
\newblock {\em Bull. Amer. Math. Soc.}, 40(12):837--842.

\bibitem[Montoya et~al., 2022]{montoya2022estimators}
Montoya, L.~M., van~der Laan, M.~J., Skeem, J.~L., and Petersen, M.~L. (2022).
\newblock Estimators for the value of the optimal dynamic treatment rule with
  application to criminal justice interventions.
\newblock {\em Int. J. Biostat.}, 19(1):239.

\bibitem[Pfister and B{\"u}hlmann, 2026]{pfister2026extrapolation}
Pfister, N. and B{\"u}hlmann, P. (2026).
\newblock Extrapolation-aware nonparametric statistical inference.
\newblock {\em Journal of Machine Learning Research}, 27(150):1--59.

\bibitem[Robins et~al., 2007]{robins2007comment}
Robins, J., Sued, M., Lei-Gomez, Q., and Rotnitzky, A. (2007).
\newblock Comment: Performance of double-robust estimators when ``inverse
  probability" weights are highly variable.
\newblock {\em Statistical Science}, 22(4):544--559.

\bibitem[Robins et~al., 2004]{robins2004effects}
Robins, J.~M., Hern{\'a}n, M.~A., and SiEBERT, U. (2004).
\newblock Effects of multiple interventions.
\newblock {\em Comparative quantification of health risks: global and regional
  burden of disease attributable to selected major risk factors}, 1:2191--2230.

\bibitem[Rudolph et~al., 2026]{rudolph2026everything}
Rudolph, K.~E., Inose, S., Williams, N.~T., D{\'\i}az, I., Calderon, L.,
  Torres, J.~M., and Kioumourtzoglou, M.-A. (2026).
\newblock Everything all at once: On choosing an estimand for multi-component
  environmental exposures.
\newblock {\em Epidemiology}.

\bibitem[Sarvet and Stensrud, 2026]{sarvet2026natural}
Sarvet, A.~L. and Stensrud, M.~J. (2026).
\newblock The natural value of treatment and its importance for causal
  inference.
\newblock {\em Annual Review of Statistics and Its Application}, 13:439--463.

\bibitem[Schindl et~al., 2026]{schindl2026incremental}
Schindl, K., Shen, S., and Kennedy, E.~H. (2026).
\newblock Incremental effects for continuous exposures.
\newblock {\em J. Amer. Statist. Assoc.}, 0(ja):1--69.

\bibitem[Schneider, 2014]{schneider2014convex}
Schneider, R. (2014).
\newblock {\em Convex Bodies: The Brunn--Minkowski Theory}.
\newblock Cambridge University Press, 2 edition.

\bibitem[Semenova, 2025]{semenova2025debiased}
Semenova, V. (2025).
\newblock Debiased machine learning of aggregated intersection bounds and other
  causal parameters.
\newblock {\em arXiv preprint arXiv:2303.00982}.

\bibitem[Stoye, 2009]{stoye2009more}
Stoye, J. (2009).
\newblock More on confidence intervals for partially identified parameters.
\newblock {\em Econometrica}, 77(4):1299--1315.

\bibitem[Susmann et~al., 2025]{susmann2025non}
Susmann, H.~P., McClean, A., and D{\'\i}az, I. (2025).
\newblock Non-overlap average treatment effect bounds.
\newblock arXiv:2509.20206.

\bibitem[{van der Laan} and Luedtke, 2015]{vanderlaan2015targeted}
{van der Laan}, M.~J. and Luedtke, A.~R. (2015).
\newblock Targeted learning of the mean outcome under an optimal dynamic
  treatment rule.
\newblock {\em J. Causal Inference}, 3(1):61--95.

\bibitem[{van der Laan} et~al., 2007]{vanderlaan2007super}
{van der Laan}, M.~J., Polley, E.~C., and Hubbard, A.~E. (2007).
\newblock Super learner.
\newblock {\em Stat. Appl. Genet. Mol. Biol.}, 6(1):1--23.

\bibitem[Whitney, 1934]{whitney1934analytic}
Whitney, H. (1934).
\newblock Analytic extensions of differentiable functions defined in closed
  sets.
\newblock {\em Trans. Amer. Math. Soc.}, 36(1):63--89.

\bibitem[Yang and Ding, 2018]{yang2018asymptotic}
Yang, S. and Ding, P. (2018).
\newblock Asymptotic inference of causal effects with observational studies
  trimmed by the estimated propensity scores.
\newblock {\em Biometrika}, 105(2):487--493.

\bibitem[Young et~al., 2014]{young2014identification}
Young, J.~G., Hern{\'a}n, M.~A., and Robins, J.~M. (2014).
\newblock Identification, estimation and approximation of risk under
  interventions that depend on the natural value of treatment using
  observational data.
\newblock {\em Epidemiol. Methods}, 3(1):1--19.

\end{thebibliography}

\endgroup
\end{document}